\documentclass[manuscript]{aastex63}

\usepackage{gensymb}
\usepackage{amsmath}
\usepackage{array}
\usepackage{graphicx}
\newcolumntype{L}{>{\raggedright\arraybackslash}p{8.8cm}}

\graphicspath{{./}{figures/}}

\received{\today}
\submitjournal{Astrophysical Journal Supplement}
\shorttitle{CatWISE Overview}
\shortauthors{Eisenhardt et al.}
\begin{document}
\title{The CatWISE Preliminary Catalog: Motions  from {\it WISE} and {\it NEOWISE} Data}
\correspondingauthor{Peter R. M. Eisenhardt.}
\email{Peter.R.Eisenhardt@jpl.nasa.gov}

\author{Peter R. M. Eisenhardt}
\affiliation{Jet Propulsion Laboratory, California Institute of Technology, 4800 Oak Grove Drive, M/S 169-327, Pasadena, CA 91109, USA}

\author[0000-0001-7519-1700]{Federico Marocco}\altaffiliation{NASA Postdoctoral Program Fellow}
\affiliation{Jet Propulsion Laboratory, California Institute of Technology, 4800 Oak Grove Drive, M/S 169-327, Pasadena, CA 91109, USA}
\affiliation{IPAC, Mail Code 100-22, California Institute of Technology, 1200 E. California Blvd., Pasadena, CA 91125, USA}

\author{John W. Fowler}
\affiliation{230 Pacific St., Apt. 205, Santa Monica, CA 90405, USA}

\author[0000-0002-1125-7384]{Aaron M. Meisner}\altaffiliation{Hubble Fellow}
\affiliation{National Optical Astronomy Observatory, 950 N. Cherry Ave., Tucson, AZ 85719, USA}

\author[0000-0003-4269-260X]{J. Davy Kirkpatrick}
\affiliation{IPAC, Mail Code 100-22, California Institute of Technology, 1200 E. California Blvd., Pasadena, CA 91125, USA}

\author{Nelson Garcia}
\affiliation{IPAC, Mail Code 100-22, California Institute of Technology, 1200 E. California Blvd., Pasadena, CA 91125, USA}

\author[0000-0002-4939-734X]{Thomas H. Jarrett}
\affiliation{Department of Astronomy, University of Cape Town, Private Bag X3, Rondebosch, 7701, South Africa} 

\author{Renata Koontz}
\affiliation{University of California, Riverside, 900 University Ave, Riverside, CA 92521, USA}

\author{Elijah J. Marchese}
\affiliation{University of California, Riverside, 900 University Ave, Riverside, CA 92521, USA}

\author{S. Adam Stanford}
\affiliation{Department of Physics, University of California Davis, One Shields Avenue, Davis, CA 95616, USA}

\author{Dan Caselden}
\affiliation{Gigamon Applied Threat Research, 619 Western Avenue, Suite 200, Seattle, WA 98104, USA}

\author[0000-0001-7780-3352]{Michael C. Cushing}
\affiliation{Department of Physics and Astronomy, University of Toledo, 2801 West Bancroft St., Toledo, OH 43606, USA}

\author{Roc M. Cutri}
\affiliation{IPAC, Mail Code 100-22, California Institute of Technology, 1200 E. California Blvd., Pasadena, CA 91125, USA}

\author[0000-0001-6251-0573]{Jacqueline K. Faherty}
\affiliation{Department of Astrophysics, American Museum of Natural History, Central Park West at 79th Street, NY 10024, USA}

\author{Christopher R. Gelino}
\affiliation{IPAC, Mail Code 100-22, California Institute of Technology, 1200 E. California Blvd., Pasadena, CA 91125, USA}

\author{Anthony H. Gonzalez}
\affiliation{Department of Astronomy, University of Florida, 211 Bryant Space Center, Gainesville, FL 32611, USA}

\author{Amanda Mainzer}
\affiliation{Jet Propulsion Laboratory, California Institute of Technology, 4800 Oak Grove Drive, M/S 169-327, Pasadena, CA 91109, USA}

\author{Bahram Mobasher}
\affiliation{University of California, Riverside, 900 University Ave, Riverside, CA 92521}

\author[0000-0002-5042-5088]{David J. Schlegel}
\affiliation{Lawrence Berkeley National Laboratory, Berkeley, CA, 94720, USA}

\author{Daniel Stern}
\affiliation{Jet Propulsion Laboratory, California Institute of Technology, 4800 Oak Grove Drive, M/S 169-327, Pasadena, CA 91109, USA}

\author{Harry I. Teplitz}
\affiliation{IPAC, Mail Code 100-22, California Institute of Technology, 1200 E. California Blvd., Pasadena, CA 91125, USA}

\author[0000-0001-5058-1593]{Edward L. Wright}
\affiliation{Department of Physics and Astronomy, UCLA, 430 Portola Plaza, Box 951547, Los Angeles, CA 90095-1547, USA}

\begin{abstract}

 CatWISE is a program to catalog sources selected from combined {\it WISE} and {\it NEOWISE}  all-sky survey data at 3.4 and 4.6 \micron\ (W1 and W2). The CatWISE Preliminary Catalog consists of  900,849,014 sources measured in data collected from 2010 to 2016. This dataset represents four times as many exposures and spans over ten times as large a time baseline as that used for the AllWISE Catalog. CatWISE adapts AllWISE software to measure the sources in coadded images created from six-month subsets of these data, each representing one coverage of the inertial sky, or epoch. The catalog includes the measured motion of sources in 8 epochs over the 6.5 year span of the data. From comparison to {\it Spitzer}, the SNR=5 limits in magnitudes in the Vega system are W1=17.67 and W2=16.47, compared to W1=16.96 and W2=16.02 for AllWISE. From comparison to {\it Gaia}, CatWISE positions have typical accuracies of 50\,mas for stars at W1=10\,mag and 275\,mas for stars at W1=15.5\,mag. Proper motions have typical accuracies of  10\,mas\,yr$^{-1}$ and 30\,mas\,yr$^{-1}$ for stars with these brightnesses, an order of magnitude better than from AllWISE. The catalog is available in the WISE/NEOWISE Enhanced and Contributed Products area of the NASA/IPAC Infrared Science Archive. 
 
\end{abstract}

\keywords{catalogs, infrared:stars, proper motions, brown dwarfs}

\section{Introduction \label{sec:intro}}

NASA's {\it Wide-field Infrared Survey Explorer} mission \citep[{\it WISE};][]{Wright2010} revealed iconic objects, including the first Earth Trojan asteroid \citep{Connors2011}, the closest and coolest brown dwarfs \citep{Luhman2013,Luhman2014}, and the most luminous galaxy yet found in the Universe \citep{Tsai2015}. These discoveries were made using two or more infrared coverages of the sky (or epochs)  obtained from January 2010 to February 2011. Each epoch typically consists of a dozen exposures per band taken within two days at a given position. The satellite was reactivated as {\it NEOWISE} and resumed searching for near-Earth objects in December 2013 \citep{Mainzer2014}, and has continued to cover the sky every six months since then.

In November 2013, the AllWISE release \citep{Cutri2013} made available to the community an atlas from coadding the two dozen exposures per position from the initial year of WISE surveying, and a catalog of source fluxes and positions measured from those exposures. With at least two  epochs per inertial position, AllWISE also provided motion estimates, and  became the definitive catalog in the WISE bandpasses. Each year, {\it NEOWISE} releases the individual exposures from the reactivated survey \citep{Cutri2015}, corresponding to two additional epochs. With the 2019 April 10 {\it NEOWISE} release, exposures from 12 epochs are now available. \citet{Meisner2018a} used unWISE processing \citep{Lang2014} to produce an image atlas which combines the 2010 and 2011 data used for AllWISE with the 2013 through 2016 {\it NEOWISE} data. An obvious next step is to catalog the sources revealed in these combined exposures.  

\begin{figure*}
\includegraphics[width=\textwidth]{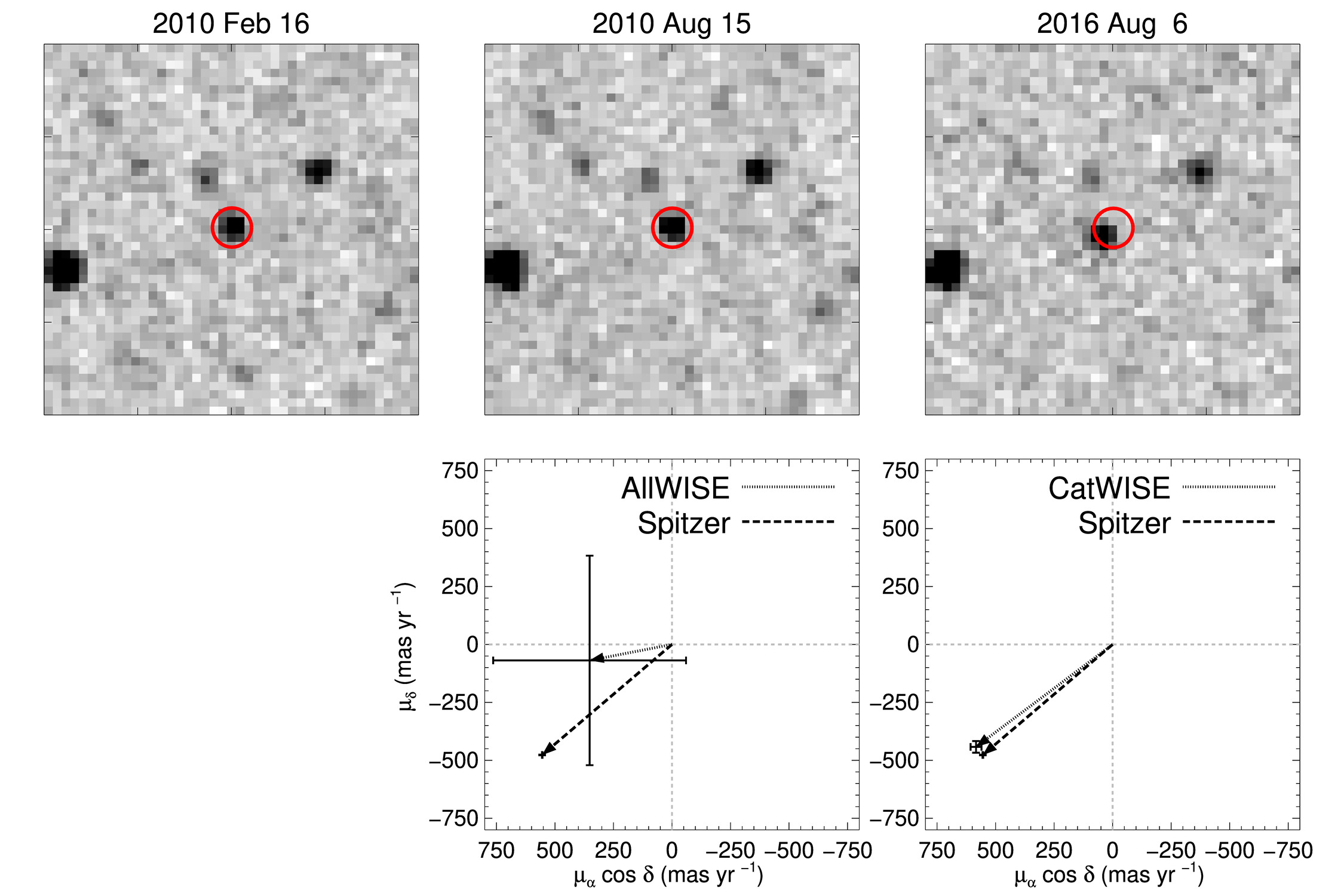}
\caption{(top) Proper motions that are insignificant in AllWISE become obvious with the increased time baseline of CatWISE. Discovered via its color in WISE data (\citet{Gelino2011}; \citet{Kirkpatrick2011}), WISEA~J161441.47+173935.4 is a T9 brown dwarf at 10 pc with a proper motion of $\mu_\alpha = 554.5\pm1.2$ mas yr$^{-1}, \mu_\delta = -476.7\pm1.2$ mas yr$^{-1}$ as determined by \textit{Spitzer} \citep{Kirkpatrick2019}. With a 6-month baseline, AllWISE (top left and center) measures a motion consistent with zero ($\mu_\alpha = 352\pm412$ mas yr$^{-1}, \mu_\delta = -69\pm452$ mas yr$^{-1}$; bottom center), while with the inclusion of \textit{NEOWISE} data from 2016 (top right), the corresponding source in the CatWISE Preliminary Catalog (CWISEP~J161441.63+173933.7) has a highly significant motion ($\mu_\alpha = 583.6\pm23.5$ mas yr$^{-1}, \mu_\delta = -442.0\pm25.0$ mas yr$^{-1}$; bottom right) consistent with the \textit{Spitzer} values. The top panels show 2$'\times2'$ cutouts in W2 centered on the \textit{WISE} position of the source at the first epoch, marked in all 3 epochs by a red circle.    \label{fig:wise1614}}
\end{figure*}

The unWISE Catalog \citep{Schlafly2019} uses a crowded-field point-source photometry code called ``crowdsource'' to do this, measuring source fluxes and positions in the coadded image, with the measurements at 3.4 \micron\ (W1) and 4.6 \micron\ (W2) carried out independently. In contrast, CatWISE has adapted AllWISE software to produce a full-sky catalog of sources selected simultaneously in both W1 and W2, and also provides motion estimates. For the CatWISE Preliminary Catalog described in this paper, sources were selected from the ensemble of 8 epochs of {\it WISE} and {\it NEOWISE} data from \citet[][the ``full coadd"]{Meisner2018a}, and the least-squares best-fit solution for point source flux, position and motion were determined from measurements on the individual ``epoch coadd" images \citep{Meisner2018c} rather than individual exposures, to reduce computational cost. Because the total time spanned by CatWISE epochs at a given inertial position is over 6 years, compared to a typical value of 6 months for AllWISE, the CatWISE motion estimates are far more accurate. This in addition to the greater depth of the CatWISE catalog relative to AllWISE are being used to extend the census of the coldest brown dwarfs in the solar neighborhood and enable definitive measurement of the form of the low-mass end of the star formation process \citep{Kirkpatrick2019}. Figure~\ref{fig:wise1614} illustrates an example of the potential of CatWISE to progress in this area.

In \S\ref{sec:obs} of this paper we summarize relevant aspects of the {\it WISE} and {\it NEOWISE} mission phases.  \S\ref{sec:processing} describes the CatWISE processing steps.  \S\ref{sec:performance} assesses the astrometric and photometric performance of CatWISE using comparisons to {\it Gaia} and {\it Spitzer} data.  \S\ref{sec:science} provides some initial examples of science results enabled by CatWISE, and \S\ref{sec:access} provides information on accessing CatWISE data products. Appendices provide additional information on CatWISE column entries, cautions on known issues in the CatWISE Preliminary Catalog, and details on how positions were combined from alternating survey scan directions (\S \ref{sec:obs}). The CatWISE website is https://catwise.github.io.

\section{Observations \label{sec:obs}}

{\it WISE} was launched on 2009 December 14, with its 40-cm telescope cooled to 12 K by an outer cryostat tank, and the W1 and W2 detectors operated at 32 K. The 12 and 22 \micron\ (W3 and W4) detectors were cooled to 7.8 K by an inner cryostat tank. Both tanks were filled with frozen hydrogen. The cryostat cover was ejected and the first images obtained on 2009 December 31, and science survey data were taken starting on 2010 January 7. 

 {\it WISE} uses dichroics to image the same $47 \times 47$ arcmin region of sky simultaneously in all bands using $1024 \times 1024$ pixel arrays with 2\farcs75 pixels, and obtaining exposures every 11 seconds. The duration of each W1 and W2 exposure is 7.7 seconds due to readout time and scan mirror settling, and discarding the initial readout.
 
 The Sun-synchronous polar {\it WISE} orbit was designed to precess so that the satellite stays over the Earth's terminator. The basic {\it WISE} survey strategy is to point near the zenith, scanning at the orbital rate along lines of ecliptic longitude, with the image motion compensated by a scan mirror that flies back for the start of each exposure.  The precession rate sweeps across the imaging field of view in approximately 12 orbits (less than one day) near the ecliptic, although the detailed survey strategy extends this time to more than a day. Because images are obtained on both sides of the orbit, the same region of inertial sky is covered every six months, with the direction of the survey scans alternating between ascending and descending in ecliptic latitude. CatWISE processes ascending and descending scans separately for source measurement.  

\begin{figure}
    \centering
    \includegraphics[width=1.\textwidth,trim={0 5cm 0 10cm}, clip]{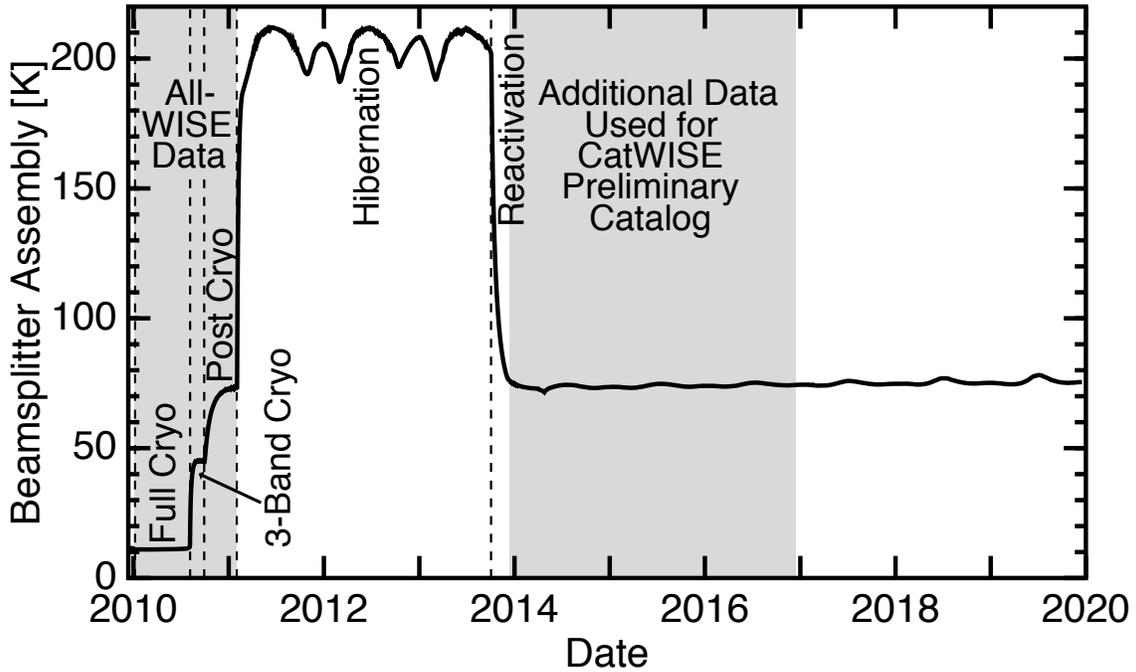}
    \caption{Temperature of the {\it WISE} beamsplitter assembly vs. date.  This temperature is close to that of the telescope and W1 and W2 detectors. Dashed lines indicate the transitions from full cryogenic to 3-band cryogenic phases, to the post-cryogenic phase, and to the start of hibernation and reactivation. Grey shading indicates the date ranges of data used for the CatWISE Preliminary Catalog.  AllWISE included only the left portion of the shaded range. }
    \label{fig:teltemp}
\end{figure}

The fully cryogenic survey continued until the hydrogen in the outer tank evaporated on 2010 August 6, exceeding the 7-month cryogen lifetime requirement and covering the sky 1.2 times (Figure \ref{fig:teltemp}). WISE then surveyed an additional 30\% of the sky during its 3-Band (W1, W2, and W3) cryogenic phase, with the W3 detector operating at reduced sensitivity. After the cryogen in both tanks was exhausted on 2010 September 29, the post-cryogenic {\it NEOWISE} survey \citep{Mainzer2011} in W1 and W2 began, as the telescope and detectors warmed to $\sim73$ K. Surveying continued until 2011 February 1, completing a second pass over the sky in W1 and W2, after which the satellite was placed into hibernation. In September 2013, the spacecraft was brought out of hibernation, where it had equilibrated at $\sim200$ K due to thermal radiation from the Earth, and renamed {\it NEOWISE}.  On 2013 December 13, with the telescope and detectors passively cooled below 76 K, {\it NEOWISE} resumed surveying the sky every 6 months in W1 and W2 \citep{Mainzer2014}.

The CatWISE Preliminary Catalog is based on the combination of W1 and W2 exposures in the two sky coverages used for the AllWISE data release \citep{Cutri2013} and in the six additional sky coverages from the 2017 {\it NEOWISE} data release.  For PSF purposes (\S\ref{sec:psf}), we use MJD 55480 (2010 Oct. 11) as the dividing point between cryogenic and post-cryogenic data. The average observation date is closer to MJD 56700 (2014 Feb. 12), which we adopt as the epoch for reporting positions when solving for source motion in the Preliminary Catalog. Source positions, whether incorporating source motion or not, are given in the equinox J2000 coordinate frame. Magnitudes in the catalog and throughout this paper are on the Vega system.

\section{CatWISE Processing \label{sec:processing}}

CatWISE adapts the AllWISE pipeline to detect and measure source fluxes in the combined WISE and {\it NEOWISE} images provided by unWISE. A full description of the AllWISE pipeline is provided in \citet{Cutri2013}. 

CatWISE processing works in the atlas tile footprint established by the WISE All-Sky Release, dividing the inertial sky into 18,240 overlapping square images (tiles), each approximately 1.56\degree\ on a side, aligned with the local right ascension and declination. Except for the ``primary flagging" step (\S \ref{sec:primary}), processing steps were carried out independently on each tile. For the Preliminary Catalog, CatWISE uses the full depth unWISE coaddition of 8 epochs \citep{Meisner2018a}, and the 8 individual unWISE epoch coadds \citep[][after astrometric modifications described in \S \ref{sec:unwise}]{Meisner2018c}. The full depth coadds and the PSF (\S\ref{sec:psf}) were used to create  detection images.  Sources were detected in these images  (\S\ref{sec:mdet}), while source properties were determined from measurements on the epoch coadds, treating ascending and descending epochs (\S\ref{sec:obs}) separately for most tiles and then merging the results (\S\ref{sec:wphot}). Potential artifacts affecting the sources were identified (\S\ref{sec:artifacts}), and sources with multiple measurements because they were in the tile overlap region were flagged to indicate which set of measurements should be used (\S\ref{sec:primary}). Finally, sources were selected for inclusion in the catalog or reject files (\S\ref{sec:prelim}). We describe these steps in more detail below. 

\subsection{unWISE Coadds \label{sec:unwise}}

The coadded images in the atlas released with the AllWISE (and All-Sky) Catalogs were primarily intended to facilitate source detection, and for this reason they are convolved with the WISE PSF. The unWISE coadds retain the resolution of individual WISE exposures. To reduce differences from the AllWISE processing approach, for source detection ($\S$\ref{sec:mdet}) in the CatWISE Preliminary Catalog, we convolved the masked version \citep{Lang2014} of the full-depth unWISE coadds {Meisner2018a} with the PSF ($\S$\ref{sec:psf}).  For source measurement ($\S$\ref{sec:wphot}) we used the masked version of the unWISE epoch coadds \citep{Meisner2018c} without convolution. 

An adjustment was made to the world coordinate system (WCS) for the pre-hibernation unWISE epoch coadds. Although all of the released individual exposures are tied to 2MASS \citep{Skrutskie2006}, the individual pre-hibernation exposures released for AllWISE are not on the same astrometric system as the individual post-hibernation exposures released for {\it NEOWISE}.  The {\it NEOWISE} images include corrections for the motions of the 2MASS reference stars, as does the AllWISE Catalog (see \S V.2.b of the AllWISE Explanatory Supplement;  \citet{Cutri2013}), but the released pre-hibernation images do not include those corrections.  A table of corrections for these images exists, however\footnote{\url{http://wise2.ipac.caltech.edu/docs/doc_tree/sis/rex19}}, and these corrections were applied to the unWISE input, making the astrometric system of the pre-hibernation epoch coadds consistent with those from the post-hibernation epoch coadds.  

\subsection{Point Spread Function (PSF) \label{sec:psf}}

The {\it WISE} and {\it NEOWISE} PSFs have been well characterized at the individual exposure level, but CatWISE works with coadded images rather than individual exposures, so PSFs appropriate for these coadded images are needed. The W1 and W2 PSFs vary with position in the focal plane, and changed somewhat between the cryogenic and post-cryogenic phases of the mission, particularly for W1. The PSFs are also asymmetric, with an orientation that is fixed with respect to the focal plane, but the focal plane orientation with respect to equatorial coordinates varies between exposures. The focal plane orientation also flips every six months because at a given inertial location the survey scan direction alternates between ascending and descending in ecliptic latitude (\S \ref{sec:obs}). CatWISE addressed these issues as follows. 

\begin{figure}
    \centering
    \includegraphics[width=0.32\textwidth,trim={3cm 8cm 0 1cm}, clip]{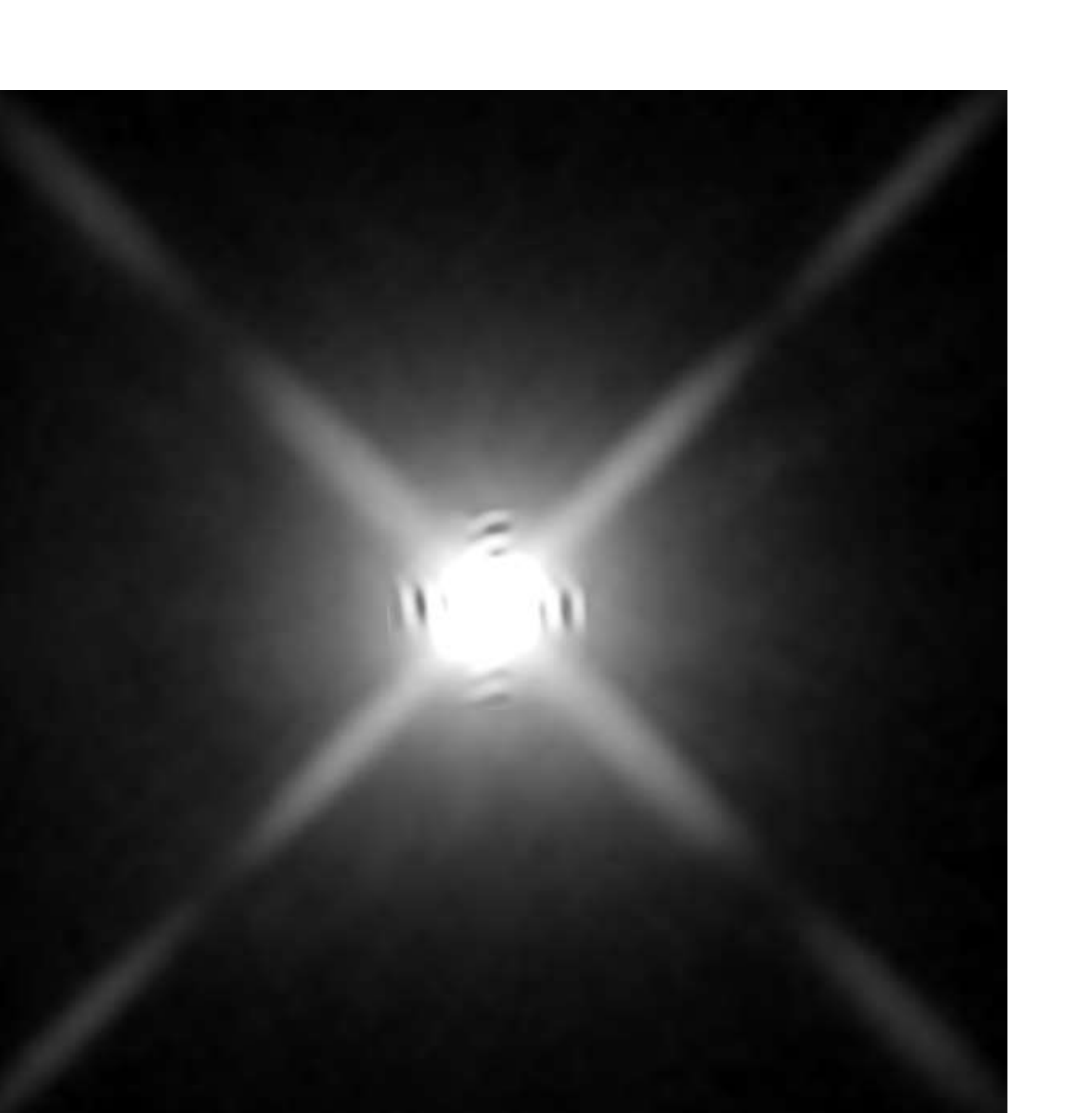}
    \includegraphics[width=0.32\textwidth,trim={3cm 8cm 0 1cm}, clip]{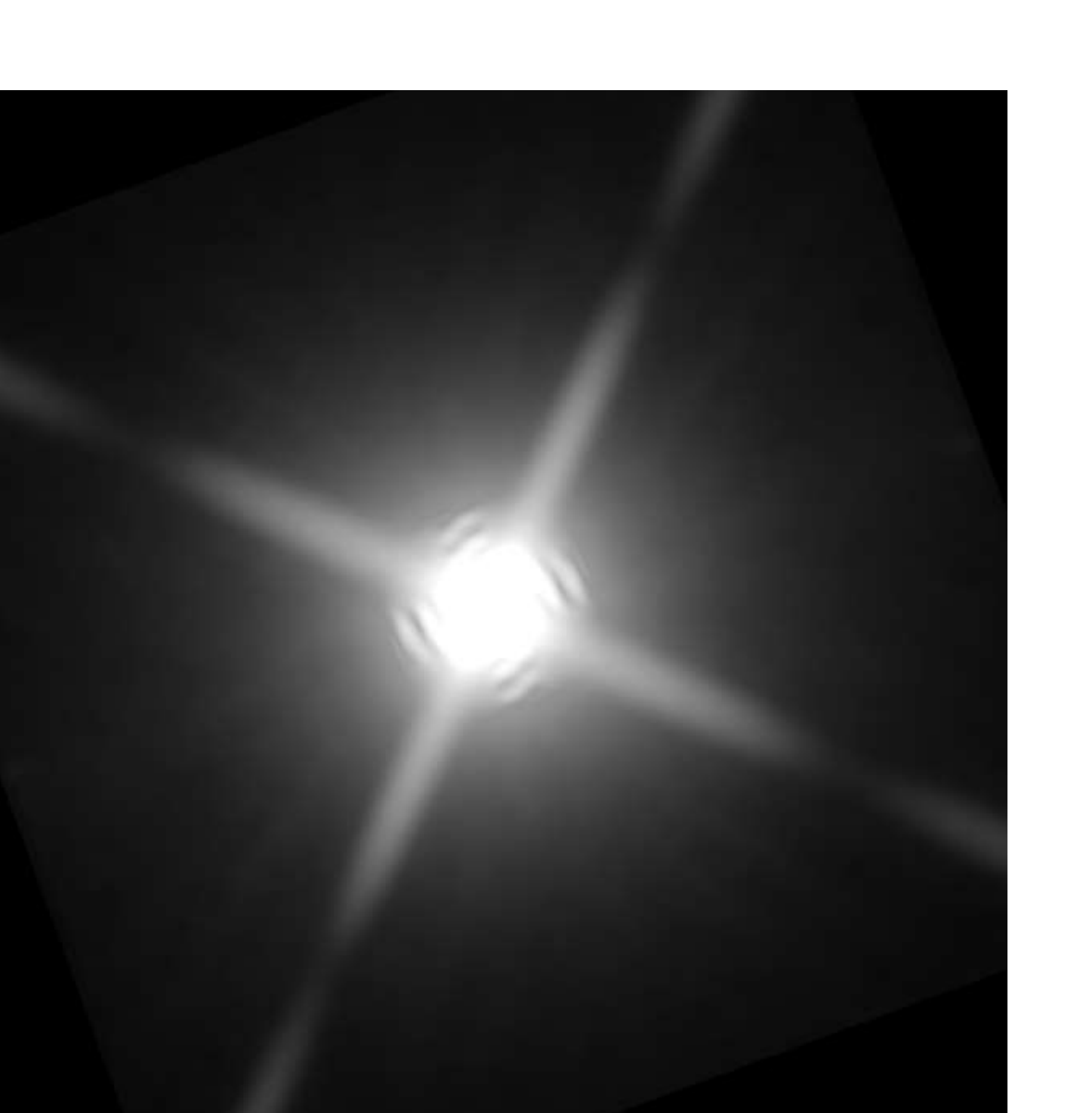}
    \includegraphics[width=0.31\textwidth,trim={0 0 0 0}, clip]{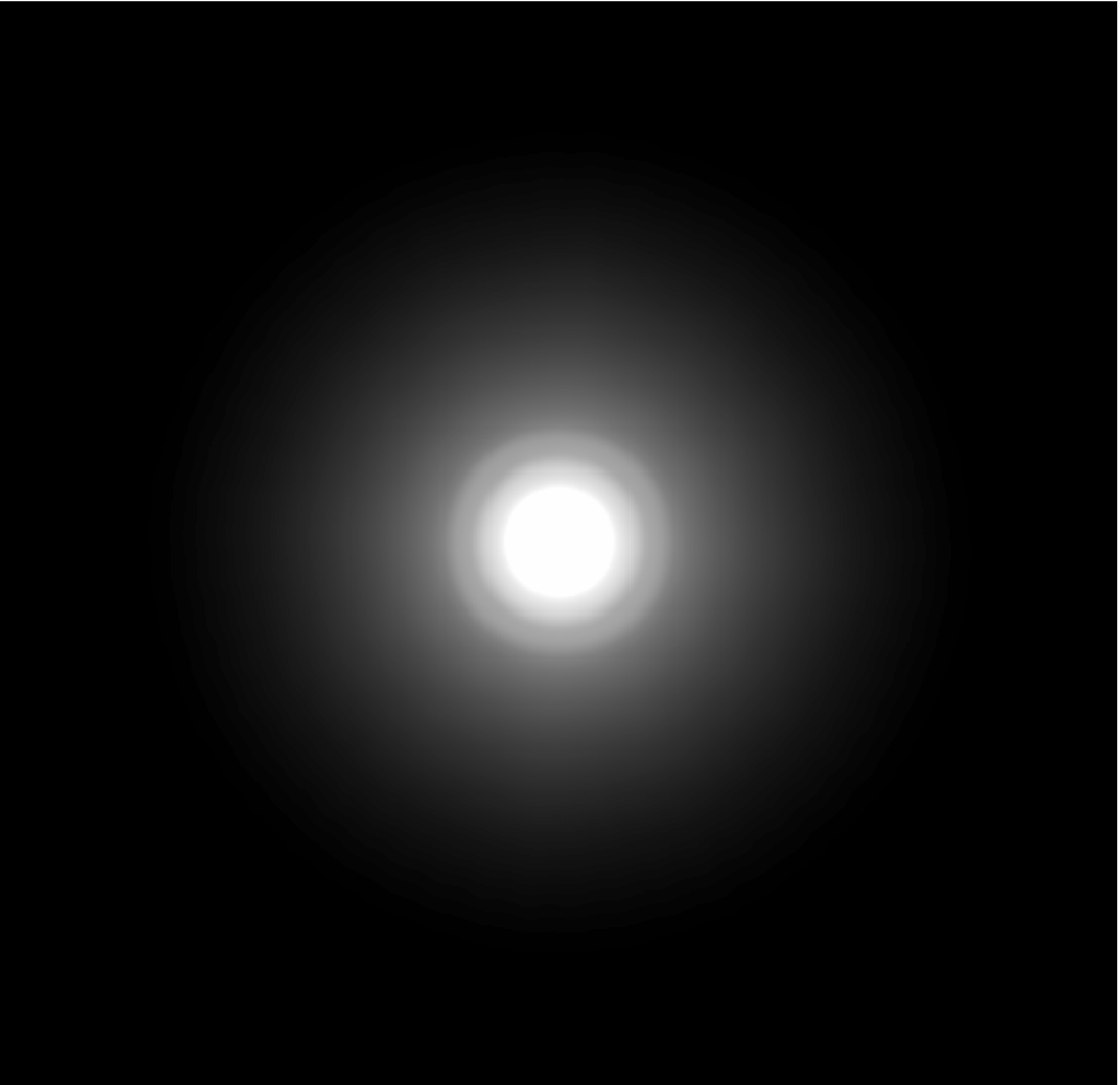}
    \caption{{\it Left}: The ``basic" (averaged over focal plane position) post-cryogenic PSF model for W1. The grey scale uses a log stretch from 1\% to 99\% of the peak value. The model extends to $\pm110''$. {\it Center}: The ascending scan W1 PSF for tile 1497p015 after averaging over cryogenic and post-cryogenic epochs, and a PA range from 199.93\degree to 200.56\degree. This is illustrative of the type of PSF used by CatWISE for source measurement, with the exception of 50 tiles near the ecliptic poles. {\it Right}: The W1 PSF averaged over all scans for tile 0890m667, which includes the south ecliptic pole and all PAs. This is illustrative of the type of PSF used by CatWISE for source measurement for 50 tiles  near the ecliptic poles.}
    \label{fig:w1psf}
\end{figure}

A $9\times9$ grid over the focal plane of cryogenic phase model PSFs and their uncertainties, $8\times$ oversampled relative to {\it WISE} pixels and covering 220'' on a side, is provided in \S IV.4.c.iii.1 of the {\it WISE} All-Sky Explanatory Supplement \citep{Cutri2012}. The post-cryogenic PSFs used for {\it NEOWISE} are given in \S IV.2.b.i of the NEOWISE Explanatory Supplement \citep{Cutri2015}. Since many focal plane positions contribute to each source in the coadded images, CatWISE averaged these model PSFs, weighted only by their partition sizes in the $9\times9$ grid. The uncertainty in the averaged PSF was taken to be the root sum square of two terms: the root mean square of the individual grid PSF uncertainties divided by $\sqrt{N - 1}$ where N is the 81 PSF's, and the standard deviation of the grid PSF's about the averaged PSF. Figure \ref{fig:w1psf} (left) shows the resulting focal plane average post-cryogenic PSF model for W1. An analogous focal plane average cryogenic PSF was created for W1, as well as average cryogenic and post-cryogenic PSFs for W2. These are referred to below as the ``basic" PSFs.

\begin{figure}
    \centering
    \includegraphics[width=0.5\textwidth,trim={0 7cm 0 1cm}, clip]{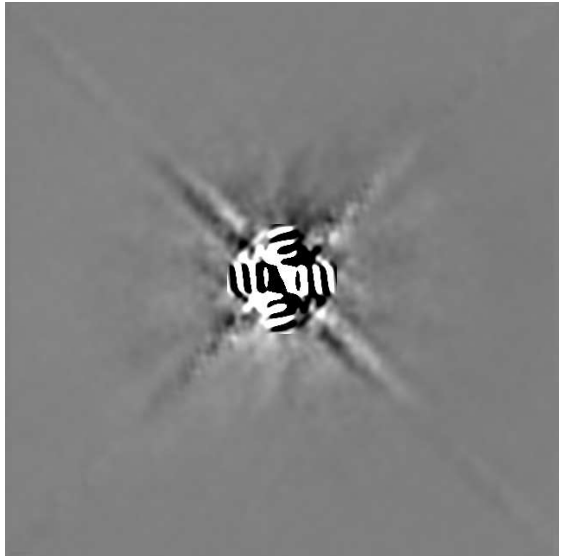}
    \caption{Difference image of W1 post-cryogenic PSFs in ascending vs. descending scans. The image extends to $\pm110''$.}
    \label{fig:asymmetry}
\end{figure}

As can also be seen in the AllWISE PSF's plotted in Figure 1 of \citet{Lang2016}, the PSFs contain fine structure in the core that is qualitatively consistent with the expected diffraction pattern for WISE. This structure is not rotationally symmetric.  Figure \ref{fig:asymmetry} shows the difference between ascending and descending orientations of the W1 post-cryogenic PSF. The bright and dark stripes in and around the core correspond to variations of roughly 25\% of the peak of the PSF. The center of the PSF is the most crucial region for position estimation, so inaccuracies in this part of the PSF have serious consequences for astrometry, as well as photometry.

These structures are not due to misalignment of the ascending and descending PSF's, which would need to be $\sim 1\farcs0$ to account for the features -- far greater than the difference in astrometry from the ascending vs. descending PSF's. For example, the median difference in source astrometry for tile 0895p227 (where ecliptic longitude and right ascension align) was 26.5 mas in ecliptic longitude, and 0.0 mas in ecliptic latitude. With the PSF's intentionally swapped, these values became -1216.0 mas in longitude and 633.6 mas in latitude. Over the sky, the tile median differences using the correct PSF's are distributed with a median value of 78.1 mas and  $\sigma=33$ mas in ecliptic longitude, and -14.4 mas and $\sigma=16$ mas in ecliptic latitude.

The structure revealed here has no impact on the measurements carried out by the {\it WISE} and {\it NEOWISE} projects, because those measurements were made (and continue to be made) on individual exposures. The PSF structure is captured in the models, and the PSF orientation is always the same in the focal plane. This is not the case for CatWISE, which uses coadds in place of individual {\it WISE} exposures. The coadded image tile orientation is aligned with the local equatorial coordinates.  Therefore CatWISE rotated the basic PSFs by an appropriate set of position angles (PAs) between focal plane and equatorial coordinates for each image tile.

The coadded images were assembled from individual exposures with a range of PA values, so histograms of exposure PA values were constructed for each tile, using a bin width of 0.1 degrees. Separate histograms were constructed for ascending vs. descending and cryogenic vs. post-cryogenic exposures. The basic PSF was rotated by each PA bin center value and these rotated PSFs averaged together with weights equal to the histogram bin count.  Those averages were then combined for each tile with weights given by the number of corresponding cryogenic and post-cryogenic exposures to make ascending and descending PSFs for each tile. 

Tiles in the vicinity of the ecliptic poles have large PA ranges, and so the tile PSF becomes very smeared (see Figure \ref{fig:w1psf}, right). For the 50 tiles nearest the ecliptic poles (listed in Table \ref{table_polar_tiles}), the ascending and descending PSFs were averaged together, weighted by the number of ascending and descending exposures. 

The description above applies to the PSF's used by CatWISE for source measurement. For the detection step in the Preliminary Catalog (\S\ref{sec:mdet}), PSF's analogous to the measurement PSF's of the 50 tiles nearest the ecliptic poles were created for every tile, but with $4\times$ oversampling relative to the unWISE images, rather than the $8\times$ oversampling of the measurement PSF's. These detection PSF's were made by resampling the central portion of the weighted average of ascending and descending PSF's for that tile. The detection PSF's have $27\times27$ pixels with each pixel spanning 0\farcs6875, i.e. covering an area of 18\farcs5625 on a side (6.75 native WISE pixels).   

\vspace{-5mm}
\startlongtable
\begin{deluxetable*}{llllll}
\tabletypesize{\footnotesize}
\tablecaption{Tiles Processed Using a Single PSF per Band \label{table_polar_tiles}}
\tablehead{
\multicolumn{5}{c}{RA range in decimal degrees of tiles in column} \\
\colhead{79.1:87.9} &                          
\colhead{89.0:95.7} &
\colhead{96.2:264.2} &
\colhead{265.0:272.1} &                          
\colhead{272.6:280.8}
}
\startdata
0791m682 & 0890m667 & 0962m697 & 2650p681 & 2726p651 \\
0803m652 & 0891m637 & 0964m667 & 2656p651 & 2729p681 \\
0830m682 & 0908m652 & 0978m652 & 2672p666 & 2741p636 \\
0837m697 & 0909m682 & 0988m682 & 2675p636 & 2746p666 \\
0838m652 & 0920m697 & 1002m667 & 2679p696 & 2761p651 \\
0853m667 & 0924m637 & 2610p681 & 2690p681 & 2762p696 \\
0858m637 & 0927m667 & 2621p651 & 2691p651 & 2769p681 \\
0870m682 & 0943m652 & 2635p666 & 2708p636 & 2783p666 \\
0873m652 & 0949m682 & 2637p696 & 2709p666 & 2796p651 \\
0879m697 & 0957m637 & 2642p636 & 2721p696 & 2808p681 \\
\enddata
\end{deluxetable*}

%\newpage
\subsection{Detection\label{sec:mdet}}

Source detection for the CatWISE Preliminary Catalog was done simultaneously in W1 and W2 with the Multiband Detection \citep[MDET;][]{MarshJarrett2012} and Image Co-addition with Optional Resolution Enhancement (ICORE) software \citep{Masci2013}   used in the WISE pipeline, following the process described in \S IV.4.b.iii.1 of the WISE All-Sky Explanatory Supplement \citep{Cutri2012}. The detection PSF's in each band were used as matched filters to generate images which are optimal for detection of isolated point sources in that band. ICORE generated the convolution (or more accurately, the cross-covariance) of the PSF with the full-depth unWISE coadd, resampling the $2048 \times 2048$  2\farcs75 pixel unWISE coadds to the $4095 \times 4095$ 1\farcs375 pixel detection image format employed by MDET for WISE. The resampling was done using nearest-neighbor interpolation into a temporary finer grid matching the PSF grid, before being down-sampled to the output pixel size. Similarly, ICORE created uncertainty images from the detection PSF's and the unWISE ``std" images, which are the standard deviation of the individual exposures that comprise the unWISE full-depth coadds. MDET then subtracted a local background from each ICORE image using a $21 \times 21$ pixel median filter,  normalizing the result by the ICORE uncertainty image. To account for confusion noise, the uncertainties were augmented by the root-sum-square of the estimated variance of the ICORE image from the local background. Finally, MDET generated a detection image from these normalized and background-subtracted ICORE images by summing the two bands in quadrature, identifying as sources local maxima (pixels with larger values than their neighbors) that exceeded a specified detection threshold.

%The Image Co-addition with Optional Resolution Enhancement (ICORE) software \citep{Masci2013} was used to resample the $2048 \times 2048$  2\farcs75 pixel unWISE coadds to the $4095 \times 4095$ 1\farcs375 pixel detection image format employed by MDET for WISE, using a PSF appropriate for each tile and band as the interpolation kernel. 

%\textbf{This matched filter is a $27\times27$ pixel PSF that is $4\times$ oversampled relative to the unWISE images, (i.e., with 0\farcs6875 pixels), so that the filter covers an area of 18\farcs5625 on a side (6.75 native WISE pixels). It is produced by resampling the corresponding central portion of the weighted average of ascending and descending PSFs for each tile ($\S$\ref{sec:psf}).}

%Because of the normalization, the units of the detection image are related to the signal to noise ratio (SNR) of the source \citep{MarshJarrett2012}, and we refer to it here as the detection SNR of the source. 
As was done for WISE, the detection threshold was set at an estimated source reliability of $\sim50$\%.  AllWISE, with two sky coverages, selected a threshold of 2.4, compared to 3.5 used for the single coverage WISE All-Sky Release. For CatWISE we estimated the threshold based on comparison to deeper {\it Spitzer} data from the SERVS \citep{Mauduit2012} and S-COSMOS \citep{Sanders2007} programs. These programs have a source density of $\sim100,000\ \rm{deg}^{-2}$, so with the matching radius of 2\farcs5 the chance of a spurious match is not trivial (15\%), but still well below the 50\% criterion for the detection threshold. The estimated threshold  varied from 1.7 in the COSMOS field to 2.6 in the Lockman Hole, and a value of 1.8 was selected for the Preliminary Catalog.  

\begin{figure}
    \centering
    \includegraphics[width=\textwidth,trim={0 1cm 0 3cm}, clip]{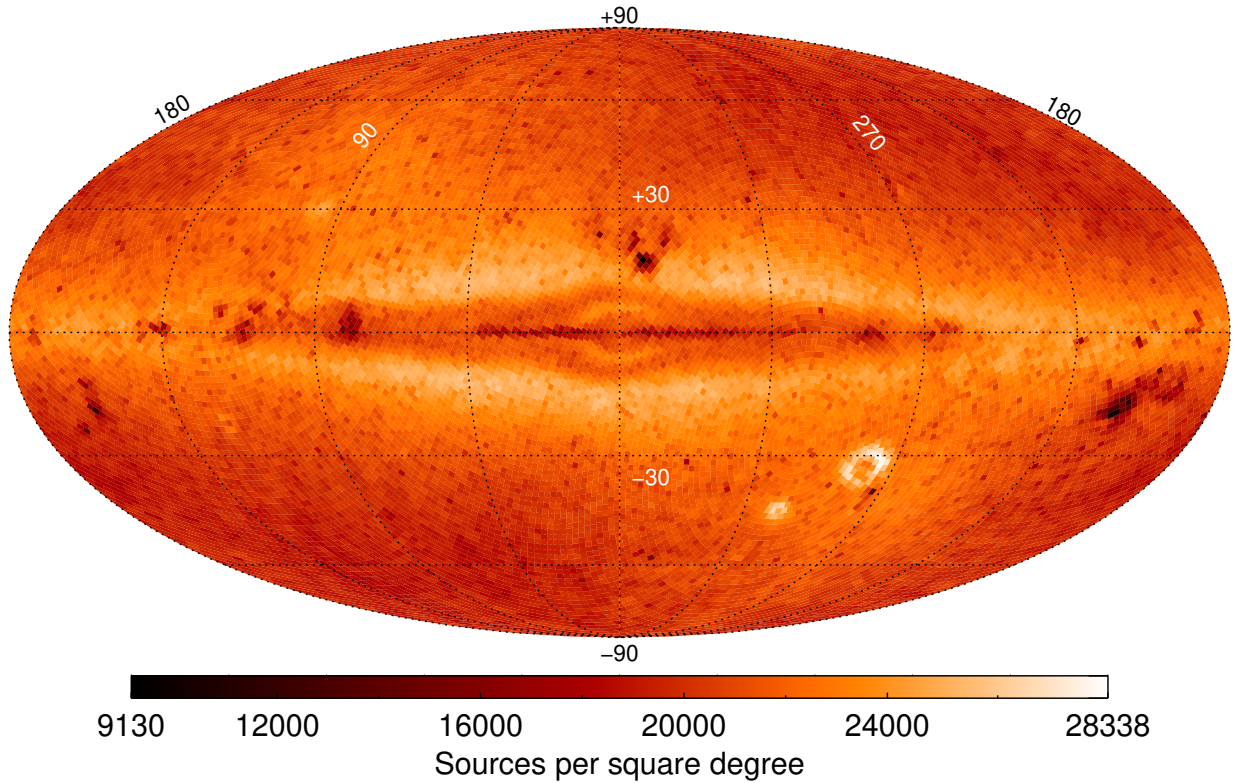}
    \caption{Map of CatWISE Preliminary Catalog source density in Galactic coordinates.}
    \label{fig:source_density}
\end{figure}

At the high source densities typical for CatWISE in coadds of $\sim 100$ or more individual WISE exposures, this matched filter  detection methodology yields an asymptotic number of $\sim 60,000$ detected sources per tile for the Preliminary Catalog (or $\sim 25,000\ {\rm deg}^{-2}$, Figure \ref{fig:source_density}) and hence becomes incomplete, particularly in W1.  The detected source density declines slightly in the Galactic plane, for reasons not well understood at present. 
%, although overlapping saturated regions from bright stars may be responsible. 
The limiting source density corresponds to an exclusion radius of 13''. This is similar to the decline in sources separated by less than 10'' shown in Figure 27 of \S VI.2.c.iv in the WISE All-Sky Explanatory Supplement \citep{Cutri2012}, so is likely a consequence of the source detection methodology used in the WISE pipeline. \citet{Schlafly2019} present an alternative approach (``crowdsource") which results in higher detected source densities, but does not provide motion estimates. We are using the \citet{Schlafly2019} catalog as the detection list for an updated version of the CatWISE catalog that is expected to be available in 2020. The completeness and reliability of the CatWISE Preliminary Catalog are discussed further in \S\ref{sec:CandR}. 

\subsection{Source Measurement  \label{sec:wphot}}

Source photometry, astrometry, and motion estimation for CatWISE use an adapted version of the WPHOT software developed for the AllWISE pipeline.  WPHOT carries out point-source extraction, solving for both photometry and astrometry simultaneously.  Source positions detected in each coadded tile image were propagated to individual exposures, and the flux and $\chi^2$ evaluated from fitting the PSF.  The source position and flux were refined by searching for a local minimum in $\chi^2$.  For AllWISE, a linear motion was also solved for, based on the observation time of the images. 

CatWISE did not have the computational resources to perform the source fitting on all of the many hundreds of individual exposures that contribute to a coadded tile image. Because the ~12 exposures for a given inertial position in each sky coverage are obtained within less than two days, the motions of sources beyond the solar system can be assumed to be fixed for each sky coverage (or epoch). Therefore CatWISE ran WPHOT treating unWISE epoch coadd images as the individual images. 

The other significant modification to the AllWISE WPHOT process implemented by CatWISE addressed the alternating scan direction of each epoch and hence varying PSF orientation.  WPHOT was not designed to use a time-dependent PSF, but with 8 or more epoch coadd images per position, CatWISE elected to measure source properties with WPHOT separately for the groups of four or more epoch coadds in each of the two scan directions, and then merge the two results. This methodology was used for all but the 50 tiles nearest the ecliptic poles, listed in Table \ref{table_polar_tiles}. For those tiles, a tile-specific average PSF over all epochs was used for both scan directions, and all epochs were processed together with WPHOT. 

WPHOT performs PSF-fit photometry and astrometry assuming sources are inertially fixed (the ``stationary fit"), and also searches for a solution assuming a linear motion with time (the ``motion fit"). The $\chi^2$ minimization fit can be performed at any location in the flux image.  The goal of astrometry is to find the location at which $\chi^2$ has a local minimum.

The stationary fit uses the gradient descent method to minimize the total $\chi^2$ (i.e., the combined W1 and W2 $\chi^2$) in the  two-dimensional space of right ascension ($\alpha$) and declination ($\delta$). As one moves around in the space, $\chi^2$ decreases and increases as the local flux distribution appears more or less like a point source. 
%In the vicinity of a real point source, $\chi^2$ tends to be concave, although with some noise fluctuations. At any point, both $\chi^2$ and its gradient can be calculated. The astrometry solution is found using the gradient descent method, by searching along the opposite direction of the gradient vector until the gradient flattens out and the local minimum has been found. 
The $(\alpha, \delta)$ where $\chi^2$ is estimated to be minimal is the position assigned to the source, and the scale factor on the PSF that minimizes $\chi^2$ there is the flux estimate. The value of $\chi^2$ is recorded, as are the individual $\chi^2$ for the W1 and W2 flux fits. Under certain conditions such as low coverage or negative fluxes, the gradient descent algorithm may be unable to compute estimates, in which case nulls are reported. Standard Gaussian error analysis provides uncertainties for $(\alpha, \delta)$ and flux. The error model includes both PSF error and image flux error, and is discussed further in \S\ref{sec:uncertainties}. More details on WPHOT are given in \S IV.4.c.iii of the {\it WISE} All-Sky Explanatory Supplement \citep{Cutri2012}.

The model used for the motion solution replaces the single location of the stationary solution with locations along a linear function of time. The slopes of this line in the $(\alpha, \delta)$ directions are the angular motion rates ($\mu_\alpha, \mu_\delta$) in $(\alpha, \delta)$. The motion fit  minimizes the total $\chi^2$ over the four-dimensional space of $(\alpha, \delta, \mu_\alpha , \mu_\delta)$.  When the motion fit is based on both ascending and descending epochs, the motions will include parallax effects \citep[see][]{Kirkpatrick2014}, and so they are not true proper motions. However, except for the 50 tiles near the ecliptic poles, CatWISE processed ascending and descending scans separately, so that the measured motions are  close to true proper motions.

\subsubsection{Uncertainties \label{sec:uncertainties}}

WPHOT estimates uncertainties in photometry, position, and motion based on error propagation from uncertainty images corresponding to the PSF and exposure images. The PSF uncertainty images from \textit{WISE} and \textit{NEOWISE} appropriate to each PSF were combined using the methodology described in \S\ref{sec:psf}, and these dominate the uncertainties estimated for brighter stars ($\lesssim 12$ mag). For fainter sources ($\gtrsim 15$ mag) the exposure image (or for CatWISE, epoch image) uncertainties dominate. For these, the ``std" unWISE images were used, which are the sample standard deviation at each pixel of the individual WISE exposures divided by $\sqrt{N-1}$, where N is the number of exposures. 

Reduced $\chi^2$ values from the CatWISE PSF fitting measurements showed that the uncertainties needed adjustment. 
%Adjustments were made to the PSF uncertainties, with the goal of making} reduced $\chi^2$ values come close to 1.0 for non-saturating sources. 
The PSF uncertainties were scaled down by 0.9 in W1 and 0.64 in W2 based on bright star reduced $\chi^2$ values. Cryogenic PSF uncertainties for W1 were scaled down by an additional factor of 0.58.   Based on faint star reduced $\chi^2$ values, the image uncertainties were scaled up by 1.4 in W1 and 1.15 in W2, and those for pre-hibernation epochs were scaled up by additional factors of 1.1 in W1 and 1.05 in W2.  The units of the unWISE images are Vega nanomaggies (nMgy), i.e. $0.306681\ \mu$Jy for W1 and $0.170663\ \mu$Jy for W2 \citep{Wright2010}, and a minimum value of 1 nMgy in W1 and 10 nMgy in W2 was also imposed on the image uncertainties.

The uncertainties for the 50 tiles near the ecliptic poles, which used averages of ascending and descending PSFs (\S\ref{sec:psf}) had somewhat different adjustments. For these tiles the PSF uncertainties were scaled up by 1.3 in W1, with no further adjustment for cryogenic values, and were left unchanged for W2.  The image uncertainties were scaled up by 1.75 in W1 and 1.25 in W2, and those for pre-hibernation epochs were scaled up further by 1.1 in W1 and 1.05 in W2.  A minimum value of 1 nMgy in W1 and 5 nMgy in W2 was imposed on the image uncertainties.

\subsubsection{Merging Measurements from Ascending and Descending Scans \label{sec:merging}}
   
As noted earlier, for nearly all tiles two independent measurements were made, one extracted from epoch coadds constructed from ascending scans and one from descending-scan epochs. These were merged into a single source list as follows.
 
Sources were matched based on their identifier in the
%with the same Each source has an identifier ({\it mdetID}) corresponding to an entry in the 
detected source list (\S\ref{sec:mdet}). A 
%, and this was used to match ascending and descending measurements of the same source. The ascending and descending sources were almost in perfect one-to-one correspondence, but a 
small number of mismatches occurred because of active deblending in WPHOT. Active deblending inserts a new source (not from the detection program) into the fitting region if the  $\chi^2$ value exceeds a threshold and is reduced by the insertion by a minimum required amount. Full details may be found in \S IV.4.c of the {\it WISE} All-Sky Explanatory Supplement \citep{Cutri2012}.  Actively deblended sources have the same identifier as the parent source in the detection list.  Because active deblending may proceed differently in the ascending and descending data, when the identifier was not unique, a nearest-match criterion was also applied. If the number of ascending sources did not equal the number of descending sources, the left over sources were discarded, leaving a one-to-one association list.

Corresponding parameters for each matched ascending-descending source pair were combined when the ascending and descending apparitions have non-null values (\S\ref{sec:wphot}), otherwise the single non-null values (if any) were retained. Aperture magnitudes for the ascending and descending scans were measured from the same full-depth coadd, differing only slightly in the position of the aperture centers.  Hence, the fluxes and flux uncertainties corresponding to the aperture magnitudes and their uncertainties were simply averaged without weighting.
Positions were averaged using inverse-covariance weighting, with the averaging done in a local Cartesian projection consistent with the uncertainty representation. Additional details are provided in Appendix \ref{sec:mergepos}.
PSF-fit photometry was combined by averaging the flux values using inverse-variance weighting. This yields refined flux values and reduced uncertainties that are used to recompute the magnitudes, magnitude uncertainties, and signal-to-noise ratios.

\newpage
\subsection{Artifact Flagging  \label{sec:artifacts}}

Bright stars create a variety of scattered light effects and electronic charge issues that require special handling by the software. These effects include scattered light halos, diffraction spikes, glints from off-frame bright stars, optical ghosting from internal reflections within the optical system, and charge persistence on the arrays. These can create false detections, hereafter called artifacts, or can contaminate detections of real, astrophysical sources. The goal of artifact flagging is to label spurious or affected sources so that the user can easily create source lists for which most of these problems are eliminated.

The CatWISE Preliminary Catalog employs two types of artifact flagging. One is the $cc\_flags$ values copied directly from AllWISE processing. If a source was found in either the AllWISE Catalog or Reject Table within 2$\farcs$75 of a CatWISE source, its $cc\_flag$ was included in the entry for the CatWISE source; if no AllWISE source was found within this radius, then $cc\_flags$ contains a null value.  For a CatWISE source having multiple AllWISE sources within the 2$\farcs$75 radius, the most pessimistic value per band was retained, as described in \S\ref{sec:cc-flags}. These $cc\_flags$ values indicate whether a source is likely to be spurious because it is dominated by an artifact (encoded as an upper-case letter) or 
is a real source contaminated by an artifact (encoded as a lower-case letter). Each of the four characters in $cc\_flags$ corresponds to artifacts in one of the four WISE bands, so the first two characters are most relevant for CatWISE, which does not include W3 or W4 data. The possibilities are ``0'' for no artifact or contamination, ``D'' or ``d'' for a diffraction spike, ``H'' or ``h'' for a scattered light halo, ``O'' or ``o'' for an optical ghost, or ``P'' or ``p'' for charge persistence. The $cc\_flags$ field conveys only the main features of the full artifact flag information contained in the AllWISE $w1cc\_map$, $w1cc\_map\_str$, $w2cc\_map$, and $w2cc\_map\_str$ fields,\footnote{Note that the column descriptions for these fields in \S II.1.a of the AllWISE Explanatory Supplement \citep{Cutri2013} are incorrect. Corrected descriptions are provided in Appendix \ref{sec:columns}.} 
which are also provided in the CatWISE source entry. A more detailed description can be found in \S IV.4.g of the {\it WISE} All-Sky Explanatory Supplement \citep{Cutri2012}.  

The second type of artifact flag is called $ab\_flags$, which was determined for every CatWISE source and is based on artifact flagging images provided by unWISE \citep[unWISE bit masks;][]{Meisner2019a}. The $ab\_flags$ do not attempt to distinguish between an outright artifact and a real astrophysical source that suffers from some level of artifact contamination, but are set %. However, we have adjusted the flagging  (\S\ref{sec-tuning-ab-flags}) 
so that only egregious artifacts more likely to be spurious detections are flagged. By analogy with $cc\_flags$, the $ab\_flags$ thus contain only upper-case letters ``D'', ``H'',``O'', or ``P'' or the value ``0''. Furthermore, because the CatWISE data deal only with W1 and W2 data, the $ab\_flags$ values are only two characters long. Additional details are given in \S\ref{sec:ab-flags}.

\begin{figure}
\centering\includegraphics[width=0.65\textwidth, clip]{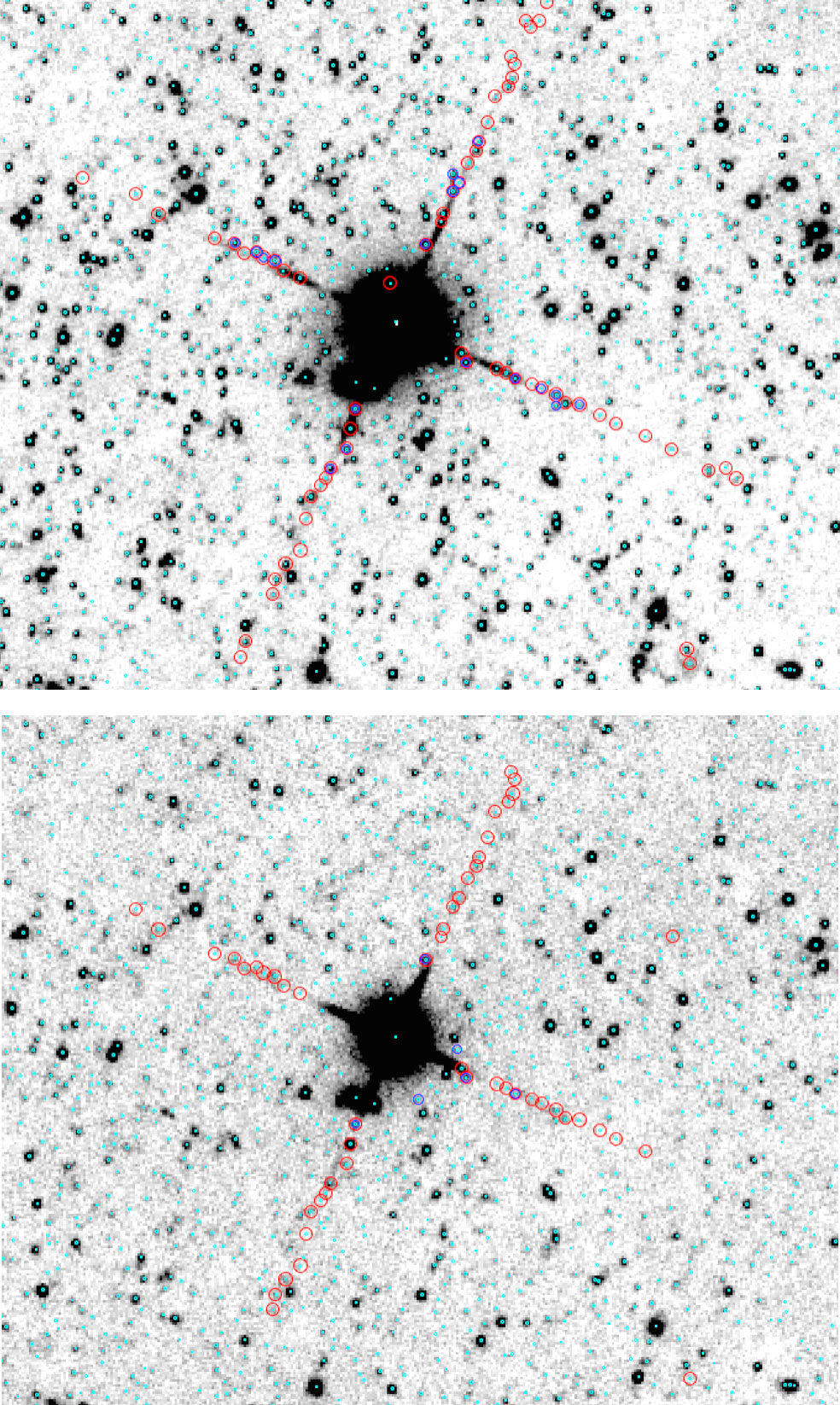}
%{figures/flagging_figure_catwise_overview_paper4b.pdf}
\caption{Example artifact flags for $16.5' \times 13.0'$ cutouts in W1 (top) and W2 (bottom) from the tile 1497p015. Detected CatWISE sources are shown by the small cyan circles. Sources with upper-case $cc\_flags$ are marked with large red circles, and sources labeled as artifacts by $ab\_flags$ are marked in medium-sized blue circles.
\label{fig:artifacts}}
\end{figure}

Figure~\ref{fig:artifacts} shows examples of both types of artifact flags. These $cc\_flags$ and $ab\_flags$ values should be regarded as two different yet complementary methods for tagging sources of special concern to the user. As described in \S~IV.4.g of the {\it WISE} All-Sky Explanatory Supplement \citep{Cutri2012}, the $cc\_flags$ are known to overflag; their purpose is to produce a reliable catalog of source extractions free of contamination by artifacts, but they do so at the expense of completeness. AllWISE $cc\_flags$ are not available for sources detected only in CatWISE, so requiring a source to have ``0'' in its $cc\_flags$ entry again emphasizes reliability over completeness. In contrast the $ab\_flags$, which are available for every CatWISE source, are used to tag only the more egregious artifacts, thereby emphasizing completeness, albeit at the expense of reliability. Users of the CatWISE data products can thus query against these two sets of artifact flags to best fit their needs. 

\subsubsection{Setting the $cc\_flags$ Values When There are Multiple Matches\label{sec:cc-flags}}

The values of $cc\_flags$ were taken directly from a join of the AllWISE Point Source Catalog and Reject Table. When more than one of these AllWISE sources matched to within 2$\farcs$75 of a CatWISE source, the band-by-band $cc\_flags$ value of the CatWISE source was determined as follows. If there was an upper-case letter for any matching source, it took precedence over any lower-case letter. The letter ``D'' is the highest priority followed, in order, by ``P'', ``H'', and ``O''. If there were no upper-case letters for any matching source, lower-case letters had the next highest priority, in the same order (``d'', ``p'', ``h'', and ``o''). If all matching sources had a value of ``0'' in that band, then ``0'' was used.

For the binary bit encoding of AllWISE artifacts given by $w1cc\_map$ and $w2cc\_map$, when there was more than one AllWISE match to a CatWISE source, a logical `OR' was performed over all the matches for each bit. The $w1cc\_map\_str$ and $w2cc\_map\_str$ values were then constructed using the priority rules given in the previous paragraph.

\subsubsection{Translating the unWISE Bit Mask Values to $ab\_flags$\label{sec:ab-flags}}

The unWISE bit mask values at the (x,y) pixel location of the source were used to set the value of $ab\_flags$, as shown in Table~\ref{table_abflags}, where we use the prefix ``b" to indicate specific unWISE mask bits in column 3. Bits 0--3 and 7--8 are related to a nearby bright star; bits 11, 12, 25, and 26 indicate optical ghosts; bits 13--20 flag charge persistence; bits 23 and 24 mark halos; and bits 27--30 are for diffraction spikes.  The checks against bits 21 and 22 are to ensure that the bright star is not flagged by its own diffraction spikes or scattered light halo. A complete description of the unWISE mask bits is given in \citet{Meisner2019a}. 
%A full description of the  unWISE mask bit meanings is available through the unWISE Catalog data release website\footnote{\url{http://catalog.unwise.me/files/unwise_bitmask_writeup-03Dec2018.pdf}, see Table \ref{table_abflags}.}. 

\startlongtable
\begin{deluxetable*}{lll}
\tabletypesize{\footnotesize}
\tablecaption{Logic Used in Converting unWISE Bitmask Values to {\it ab\_flags} Values\label{table_abflags}}
\tablehead{
\colhead{Value} &                          
\colhead{Band} &
\colhead{Logic} \\
%\colhead{(1)} &                          
%\colhead{(2)} &  
%\colhead{(3)}
}
\startdata
``D''   & W1 & b0 or b1 or b7 or b27 or b29 but only if not b21 \\
\nodata & W2 & b2 or b3 or b8 or b28 or b30 but only if not b22 \\
``H''   & W1 & b23 but only if not b21 \\
\nodata & W2 & b24 but only if not b22 \\
``O''   & W1 & b25 or b26 \\
\nodata & W2 & b11 or b12 \\
``P''   & W1 & b13 or b14 or b17 or b18 \\
\nodata & W2 & b15 or b16 or b19 or b20 \\
\enddata
\end{deluxetable*}

In addition to the $ab\_flags$ field, the CatWISE catalog includes fields with detailed artifact flag information ($w1ab\_map$, $w1ab\_map\_str$, $w2ab\_map$, and $w2ab\_map\_str$) for each source, analogous to the similar fields providing detailed artifact information related to $cc\_flags$. 

%\subsubsection{Adjusting the $ab\_flags$ Values\label{sec-tuning-ab-flags}}

%For the $ab\_flags$, we performed by-eye checks on selected tiles to make sure that the extent of the flagging matched expectations. These included tiles at a range of  Galactic latitudes. Several iterations were performed, primarily to adjust the extent of the halo flagging, before values were finalized. Figure~\ref{fig:artifacts} shows examples of the resultant final artifact flagging in tiles containing 47 Tucanae (0048m727, $b=-44\degree$), a portion of the Hyades (0657p151, $b=-24\degree$), the COSMOS field (1497p015, $b=+41\degree$), and Barnard's Star (2688p045, $b=+15\degree$). Note that the upper-case $cc\_flags$ slightly overflag, as designed, particularly in scattered light halos.

\subsection{Duplicate Source Measurements  \label{sec:primary}}

\textit{WISE} tiles, and, therefore, unWISE coadds, overlap each other by $\sim3'$ on the equator. As a result, sources appearing near the edge of a tile will also be measured in neighboring tiles. To remove these duplicate measurements we adopted the approach  used by \citet{Schlafly2019}. For each source in a tile, its coordinates were used to calculate the minimum distance to the edge of the tile. The same coordinates were used to calculate the minimum distance to the edge for all neighboring tiles, and the source was flagged as ``primary" if these other minimum distances were all smaller.

%\newpage
\subsection{The CatWISE Preliminary Catalog \label{sec:prelim}} 

With the necessary flagging in place, catalog generation could proceed. CatWISE Preliminary Catalog sources are required to:
 
1) be from the tile where that source is furthest from the tile edge (i.e. flagged as ``primary," \S\ref{sec:primary}) 

\noindent and

2a) have W1 SNR $\geq 5$ with no identified artifacts (a value of 0 in the left character of {\it ab\_flags}) 

\noindent or

2b) have W2 SNR $\geq 5$ with no identified artifacts (a value of 0 in the right character of {\it ab\_flags}).

There are 900,849,014 sources that meet these criteria. The 167,831,546 sources that fail to meet these criteria go into the reject file for their tile. Individual tile reject files typically contain 8,000 sources, although near the celestial poles they can contain over 30,000 sources due to large tile overlap. There are 186 formatted columns of information about each source in the tile catalog files. Reject files have one additional 
column, indicating if the source is primary in its tile. Descriptions of most of the columns are available in \S II.1.a of the AllWISE Explanatory Supplement \citep{Cutri2013}, and Appendix \ref{sec:columns} provides additional information about CatWISE columns.

 The individual Preliminary catalog and reject files for the 18,240 tiles were transferred to the NASA Infrared Science Archive (IRSA), where they were merged into the IRSA database.  Four columns (with names in italics in Table \ref{table_newcolumns}) were removed from view by IRSA, for reasons explained in that Table. Information regarding access to the catalog is provided in \S\ref{sec:access}.
 
CatWISE source designations should have the prefix CWISEP
%Jhhmmss.ss$\pm$ddmmss.s 
for objects in the CatWISE Preliminary Catalog, and CWISEPR
%Jhhmmss.ss$\pm$ddmmss.s 
for objects in the CatWISE Preliminary Reject Table. 
%\footnote{These designations conform to the naming conventions outlined by the International Astronomical Union (\url{http://cdsweb.u-strasbg.fr/Dic/iau-spec.html}), where the decimal portions of RA and Dec are truncated rather than rounded, and the "J" indicates that the coordinates are for J2000 equinox.}. 
The designation for each source, based on its coordinates for the J2000 equinox following the IAU truncation convention and without the leading CWISEP or CWISEPR prefix, is given by the field {\it source\_name} which is the first column in the files. For example, the quasar 3C~273 is CWISEP J122906.70+020308.6\footnote{The CatWISE Preliminary Catalog coordinates for 3C 273 are within 0\farcs011 of the Gaia DR2 coordinates.}

\section{Performance Characterization \label{sec:performance}}

The greatest potential for improvement of the CatWISE Preliminary Catalog over existing {\it WISE} catalogs is in providing more accurate motion measurements, due to the much longer time baseline compared to AllWISE. Hence, characterization has focused on astrometric properties of the Preliminary Catalog. Photometric depth is also improved due to the four times larger number of exposures than AllWISE.  Characterization has emphasized comparison to external truth sets, using {\it Spitzer} for photometric comparisons including completeness and reliability, and \textit{Gaia} for astrometric comparisons. We begin with the photometric assessments.

\subsection{Completeness and Reliability  \label{sec:CandR}}

\subsubsection{Bright Sources \label{sec:bright}}

CatWISE completeness and reliability for sources with W1 or W2 $<$8\,mag were assessed using an updated version of the \textit{WISE} Bright Star List (BSL) 
as a truth set. The list was generated by the \textit{WISE} team for artifact flagging (see \S 4.4.g.vi in the \textit{WISE} All-Sky Release Explanatory Supplement; \citealp{Cutri2012}). To avoid degradation of completeness or reliability due to missed matches for fast-moving sources, astrometric information for sources moving faster than 0\farcs275 yr$^{-1}$ was added to the BSL to propagate the positions of such stars to the CatWISE epoch. Astrometric information was taken from \textit{Gaia} DR2 \citep{Brown2018, Lindegren2018} or, when not available (given \textit{Gaia}'s known incompleteness for bright stars) from the \textit{Hipparcos} main catalog \citep{vanLeeuwen2007}, the Luyten Half-Second catalog \citep{Bakos2002}, or the Gliese-Jahrei{\ss} catalog \citep{Stauffer2010}. 

CatWISE completeness was determined as the percentage of sources  that have astrometric matches in CatWISE as a function of BSL magnitude.  Differential CatWISE reliability was determined as the percentage of sources  that have astrometric matches in the BSL as a function of CatWISE magnitude.  We used a relatively large matching radius of 5\farcs5 (corresponding to two {\it WISE} pixels) to account for the poorer centroiding accuracy expected for highly saturated sources.

\begin{figure*}
    \centering
    \includegraphics[width=0.49\textwidth, trim={1cm 3cm 3cm 3cm}, clip]{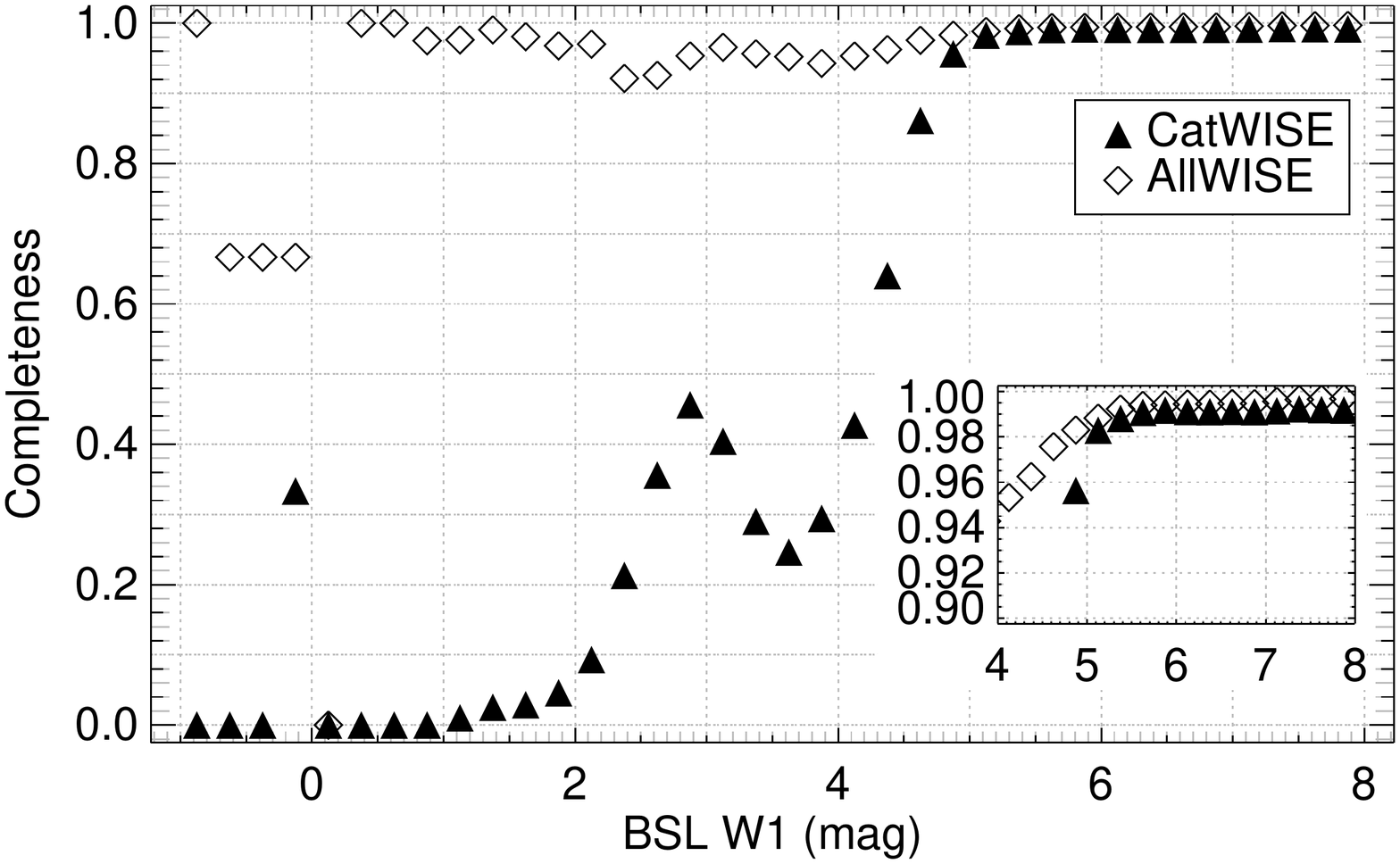}
    \includegraphics[width=0.49\textwidth, trim={1cm 3cm 3cm 3cm}, clip]{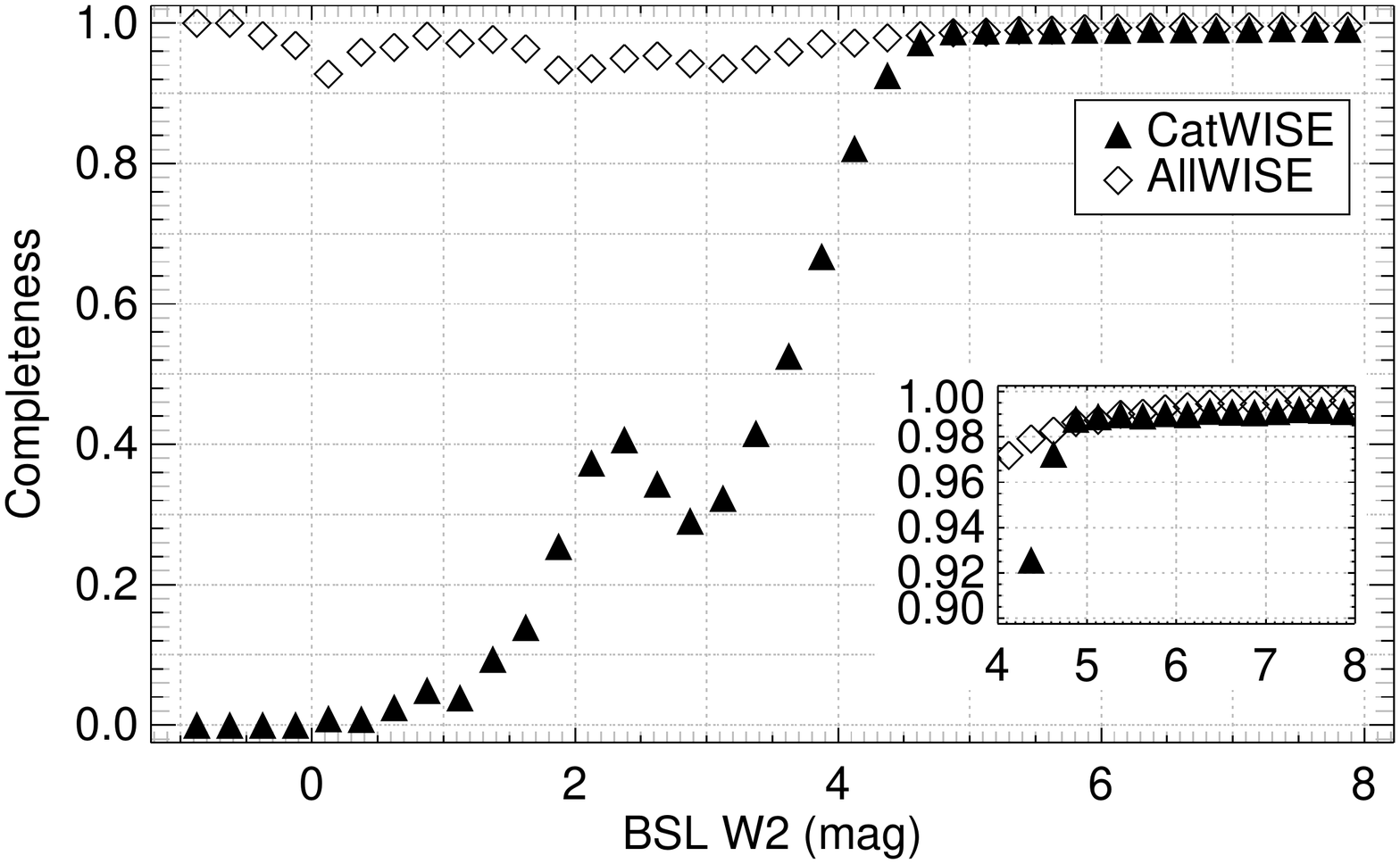}
    \caption{Differential completeness of the CatWISE Preliminary Catalog as a function of the Bright Star List's W1 (left) and W2 (right), compared to AllWISE.}
    \label{fig:brightCompleteness}
\end{figure*}

Figure~\ref{fig:brightCompleteness} shows the results for completeness, compared to those from AllWISE. CatWISE achieves $\sim99\%$ completeness in the BSL W1$\sim5.5-8$\,mag and BSL W2$\sim5-8$\,mag ranges, slightly lower than AllWISE. For reasons that we have not investigated, CatWISE completeness drops for brighter stars, falling to $\sim 50\%$ by W1$\sim4.3$\,mag and W2$\sim3.6$\,mag.  AllWISE completeness remains above 90\% even for stars as bright as 0.25 mag, making it preferable to CatWISE for investigations of the brightest stars.

\begin{figure*}
    \centering
    \includegraphics[width=0.49\textwidth, trim={1cm 3cm 3cm 3cm}, clip]{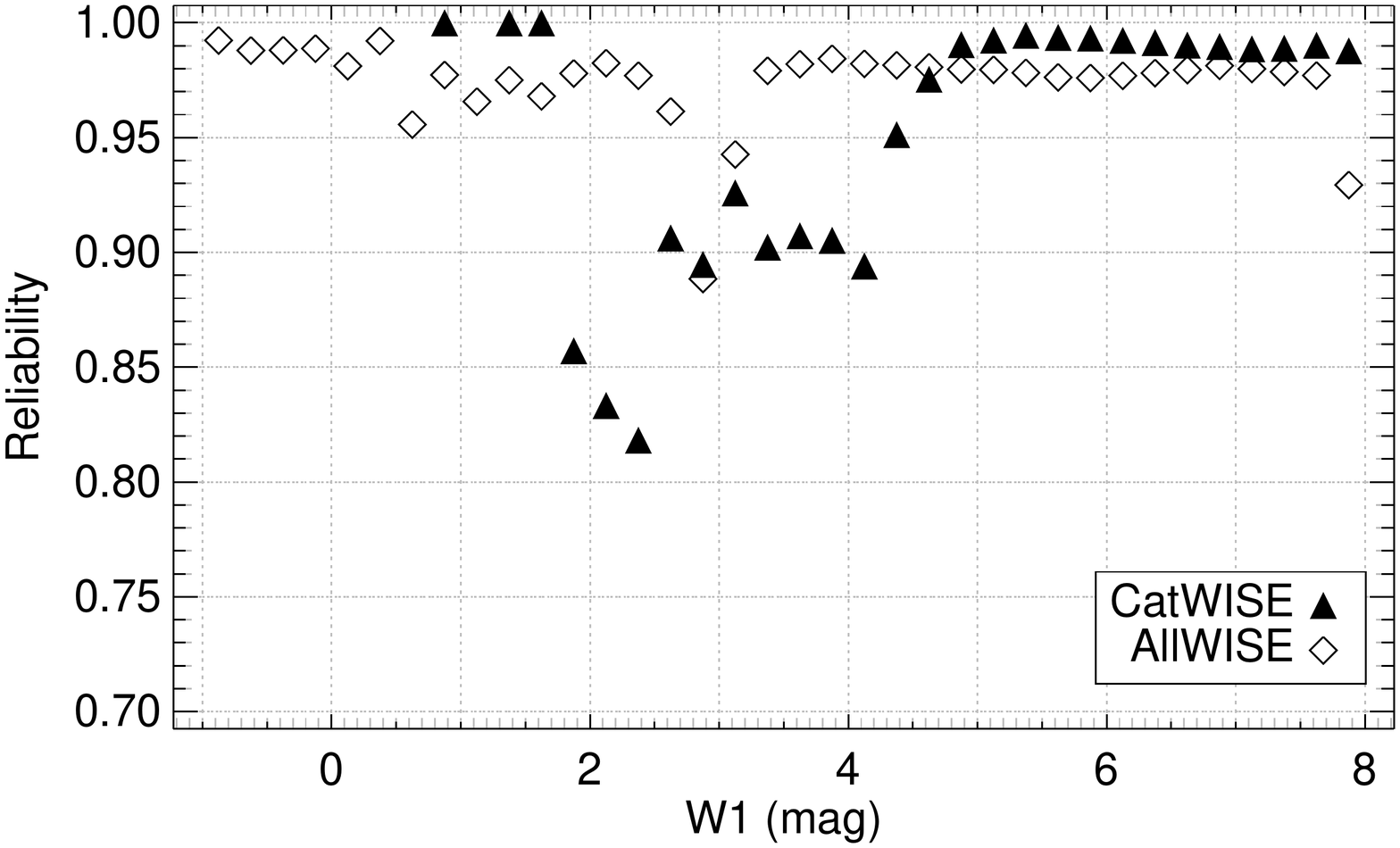}
    \includegraphics[width=0.49\textwidth, trim={1cm 3cm 3cm 3cm}, clip]{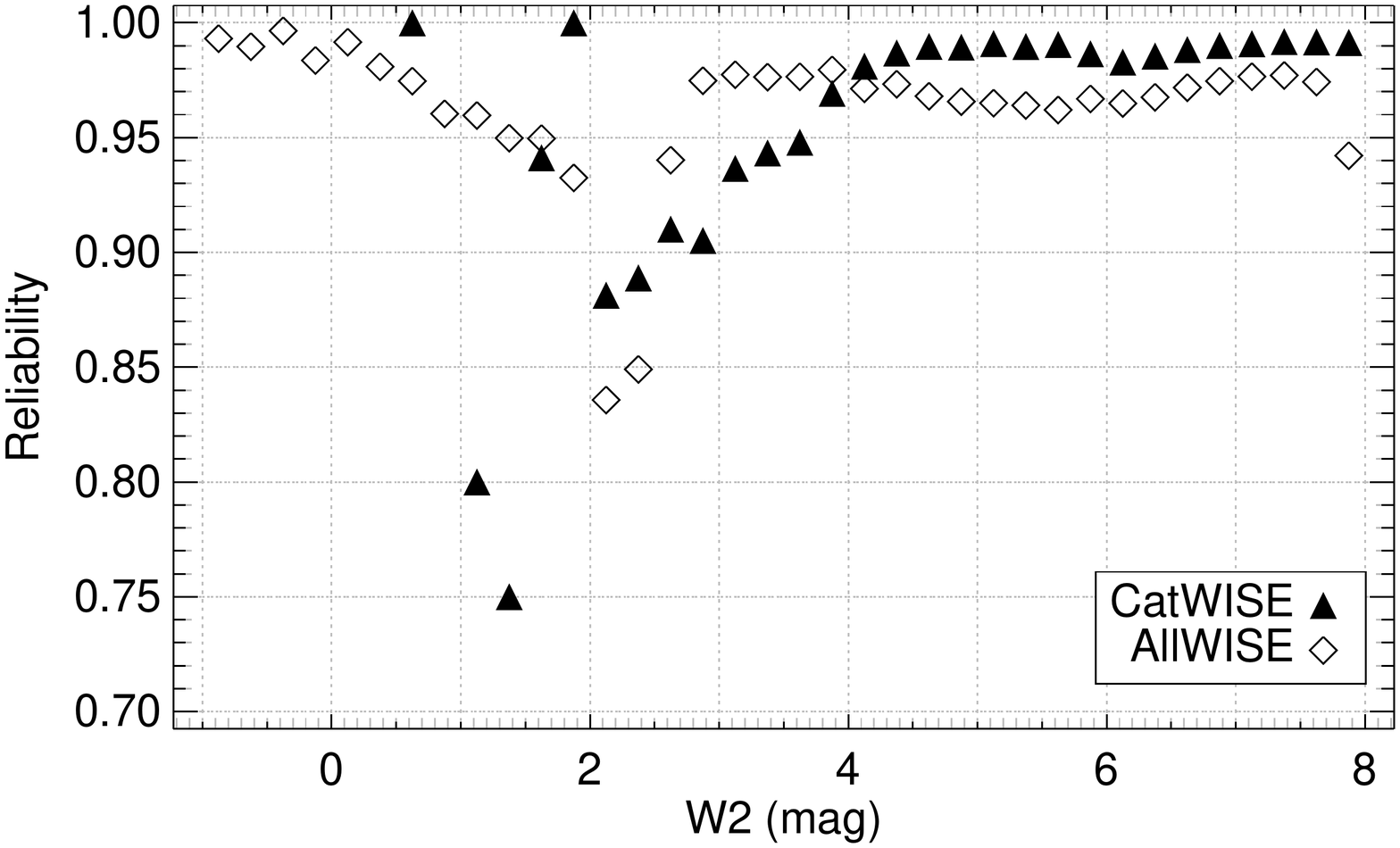}
    \caption{Differential reliability of the CatWISE Preliminary Catalog as a function of the Bright Star List's W1 (left) and W2 (right), compared to AllWISE. Not shown are points at W2=-0.75 mag and +0.75 mag with reliabilities of 0 and 0.5 respectively.}
    \label{fig:brightReliability}
\end{figure*}

CatWISE reliability is $\sim99\%$ for W1$\sim4.8-8$\,mag and  W2$\sim4.5-8$\,mag, somewhat better than AllWISE, as can be seen in Figure~\ref{fig:brightReliability}. (For sources without artifacts in either CatWISE or AllWISE, the median (CatWISE - AllWISE) value over the 2.5 to 8 mag range is 0.17 mag in W1 and 0.02 mag in W2, and the median absolute deviation is 0.08 mag in W1 and 0.02 mag in W2). Even though completeness falls rapidly for brighter stars, CatWISE reliability is $\sim90\%$ or better down to 2.5 mag.  For the brightest stars, CatWISE completeness is poor, and CatWISE reliability shows large fluctuations due in part to the resulting small number statistics.

\subsubsection{Faint Sources \label{sec:faint}}

Completeness and reliability were assessed for faint sources using the {\it Spitzer} South Pole Telescope Deep Field (SSDF) survey \citep{Ashby2013} as a truth set. The SSDF covers $\sim94$ deg$^2$ with 2 minutes of integration per position with {\it Spitzer}. {\it WISE} survey depth increases towards the ecliptic poles, and for the SSDF, with $\beta \sim -46\degree$, the typical number of CatWISE exposures is 125, or 16 minutes. Nevertheless the larger {\it Spitzer} mirror diameter (85 cm vs. 40 cm for {\it WISE}) and better image quality (2" vs. 6")  makes the SSDF data more sensitive than CatWISE.

\begin{figure*}
    \centering
    \includegraphics[width=0.49\textwidth, trim={0cm 3cm 2cm 4cm}, clip]{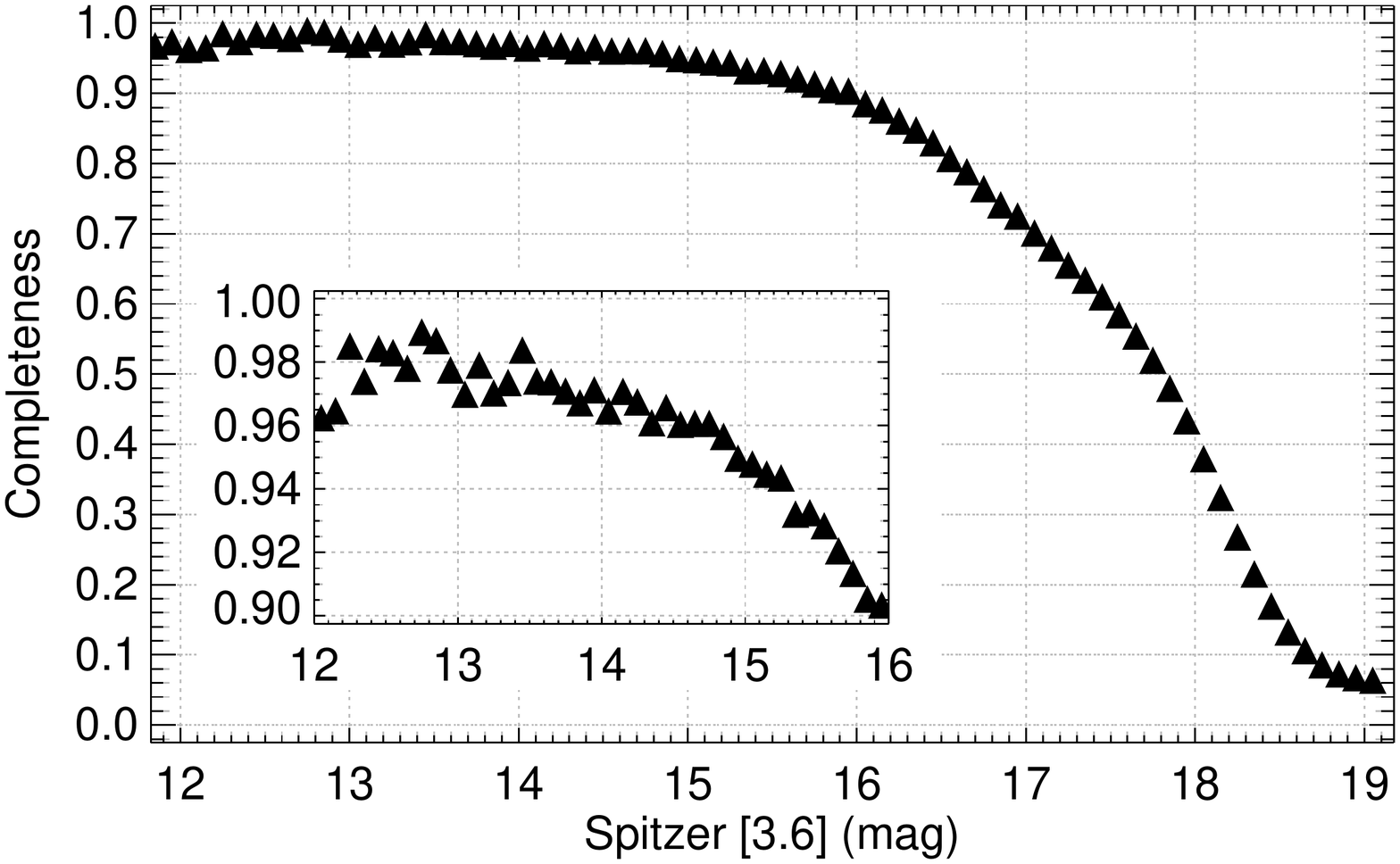}
    \includegraphics[width=0.49\textwidth, trim={0cm 3cm 2cm 4cm}, clip]{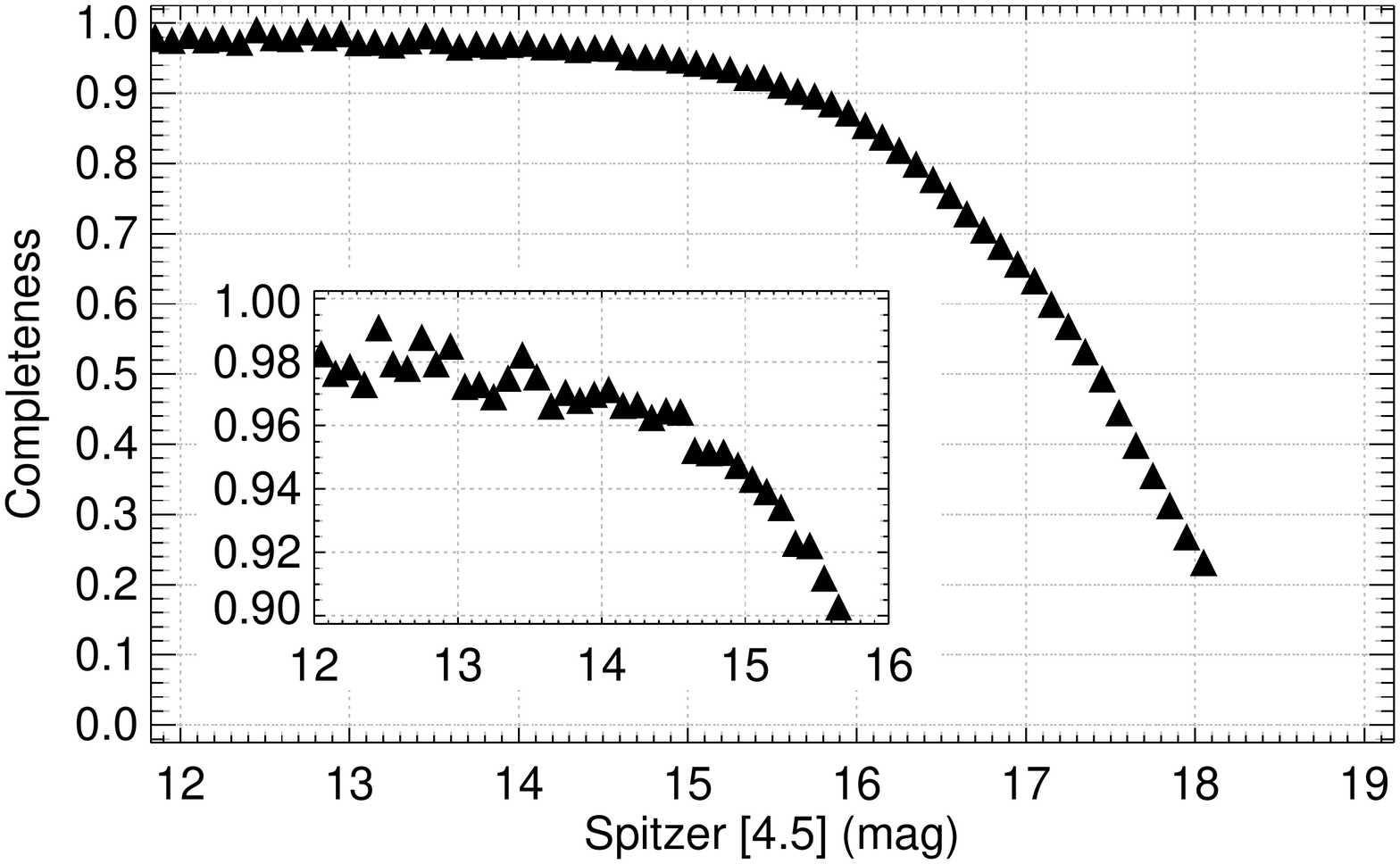}
    \caption{Completeness of the CatWISE Preliminary Catalog vs. {\it Spitzer} $3.6 \mu$m (left) and $4.5 \mu$m  (right) magnitude for sources in the SSDF.}
    \label{fig:SSDFcompleteness}
\end{figure*}

\begin{figure*}
    \centering
    \includegraphics[width=0.49\textwidth, trim={0cm 3cm 2cm 4cm}, clip]{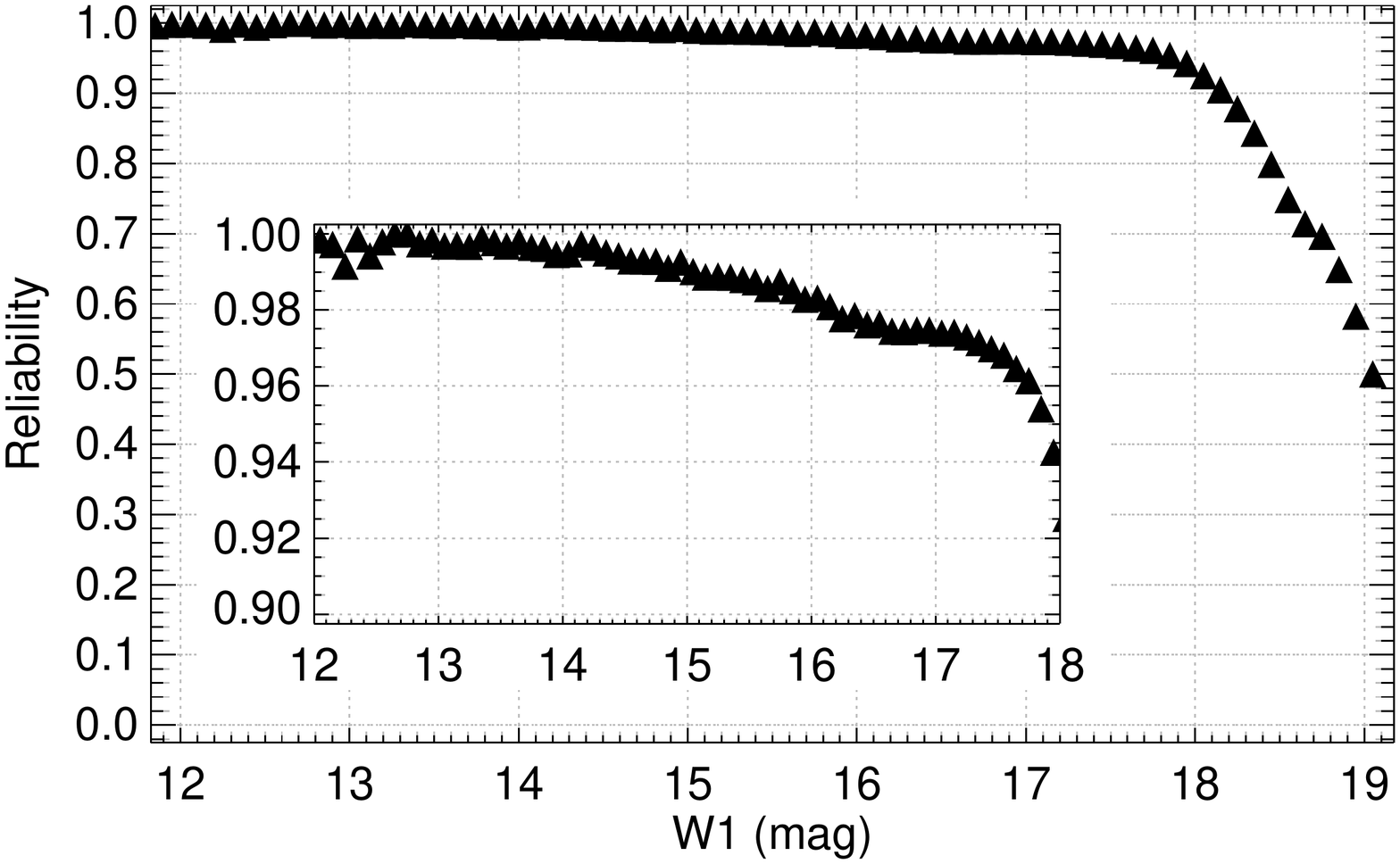}
    \includegraphics[width=0.49\textwidth, trim={0cm 3cm 2cm 4cm}, clip]{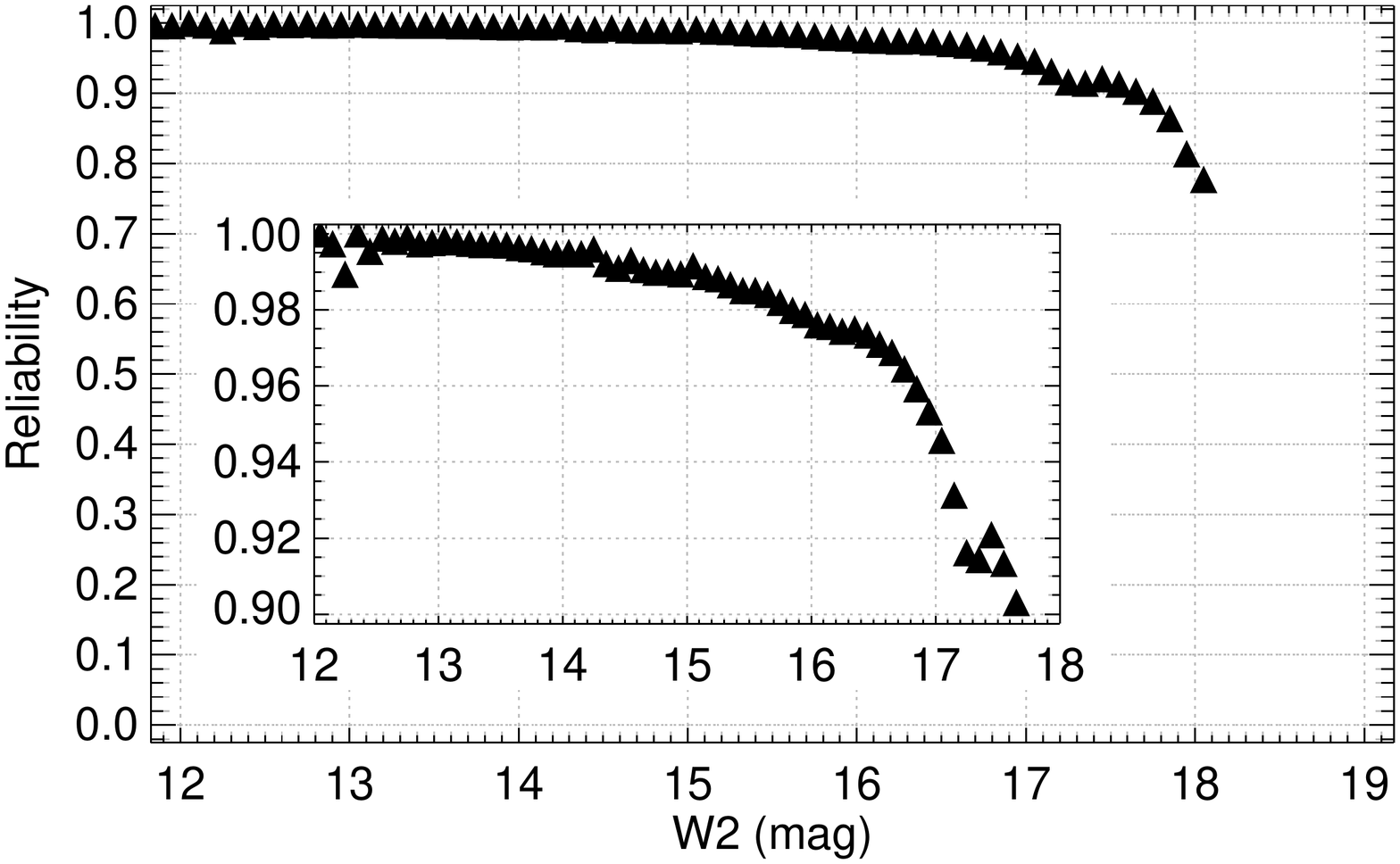}
    \caption{Reliability of the CatWISE Preliminary Catalog as a function of W1 (left) and W2 (right), for sources in the SSDF.}
    \label{fig:SSDFreliability}
\end{figure*}

The radius used to match {\it Spitzer} SSDF sources with CatWISE sources was 2\farcs5, and otherwise the methodology to determine completeness and reliability was the same as in \S\ref{sec:bright}. The results are shown in Figures~\ref{fig:SSDFcompleteness} and \ref{fig:SSDFreliability}.

At this coverage depth, the CatWISE Preliminary Catalog completeness is $\sim98\%$ for sources brighter than 13th mag, consistent with the performance seen from comparisons to the Bright Star List (\S\ref{sec:bright}), and remains above 90\% for sources brighter than [3.6] = 16 mag or [4.5] = 15.5 mag, dropping to 50\% at [3.6] = 17.8 mag and [4.5] = 17.4 mag. The SSDF achieves 55\% completeness at [3.6] = 18.5 mag and [4.5] = 18.0 mag \citep{Ashby2013}.

CatWISE reliability at this coverage depth is better than 99\% for sources brighter than 15th mag in both W1 and W2, again consistent with the Bright Star List results. CatWISE reliability remains above 90\% to W1 = 18.2 and W2 = 17.6 for this ecliptic latitude and coverage depth. Reliability at fainter magnitudes is not well determined because the SSDF data are not deep enough.

\newpage
\subsection{Photometric properties  \label{sec:photom_perf}}

We assessed CatWISE photometric depth using both the SSDF (\S\ref{sec:faint}) and the COSMOS field. The COSMOS field is an important benchmark for assessing CatWISE performance,  because it has been intensively observed, is near the ecliptic ($\beta = -11\degree$), and is at fairly high Galactic latitude ($b=41\degree$). Although confusion effects are reduced by the high Galactic latitude, the ecliptic has the highest zodiacal emission and lowest survey coverage from {\it WISE}, making COSMOS a representative base for performance estimates.  

\begin{figure}
        \centering
        \includegraphics[width=0.49\textwidth, trim={0 3cm 2cm 4cm}, clip]{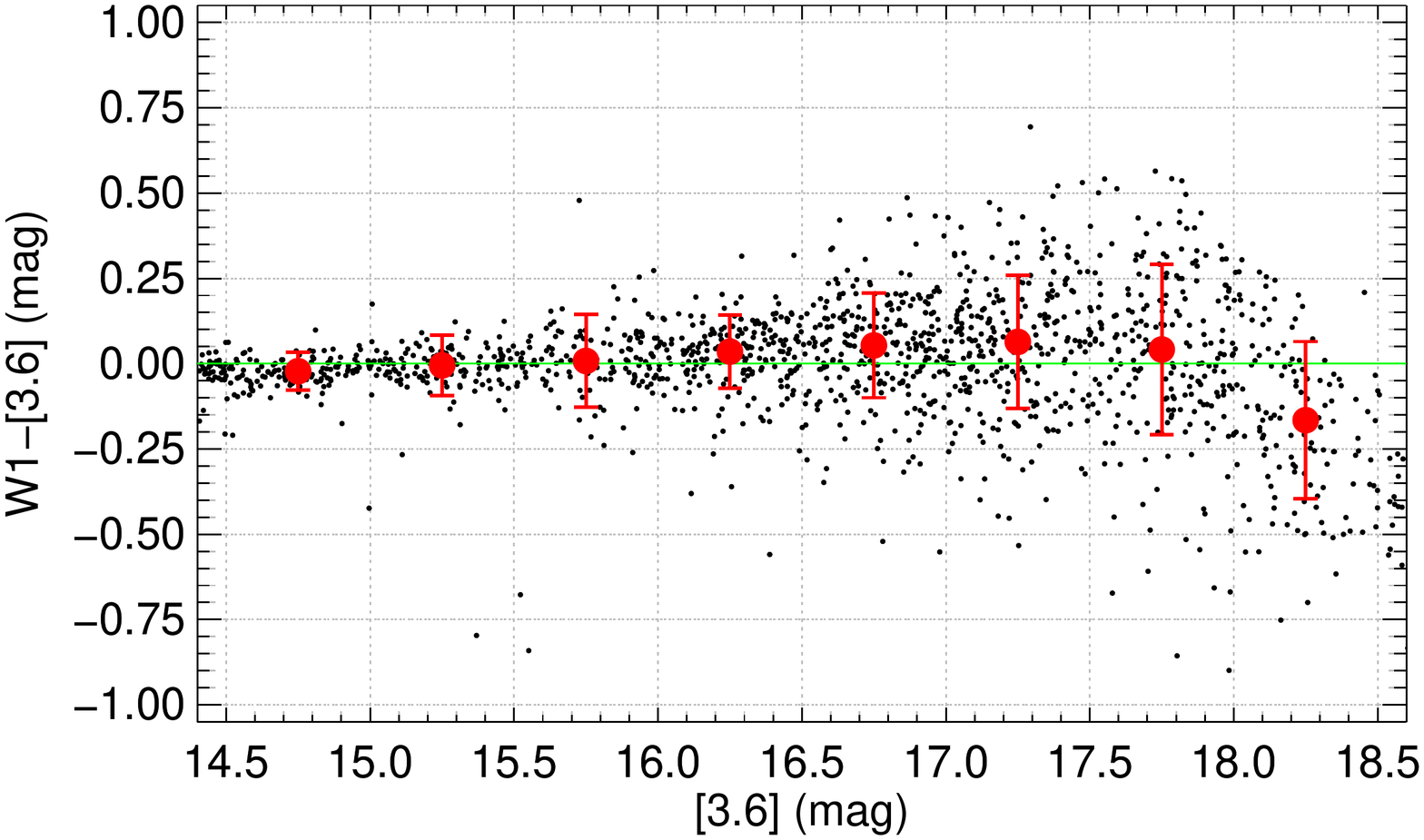}
        \includegraphics[width=0.49\textwidth, trim={0 3cm 2cm 4cm}, clip]{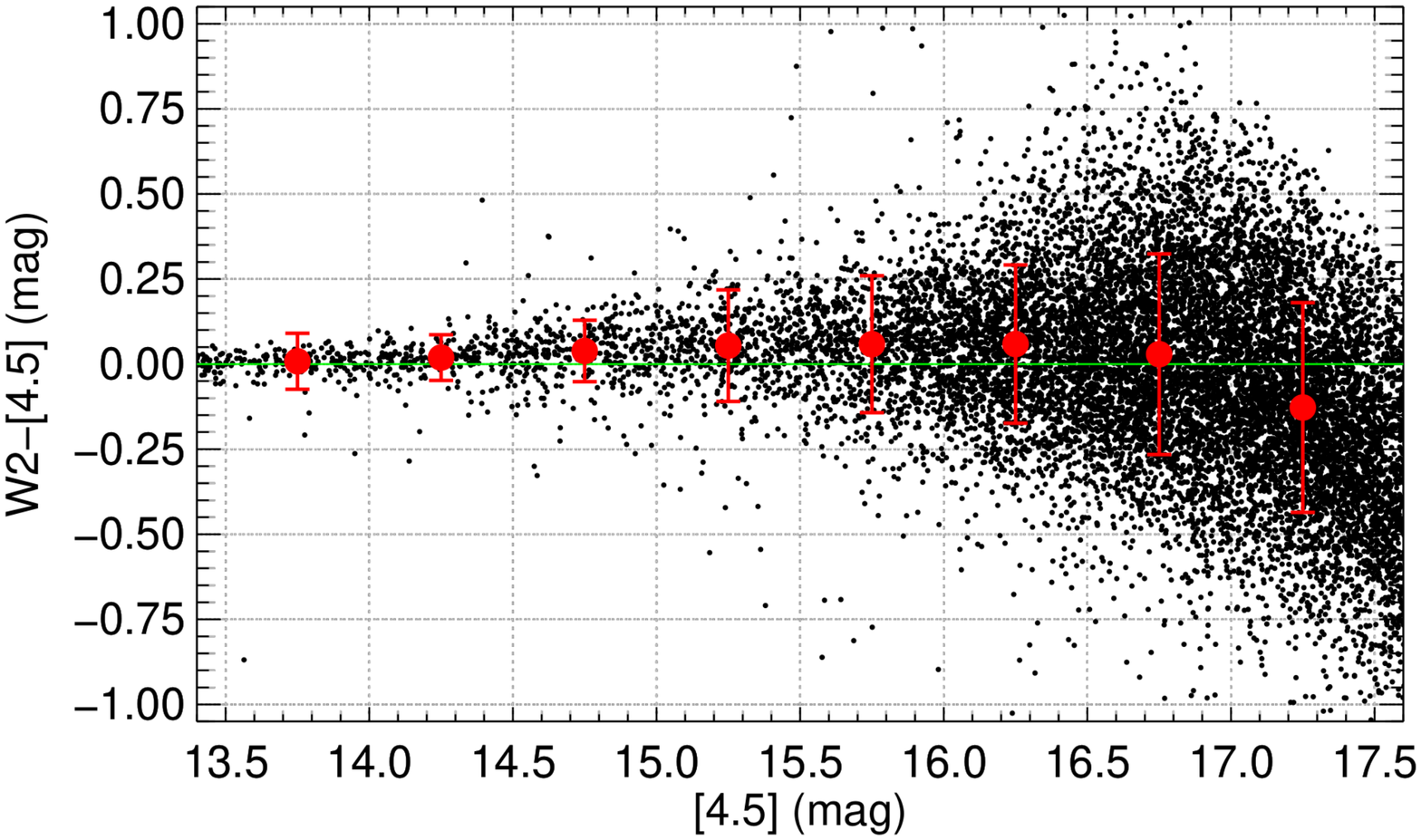}
        \caption{Comparison of CatWISE photometry to {\it Spitzer} photometry for COSMOS. {\it Left}: Difference between CatWISE W1 PSF and {\it Spitzer} S-COSMOS 2\farcs9 radius aperture photometry at [3.6], for sources with $-0.1 < [3.6] - [4.5] < 0$ and $< 10$\% flux increase from the 1\farcs9 to 2\farcs9 aperture. Median differences and standard deviations in 0.5 mag bins are shown by the red points and error bars. {\it Right}: The analogous comparison for CatWISE W2 and {\it Spitzer} [4.5] photometry, but without the restriction on {\it Spitzer} source color.}
        \label{fig:CatWISE_vs_COSMOS}
    \end{figure}

\begin{figure}
        \centering
        \includegraphics[width=0.49\textwidth, trim={0cm 3cm 2cm 4cm}, clip]{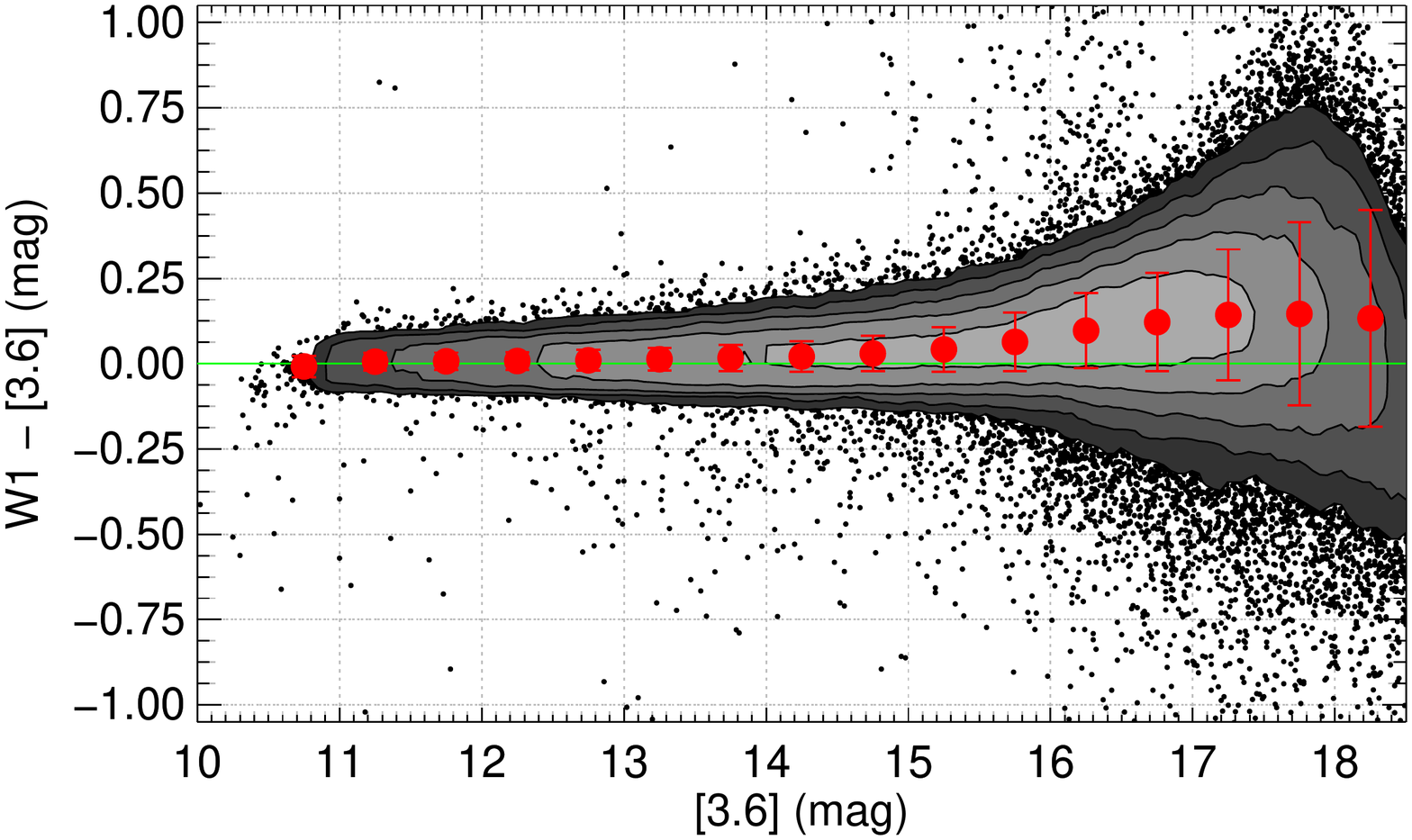}
        \includegraphics[width=0.49\textwidth, trim={0cm  3cm 2cm 4cm}, clip]{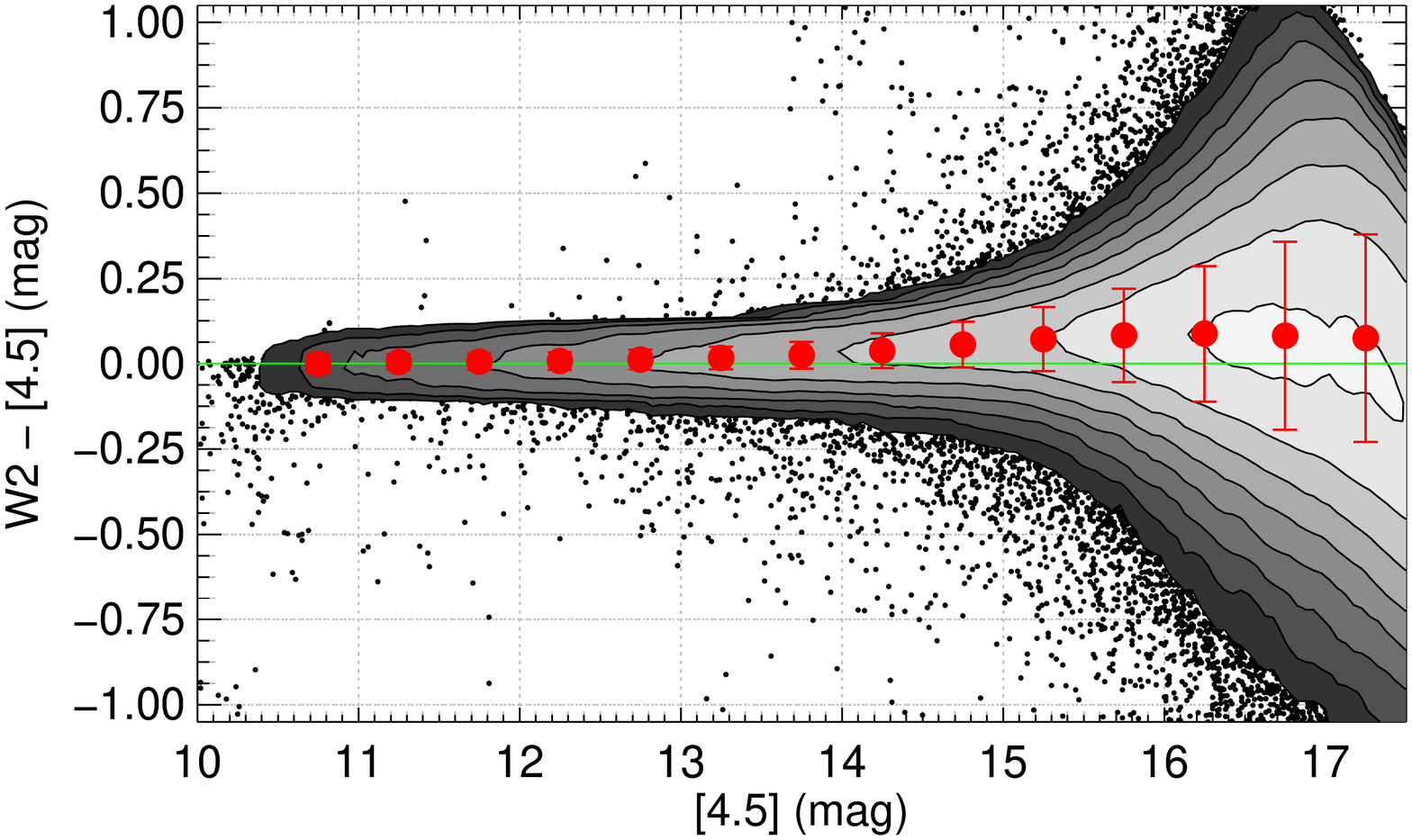}
        \caption{Comparison of CatWISE photometry to {\it Spitzer} photometry for the SSDF, using the same methodology as in Figure \ref{fig:CatWISE_vs_COSMOS}. The outer contour represents a source density of 10 sources per $0.05 \times 0.05$ mag bin, with each additional contour showing a factor of two increase in source density. Note the three additional contour levels on the right, due to the lack of the color cut used on the left.} 
        \label{fig:CatWISE_vs_SSDF}
    \end{figure}

Figure~\ref{fig:CatWISE_vs_COSMOS} compares CatWISE Preliminary Catalog PSF-fitting photometry to 2\farcs9 radius aperture photometry from the {\it Spitzer} S-COSMOS program \citep{Sanders2007}. These observations were obtained using longer integration times (20 minutes) than for the SSDF, while the CatWISE integration is lower (12 minutes) than in the SSDF, so the S-COSMOS data are much deeper than CatWISE. The closest CatWISE source within 2\farcs75 was taken as the match to the S-COSMOS source. Because the CatWISE photometry is point source fitting, S-COSMOS sources were required to have $<10\%$ flux increase between the 1\farcs9 and 2\farcs9 radius apertures. In addition, because the W1 band is significantly bluer than the [3.6] band, S-COSMOS sources at [3.6] were required to have $ -0.1 \leq [3.6] - [4.5] \leq 0$ (see Figures 2 and 3 in \S VI.3.a of the All-Sky Explanatory Supplement; \citealp{Cutri2012}). While necessary for comparing W1 and [3.6] photometry, the color cut significantly reduces the numbers of sources for comparison.   Figure~\ref{fig:CatWISE_vs_SSDF} gives the analogous comparison to photometry from the SSDF survey. 

The comparison between the CatWISE and {\it Spitzer} photometry is consistent for both fields, in both bands.  CatWISE photometry  becomes $\sim 0.1$ mag fainter than {\it Spitzer} beyond 16th mag, possibly due to the increasing incidence of extragalactic sources at faint magnitudes. The measured scatter in the SSDF reaches 0.217 mag, equivalent to an SNR of 5, at [3.6] = 17.64 mag and [4.5] = 16.49 mag. Adjusting for the mean offsets in  W1--[3.6] and W2--[4.5] at these magnitudes, and subtracting 0.14 mag to adjust the SSDF 125 exposure depth to the 96 exposure depth that we take as the baseline for the CatWISE Preliminary Catalog, we find that the SNR=5 limits for CatWISE are W1=17.67 mag and W2=16.47 mag.

%- in COSMOS

%- in SSDF

\subsection{Astrometric properties  \label{sec:astrom_perf}}

\subsubsection{Full-sky Astrometric Assessments \label{sec:astrom_full_sky}}

The astrometric performance of CatWISE was assessed by comparing to \textit{Gaia} DR2 \citep{Brown2018, Lindegren2018}. Within each tile, the 10 brightest sources in bins of 0.5\,mag over the $10<$W1$<17.5$\,mag range were selected, providing a sample of 150 sources per tile, uniformly distributed on the sky. At low Galactic latitudes, CatWISE does not reach to W1=17.5\,mag, and therefore the faintest bins are empty. The sample for astrometric comparison consists of 2,699,315 sources. These sources were cross-matched with \textit{Gaia} DR2 using a 5\farcs5 radius (corresponding to two {\it WISE} pixels), requiring the \textit{Gaia} counterpart to have measured proper motions. This returned 2,148,274 unique matches. The completeness of the match approaches $100\%$ for sources brighter than W1=14\,mag, and drops to $\sim25\%$ at the faintest magnitudes. \textit{Gaia} astrometry was used to propagate the \textit{Gaia} counterparts to the CatWISE epoch, and the standard deviation\footnote{The IDL ``robust\_sigma" function was used to calculate the standard deviation.} between the CatWISE motion-fit and \textit{Gaia} position and motion values was computed.

\begin{figure*}
\centering
\includegraphics[trim={2cm 3cm 2cm 2cm}, clip, width=0.49\textwidth]{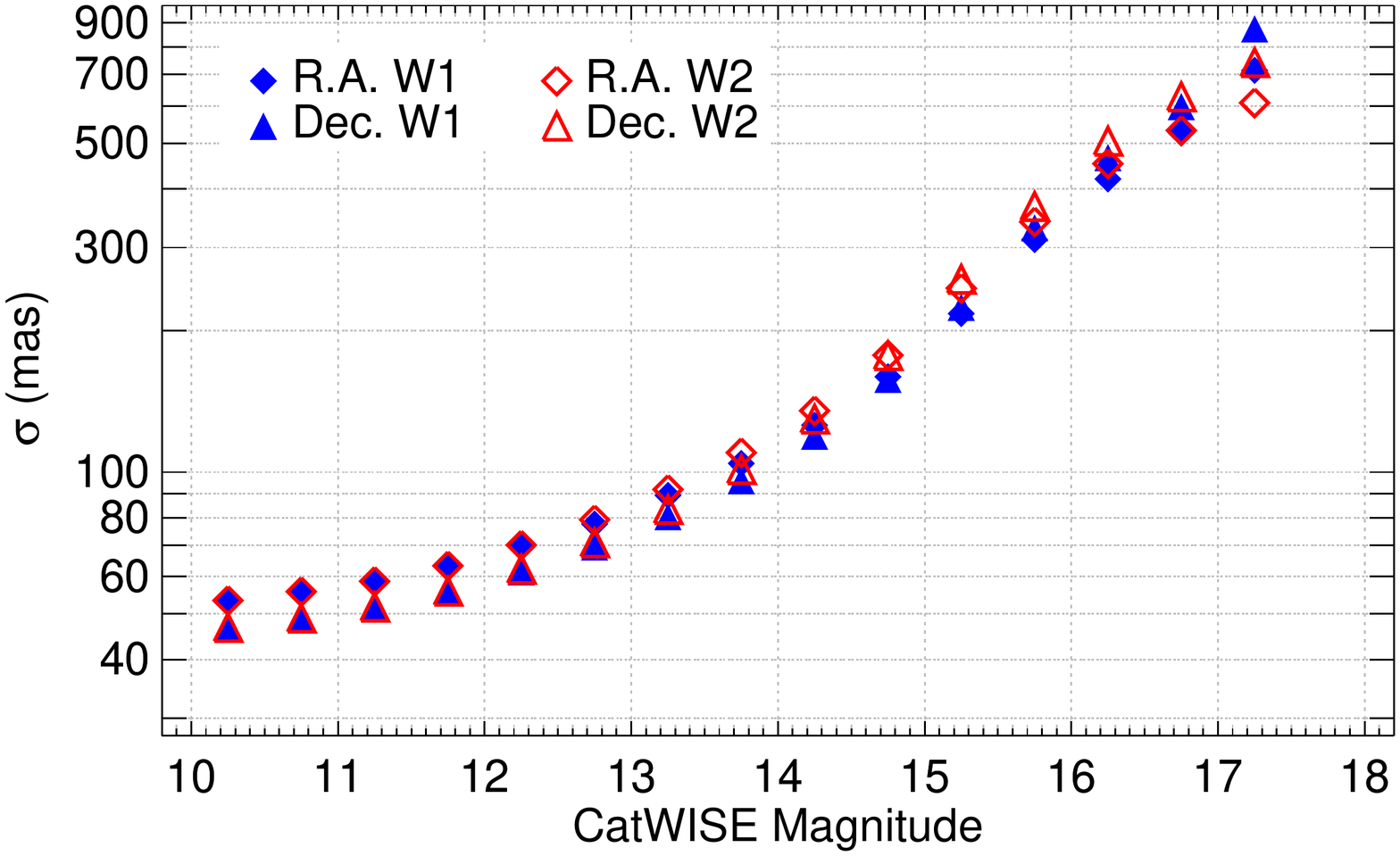}
\includegraphics[trim={2cm 3cm 2cm 2cm}, clip, width=0.49\textwidth]{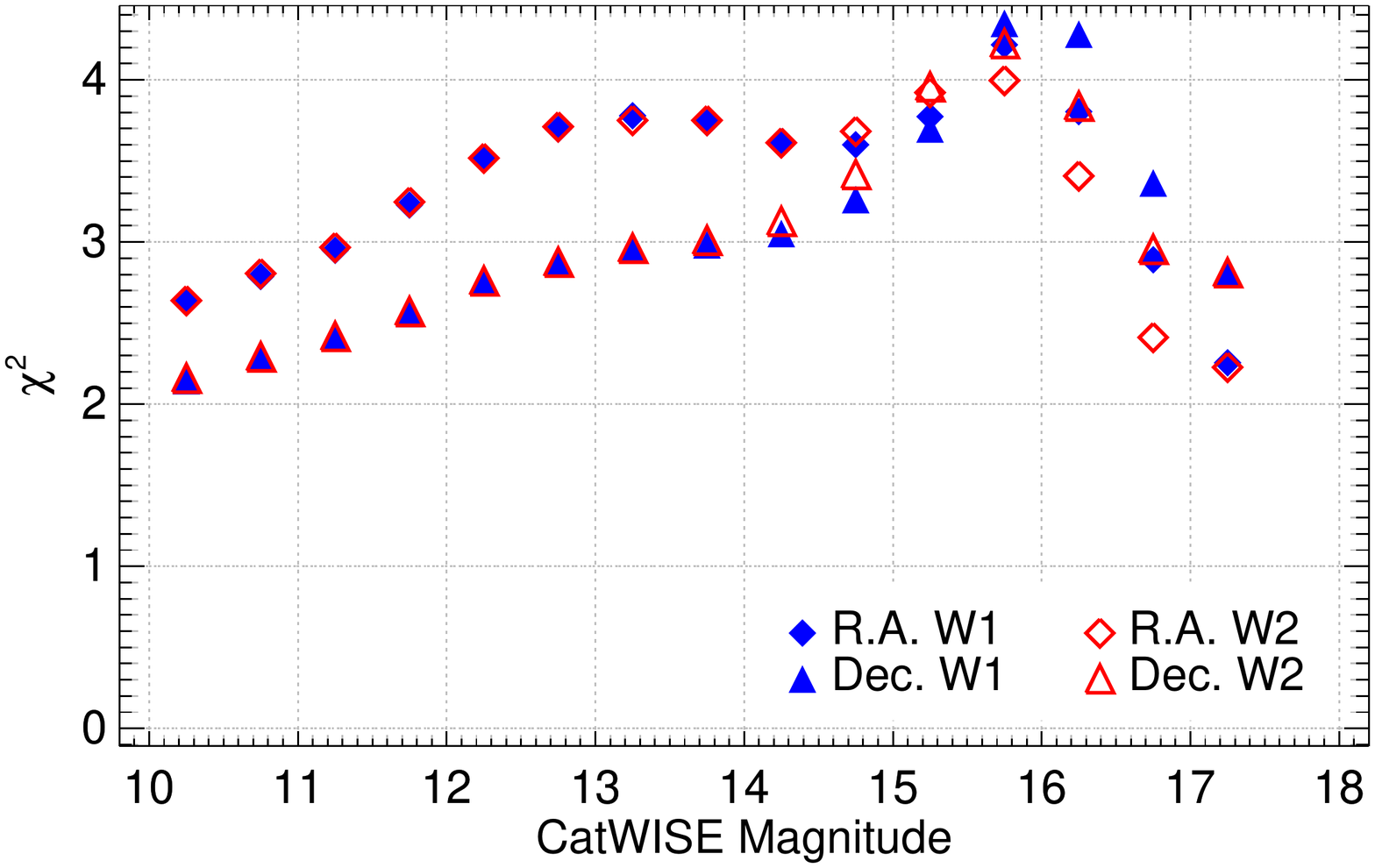}
\includegraphics[trim={2cm 3cm 2cm 2cm}, clip, width=0.49\textwidth]{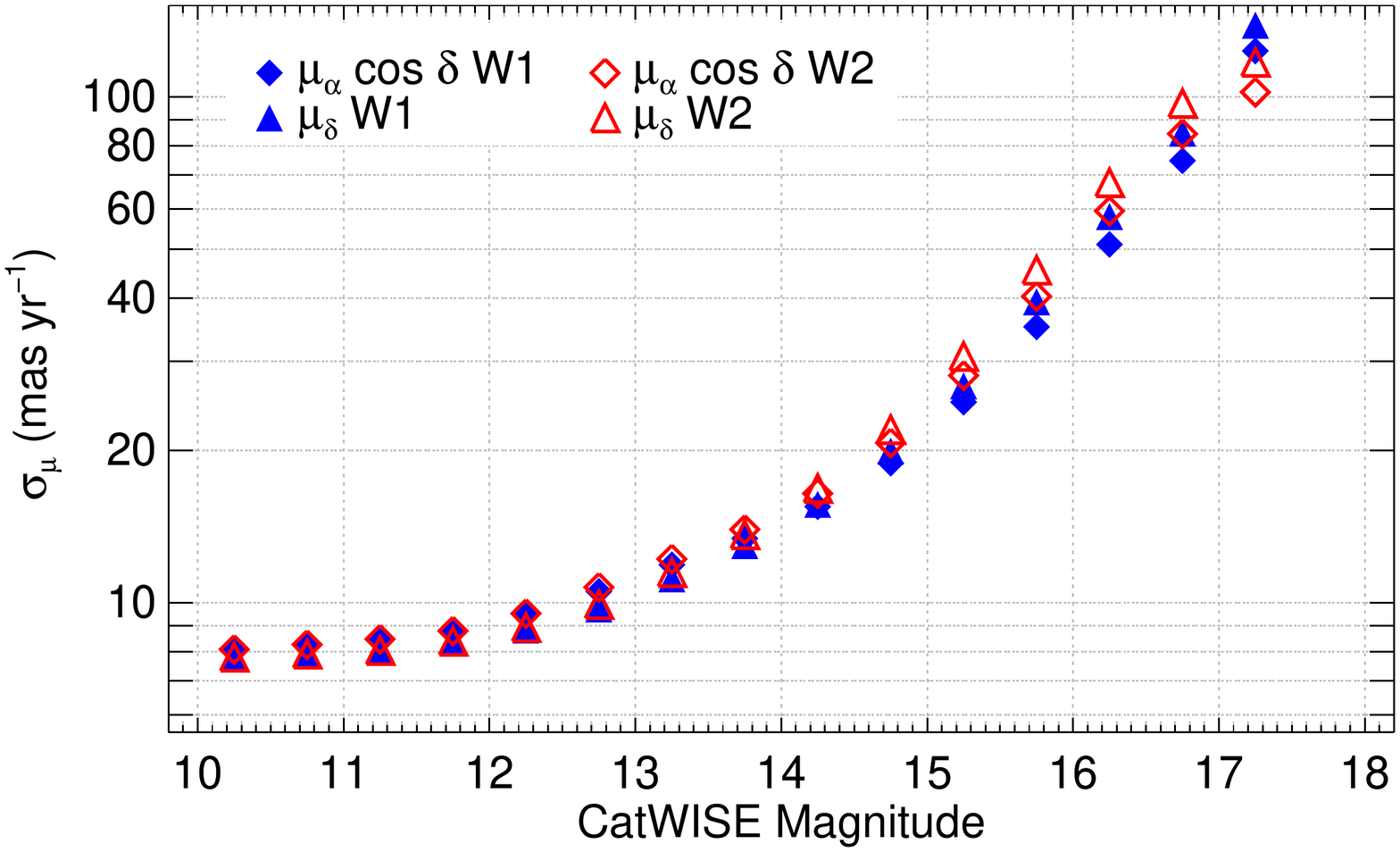}
\includegraphics[trim={2cm 3cm 2cm 2cm}, clip, width=0.49\textwidth]{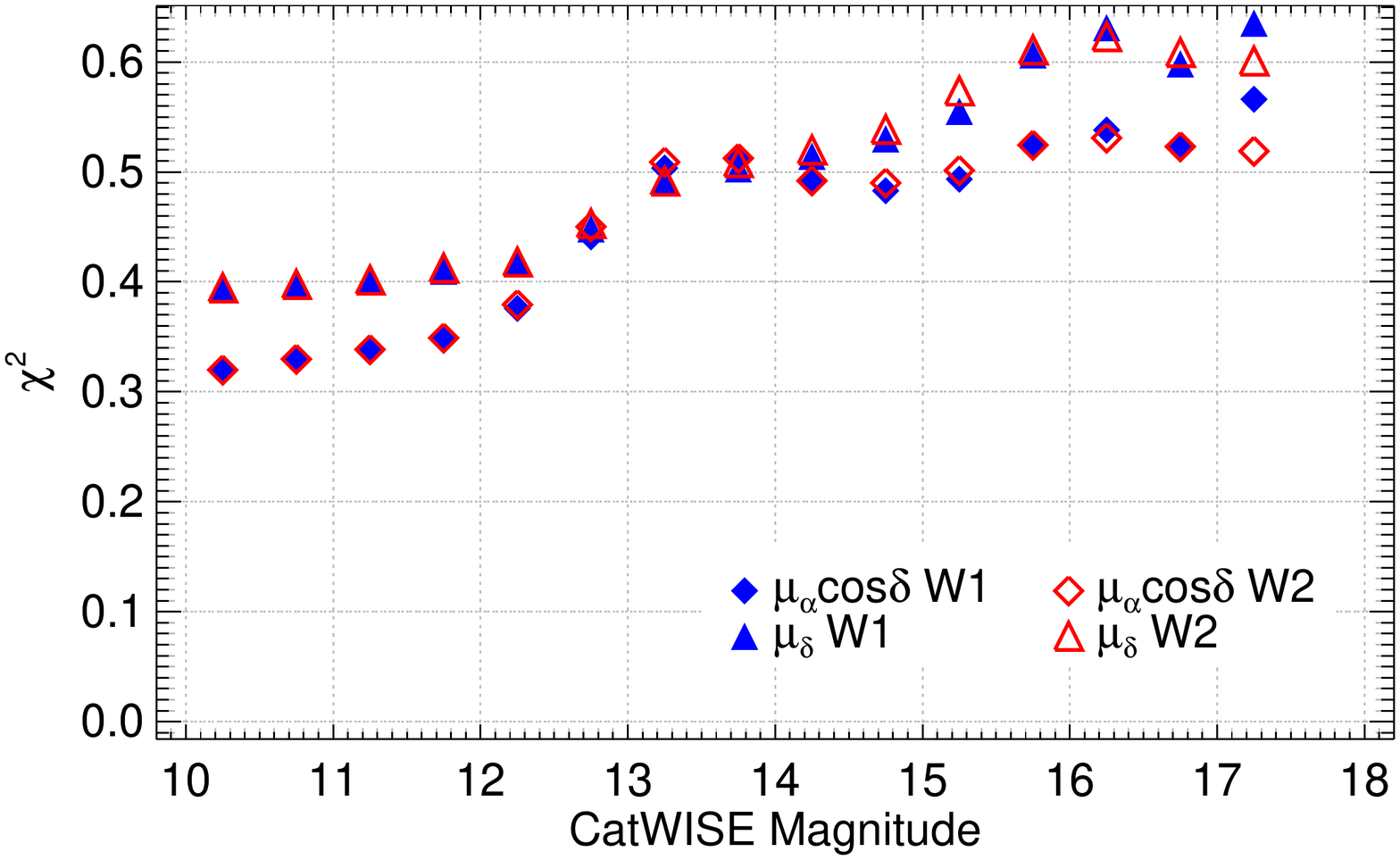}
\caption{CatWISE astrometric performace with respect to \textit{Gaia} DR2. \textit{Left:} the 1-$\sigma$ dispersion between CatWISE and \textit{Gaia} R.A. (specifically, $\Delta\alpha\ \textrm{cos}(\delta)$) and Dec. (top) and proper motion (bottom), for a subsample of $\sim$2.1 million sources in the $10<$W1$<17.5$\,mag range, uniformly distributed over the entire sky. \textit{Right:} the median $\chi^2$ computed taking into account CatWISE catalog uncertainties, \textit{Gaia} catalog uncertainties, and the uncertainty introduced by the translation of \textit{Gaia}'s positions to the CatWISE epoch. \label{fig:catwise_vs_gaia_fullsky}}
\end{figure*}

Figure~\ref{fig:catwise_vs_gaia_fullsky} summarizes the results of the full sky comparison. The positional accuracy floor for bright sources approaches $\sim50$\,mas and remains approximately constant until W1$\sim12.5$\,mag. At fainter magnitudes, the dispersion increases to 275\,mas (1/10 of a pixel) at W1$\sim15.5$\,mag, while in the faintest magnitude bin the dispersion is $\sim700$\,mas, with the dispersion in $\alpha$ being slightly better than in $\delta$. This is expected since the scan direction is closer to $\delta$ and the PSF is more elongated in the scan direction. At the bright end, however, the accuracy in $\delta$ is better than in $\alpha$, an effect not fully understood. 

The motion accuracy floor for bright stars is just under 10\,mas\,yr$^{-1}$, consistent with the positional accuracy floor -- given CatWISE's 6 year baseline, and since the motion accuracy scales linearly with time, one would expect $\sigma_\mu \sim \sigma_{\rm pos} / 6$. At W1$\sim$15.5\,mag, the motion accuracy is $\sim$30\,mas\,yr$^{-1}$, a factor of 10 better than AllWISE \citep{Kirkpatrick2014}. At the faint end, CatWISE is sensitive to proper motion of 100\,mas\,yr$^{-1}$ until W1$\sim$17\,mag.

The two panels on the right of Figure~\ref{fig:catwise_vs_gaia_fullsky} show the $\chi^2$ computed using the CatWISE catalog uncertainties, the \textit{Gaia} catalog uncertainties, and the uncertainty introduced by the translation of the \textit{Gaia} position to the CatWISE epoch. Since the total uncertainty is dominated by  CatWISE, these $\chi^2$ values are essentially a measurement of how accurate the CatWISE Preliminary Catalog uncertainties are. The expected value of the median $\chi^2$ with one degree of freedom is $\sim0.45$, so the top right panel of Figure 13 indicates that the catalog position uncertainties underestimate the actual errors by a factor of 2.5--3. In contrast, the motion uncertainties are consistent with the expected value. We have not investigated the reason for this, although it suggests an additional error in position that is stable for a given source. Confusion is one possible explanation. Although the uncertainties are underestimated, the actual CatWISE position errors are comparable to those for AllWISE (see \S II.5.c of the AllWISE Explanatory Supplement \citep{Cutri2013}.
%It is apparent from the top right panel of Figure~\ref{fig:catwise_vs_gaia_fullsky} that the catalog position uncertainties  underestimate the actual errors by a factor of 1.6--2. In contrast, the motion uncertainties are overestimated by a factor of 1.3--1.7 (lower right panel). The latter effect is due in part to imposing a minimum value of 10\,mas\,yr$^{-1}$ on the motion uncertainties, a setting used by AllWISE and left unchanged for the CatWISE Preliminary catalog.

\begin{figure*}
\includegraphics[width=0.5\textwidth]{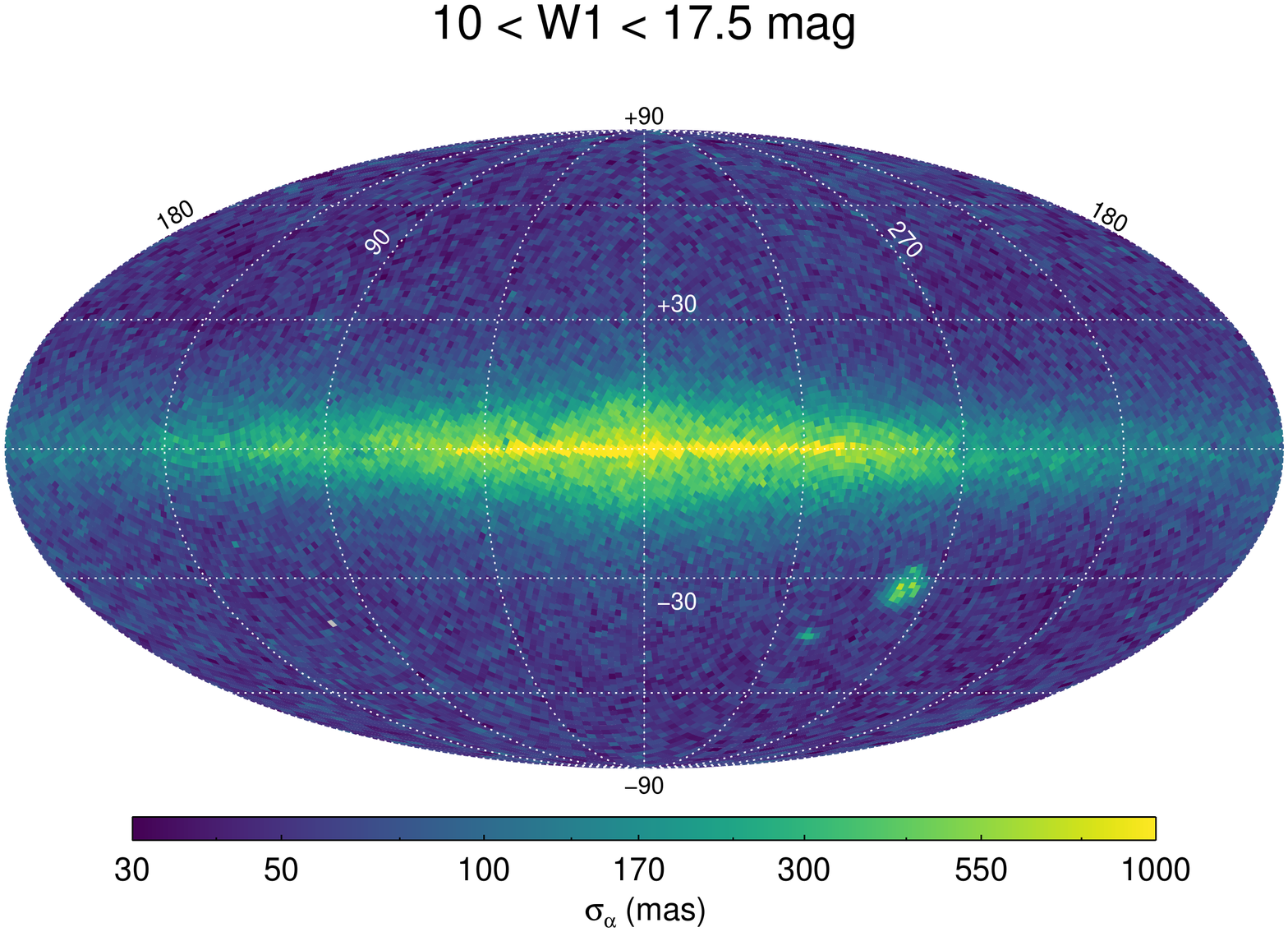}
\includegraphics[width=0.5\textwidth]{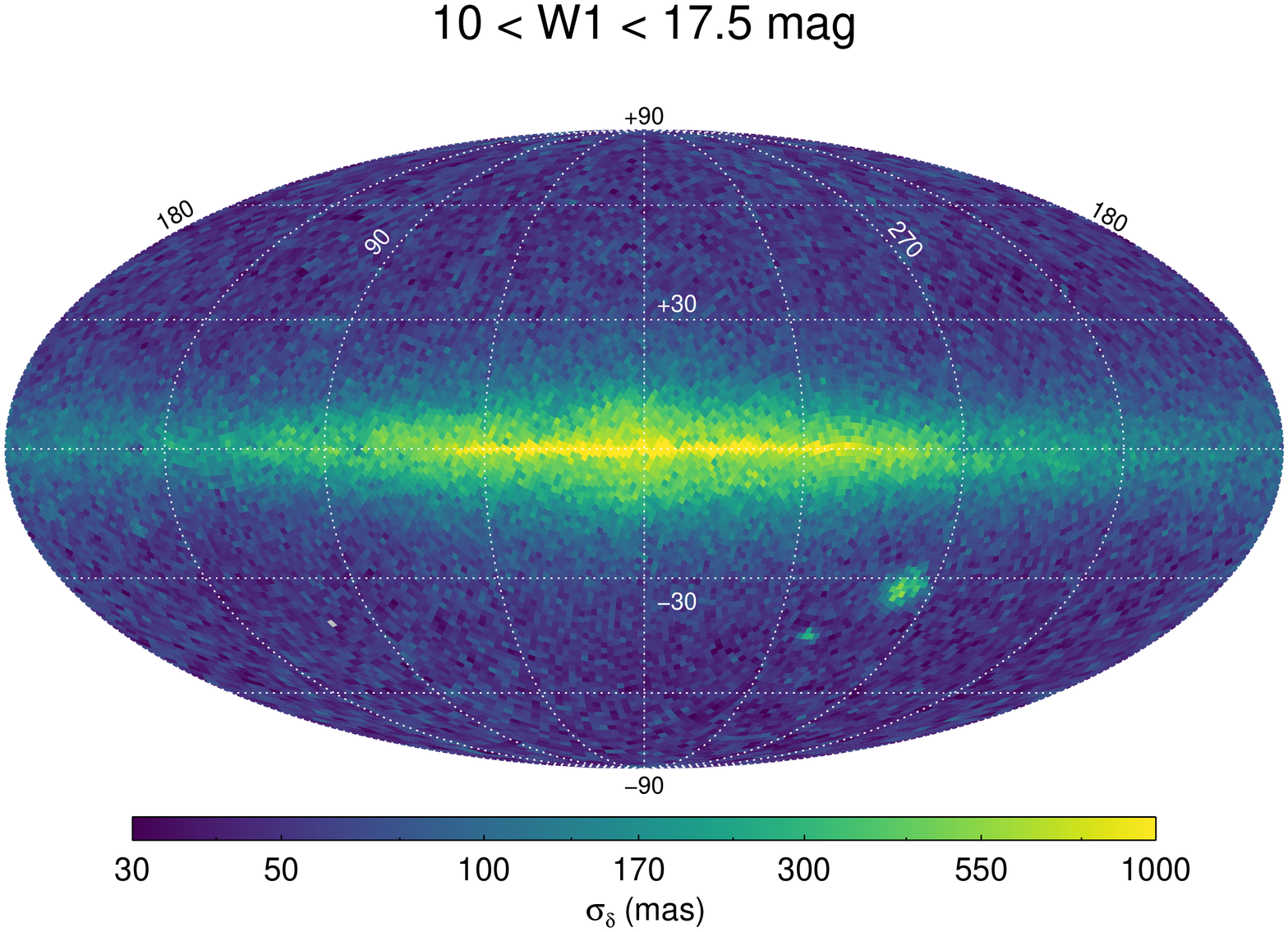}
\caption{1-$\sigma$ dispersion of the CatWISE $\alpha$ (left) and $\delta$ (right) with respect to \textit{Gaia} DR2, for sources in the $10<$W1$<17.5$\,mag range. \label{fig:map_positions}}
\end{figure*}

\begin{figure*}
\includegraphics[width=0.5\textwidth]{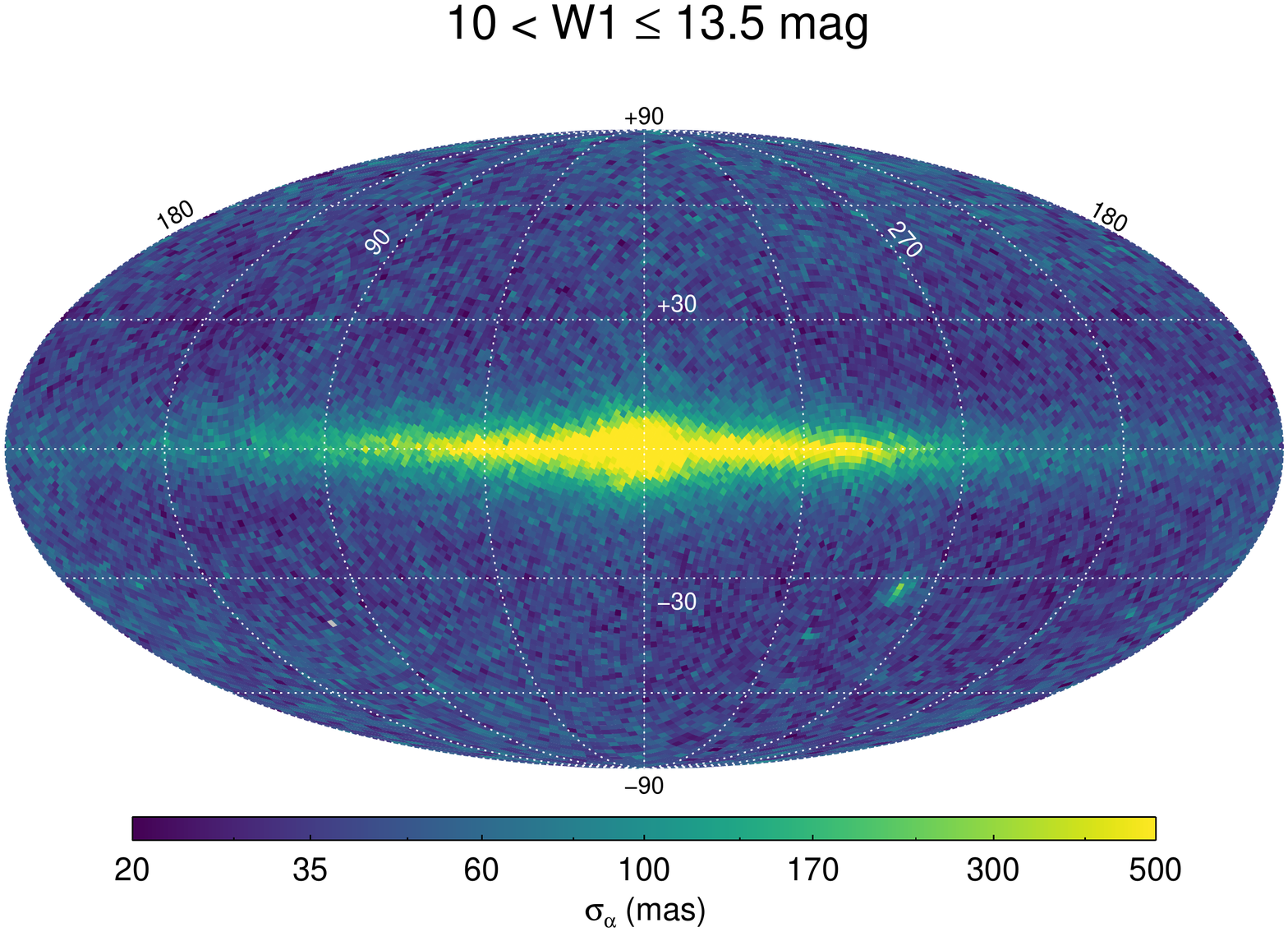}
\includegraphics[width=0.5\textwidth]{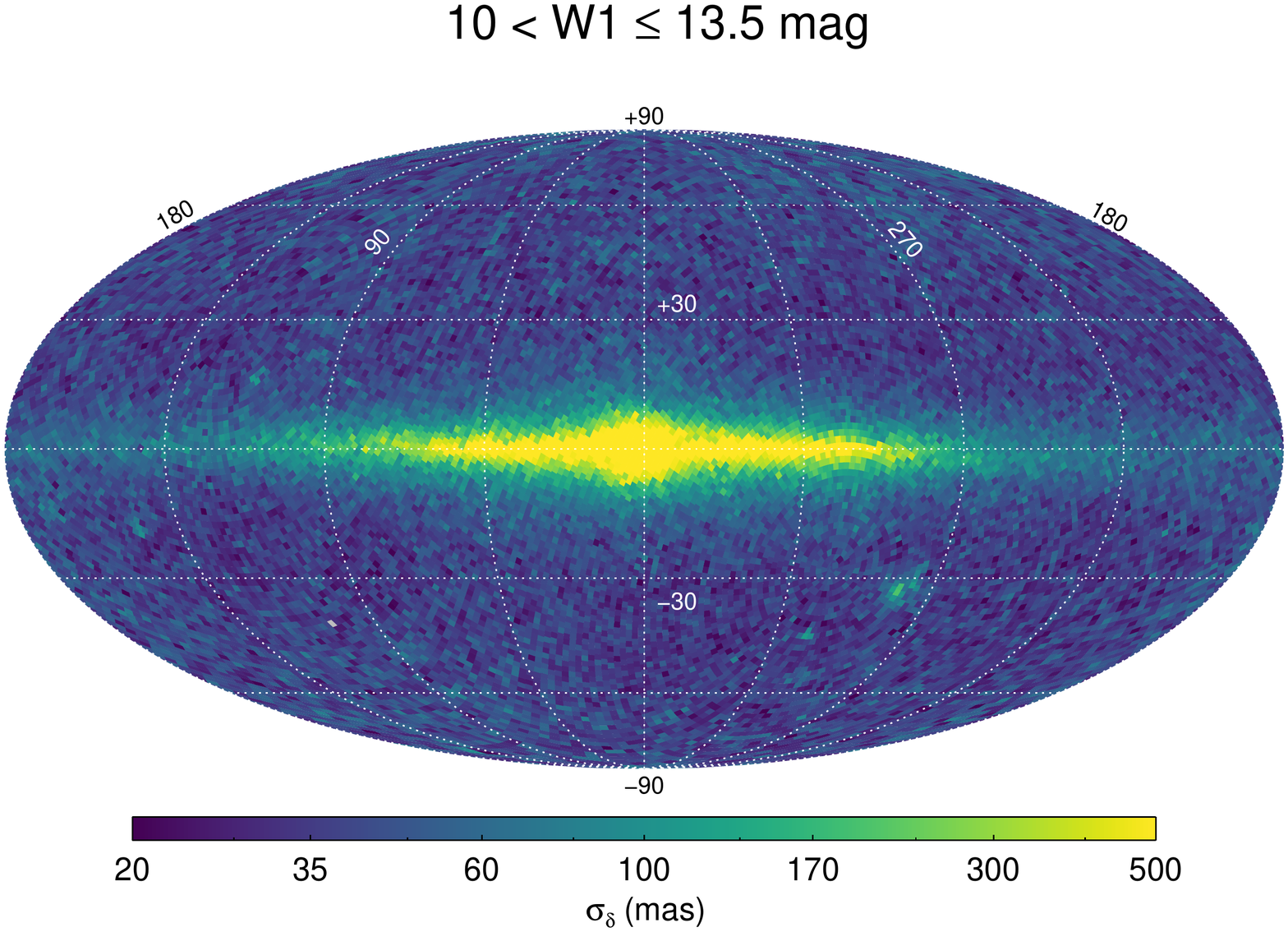}
\includegraphics[width=0.5\textwidth]{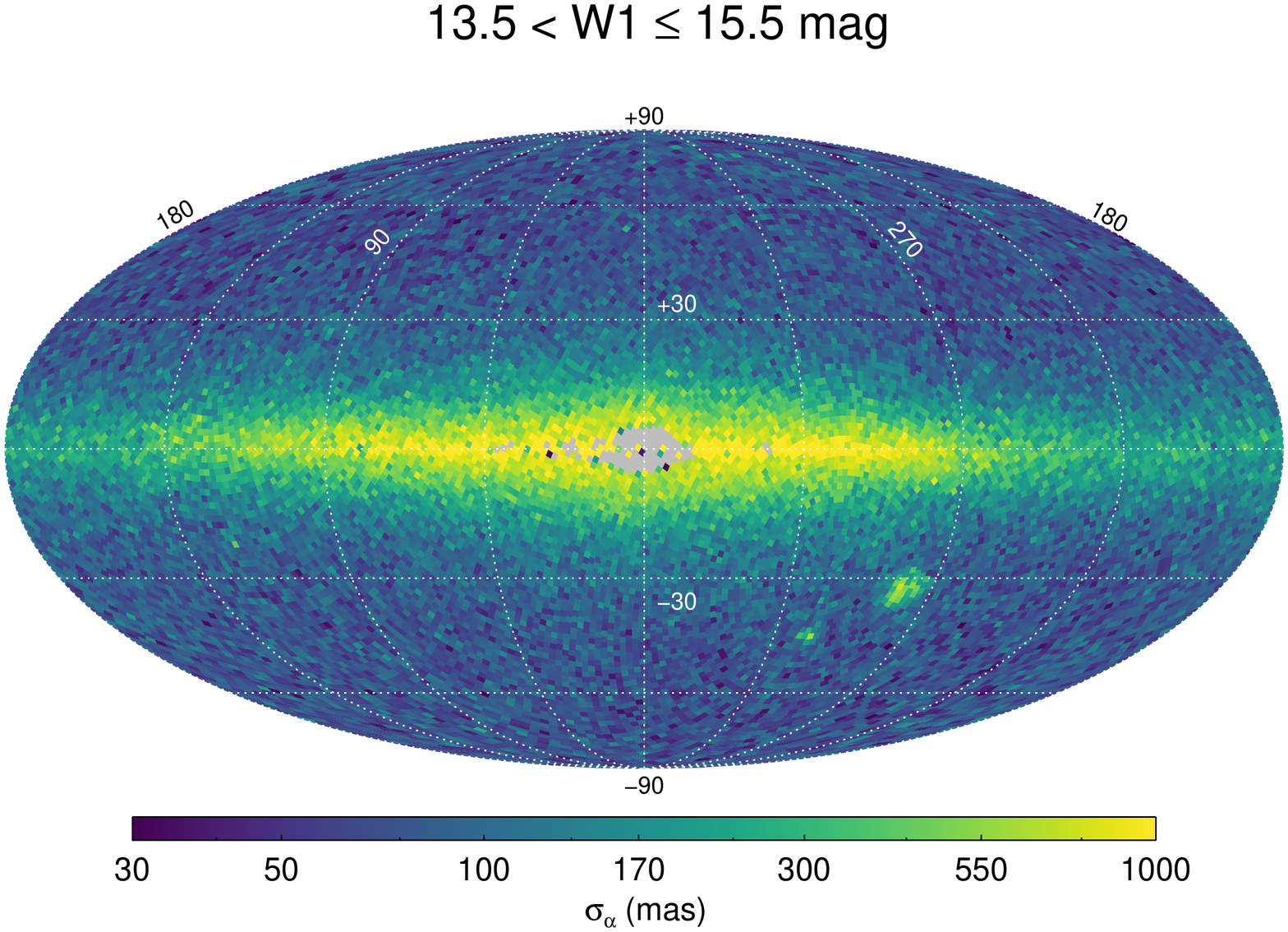}
\includegraphics[width=0.5\textwidth]{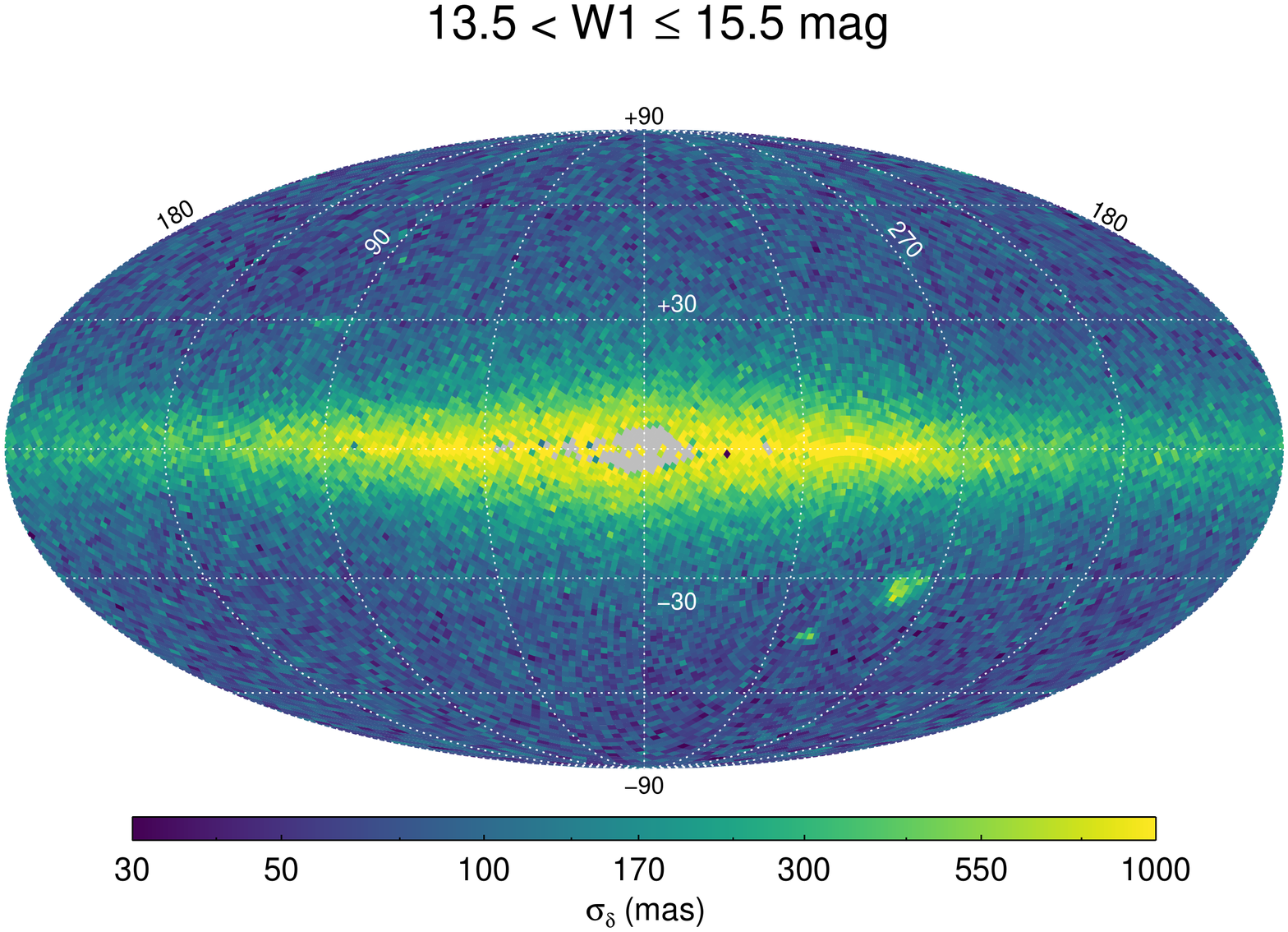}
\includegraphics[width=0.5\textwidth]{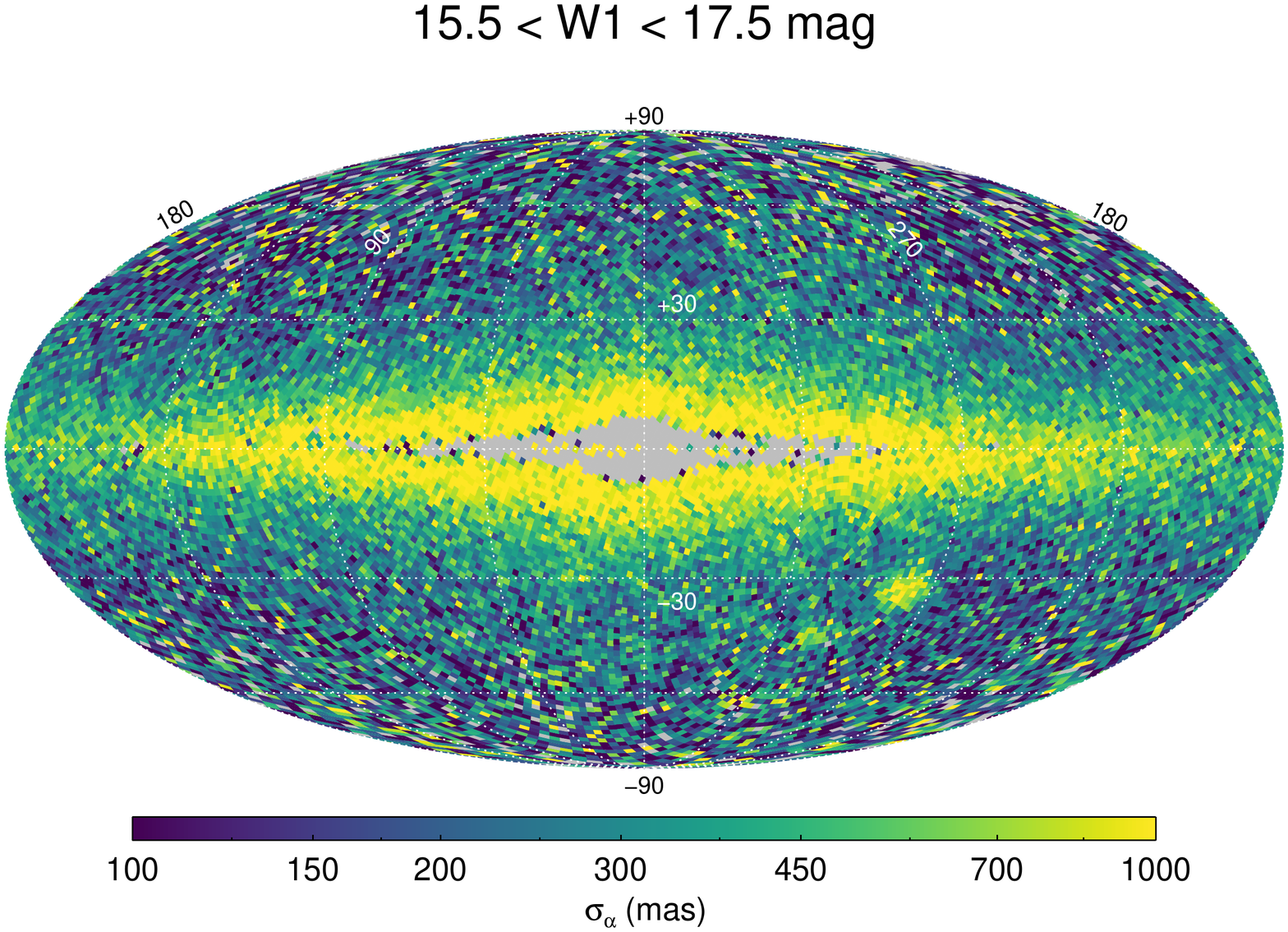}
\includegraphics[width=0.5\textwidth]{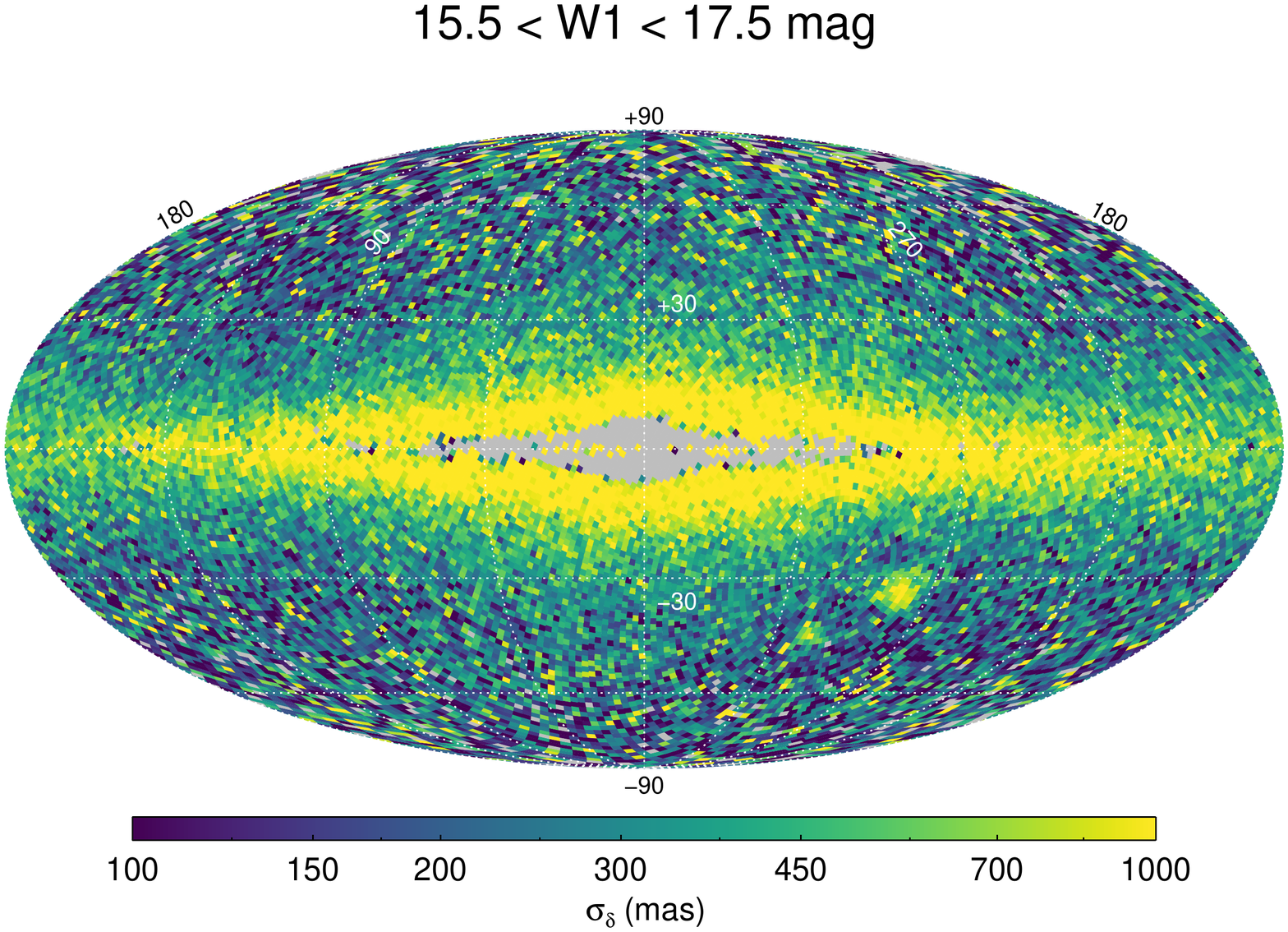}
\caption{Same as Figure~\ref{fig:map_positions}, but for three W1 magnitude ranges. Gray tiles are those where there were no sources in CatWISE in the given magnitude bin. \label{fig:map_positions_bins}}
\end{figure*}

\begin{figure*}
\includegraphics[width=0.5\textwidth]{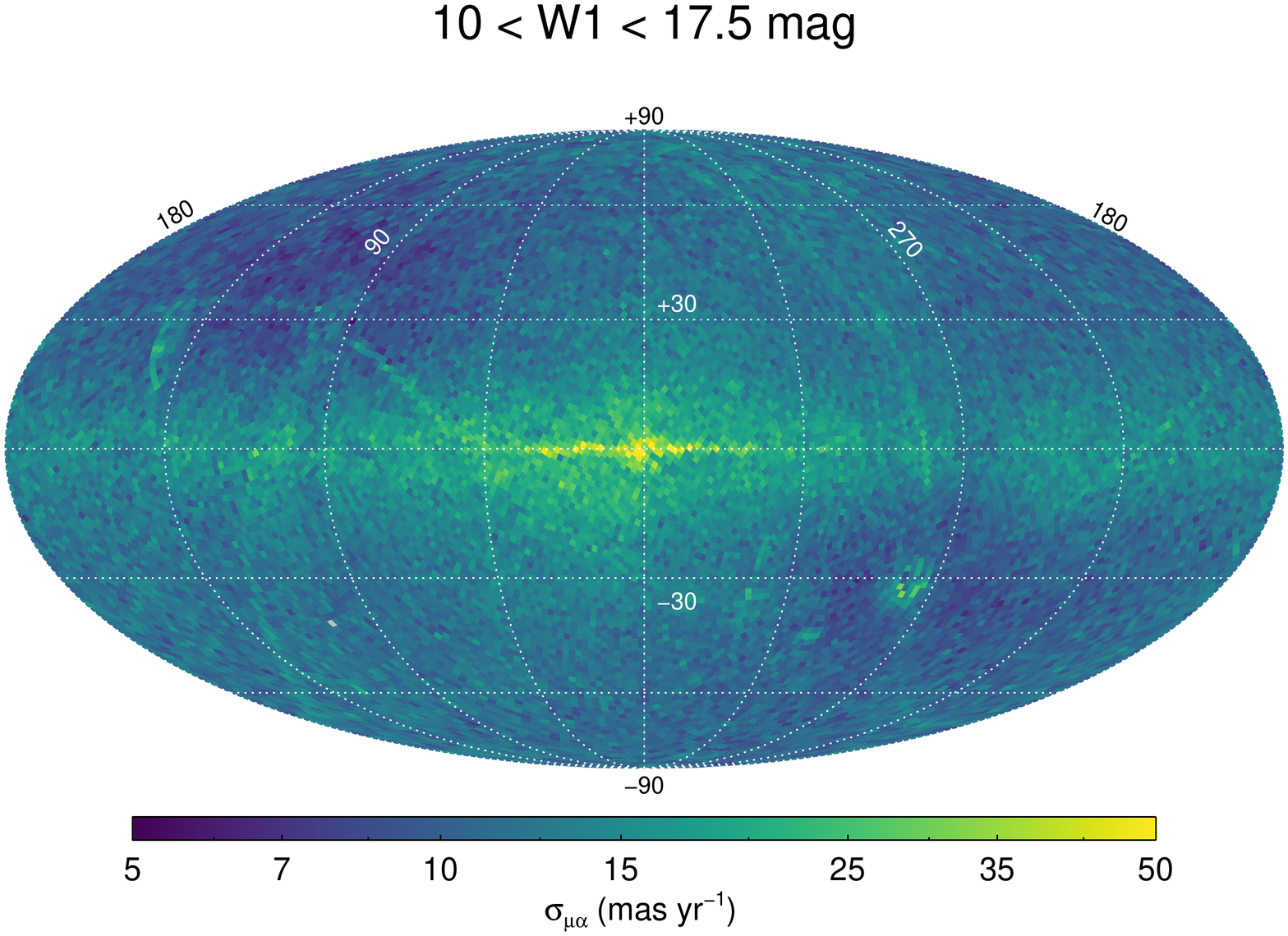}
\includegraphics[width=0.5\textwidth]{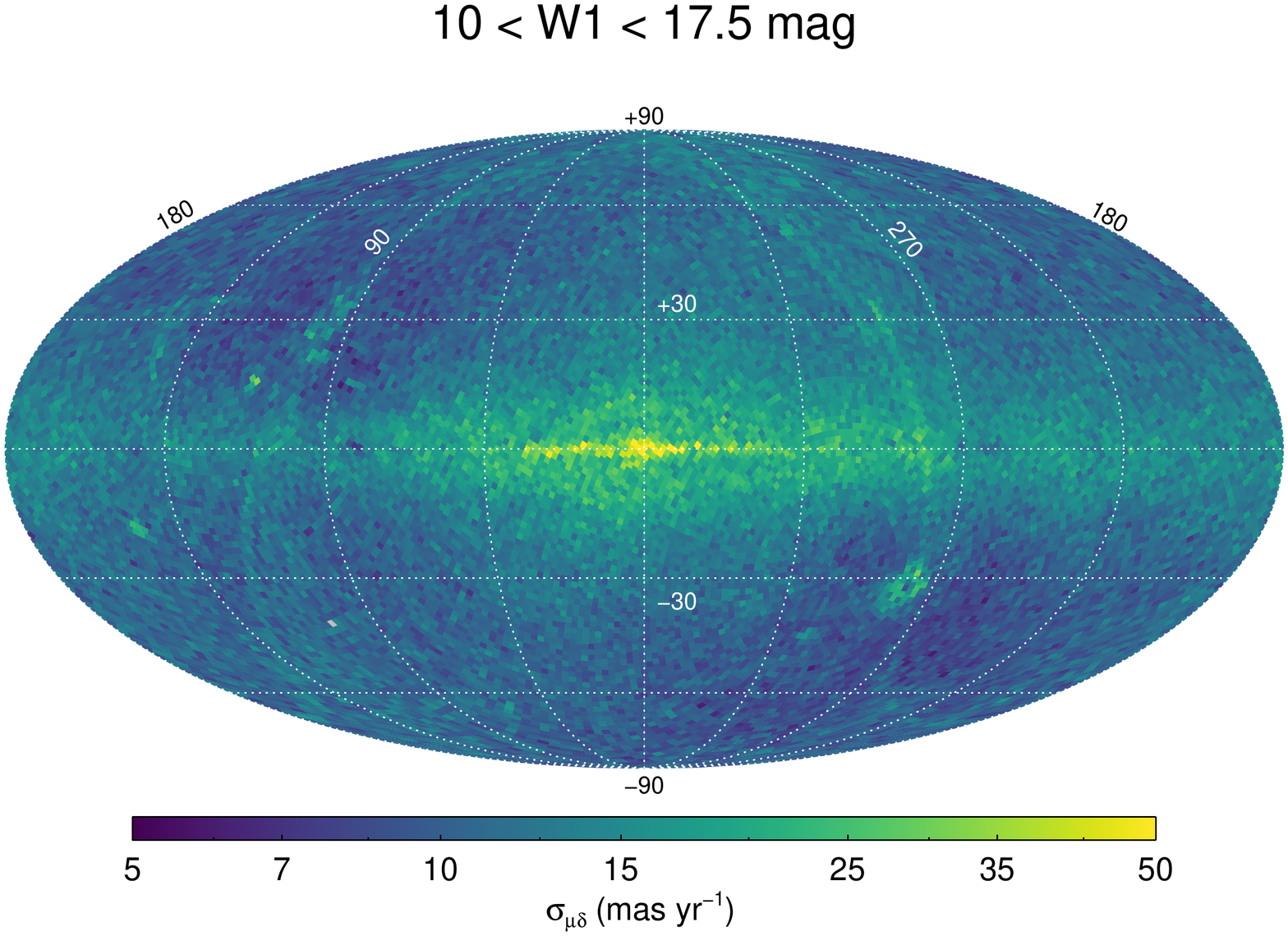}
\caption{Same as Figure~\ref{fig:map_positions}, but for the proper motion components.\label{fig:map_pms}}
\end{figure*}

\begin{figure*}
\includegraphics[width=0.5\textwidth]{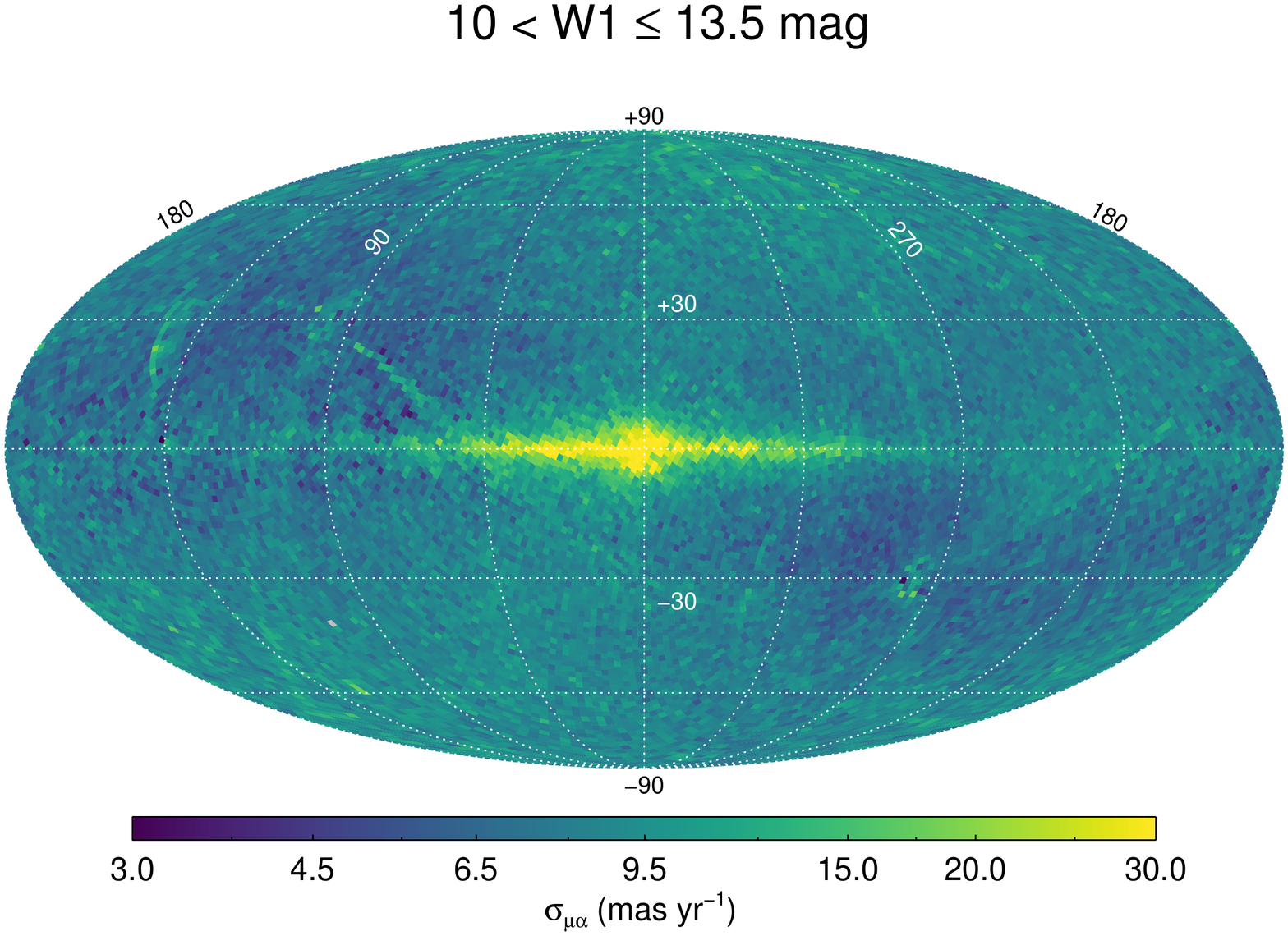}
\includegraphics[width=0.5\textwidth]{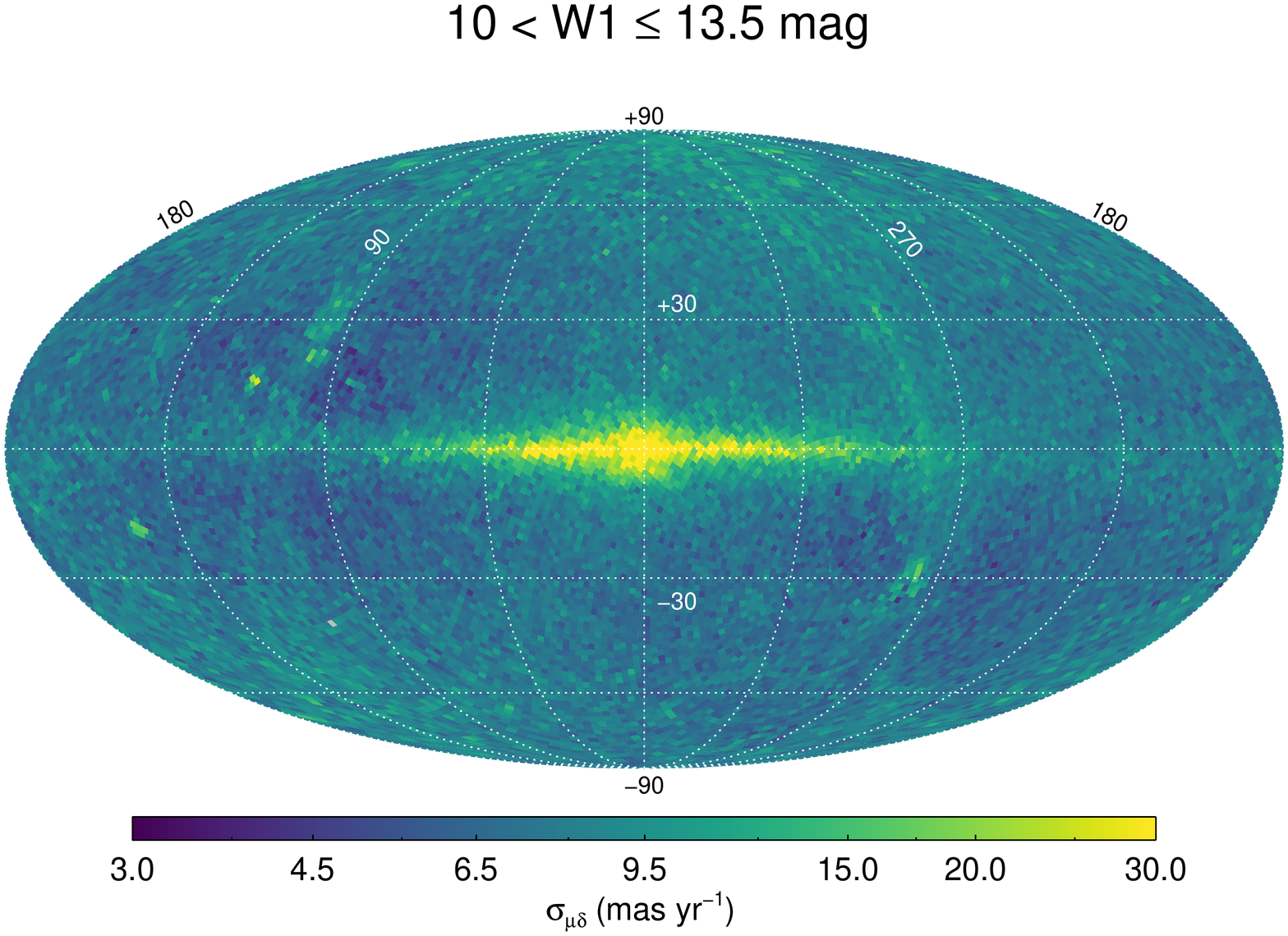}
\includegraphics[width=0.5\textwidth]{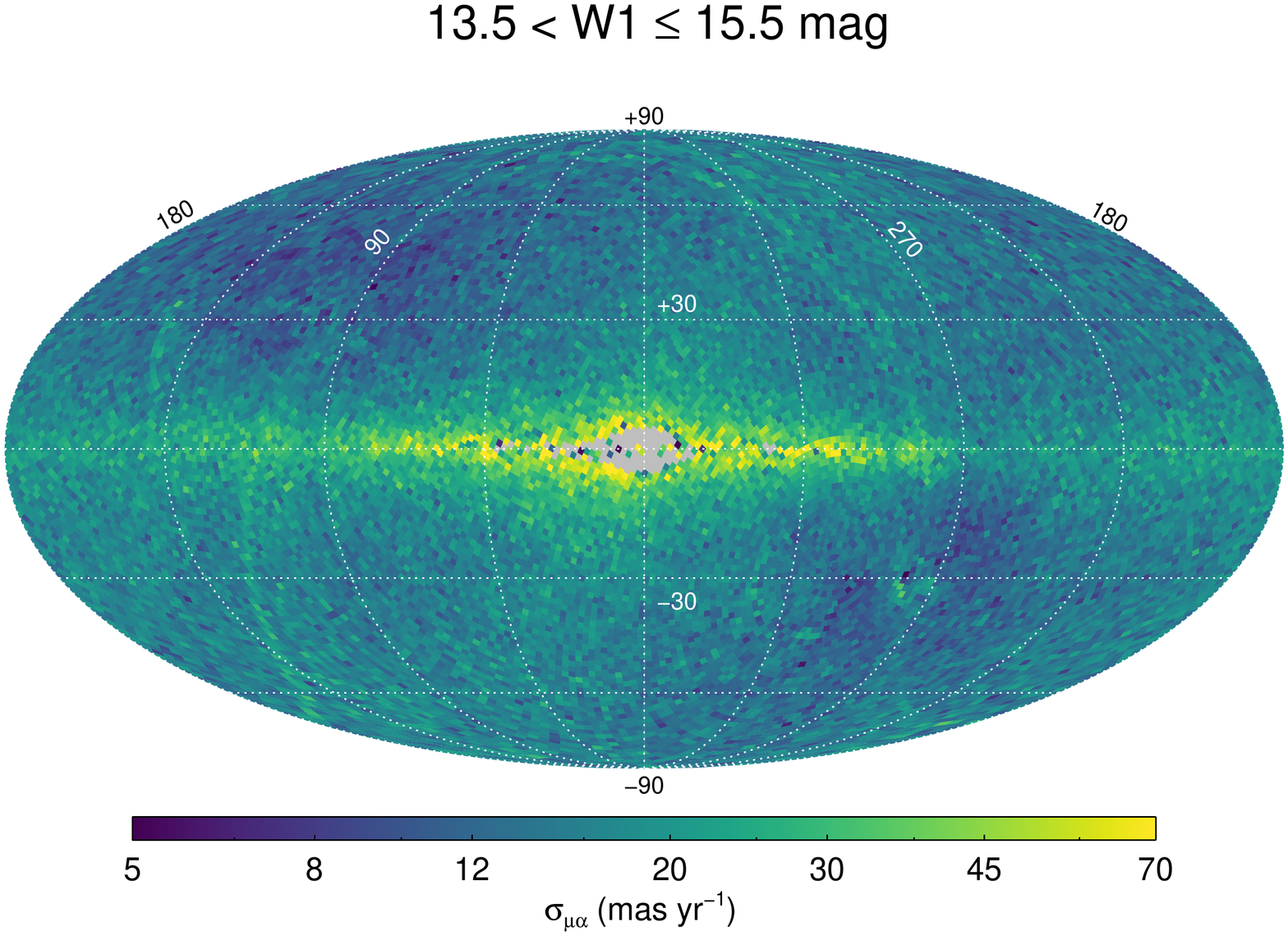}
\includegraphics[width=0.5\textwidth]{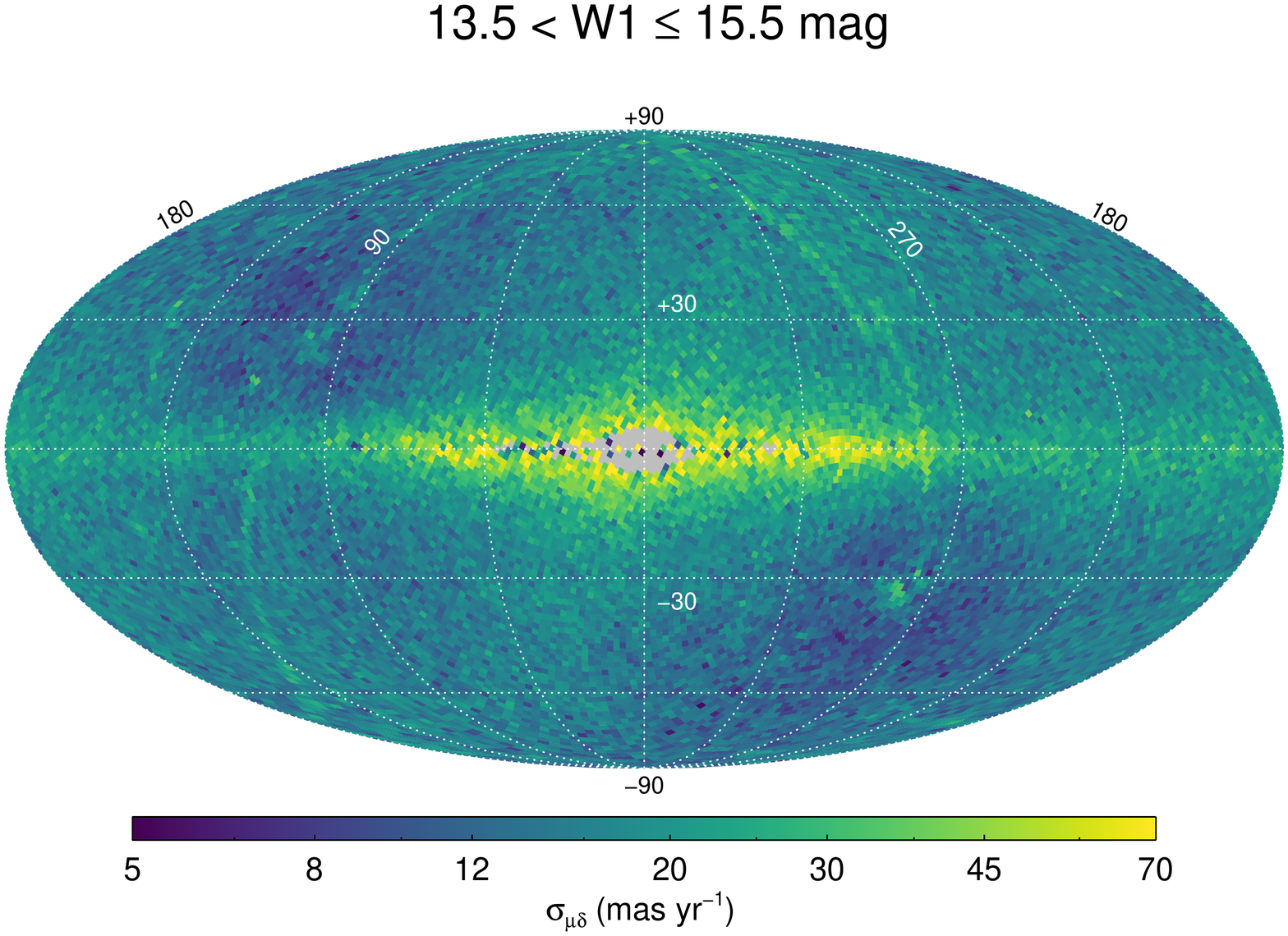}
\includegraphics[width=0.5\textwidth]{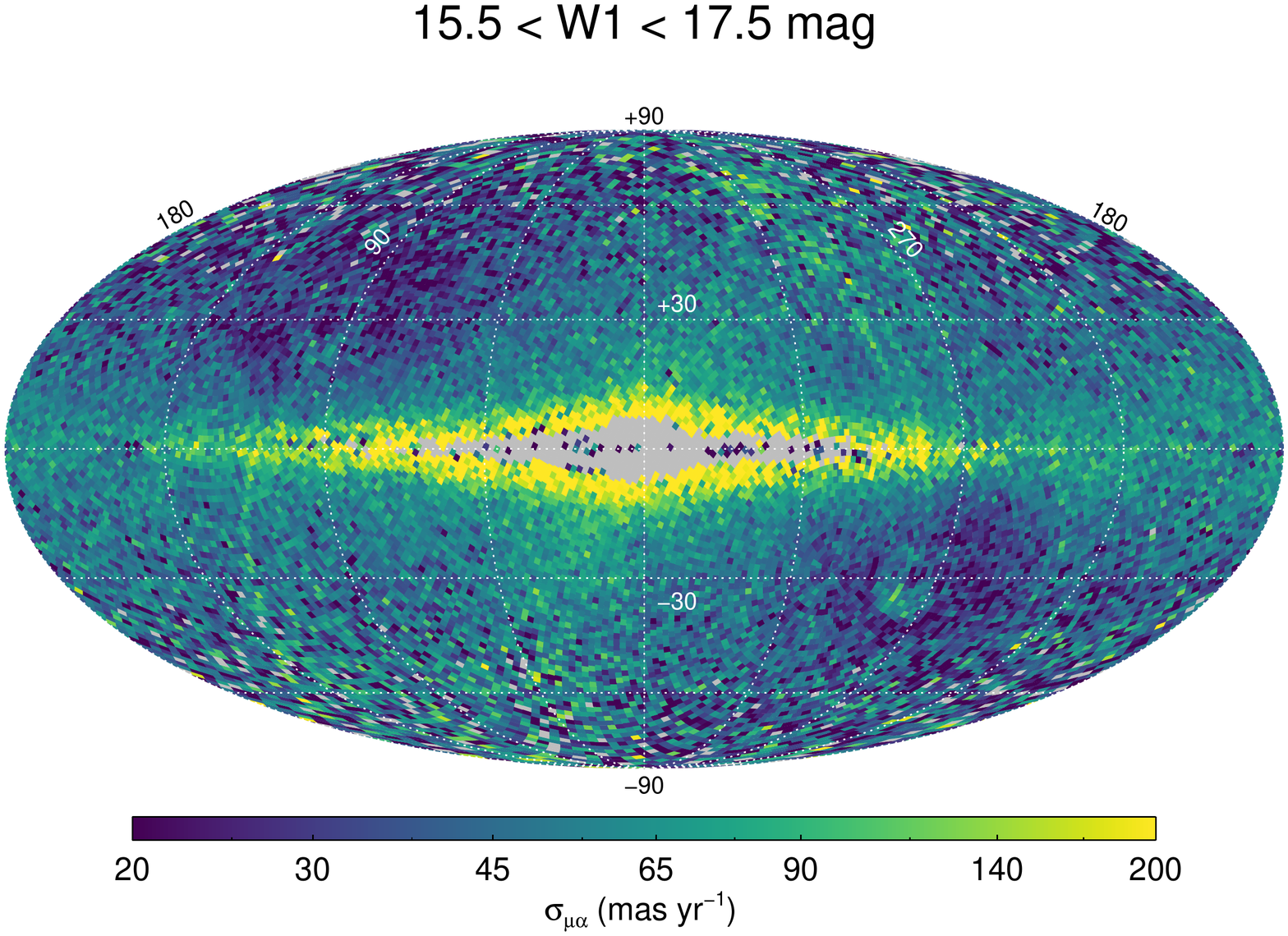}
\includegraphics[width=0.5\textwidth]{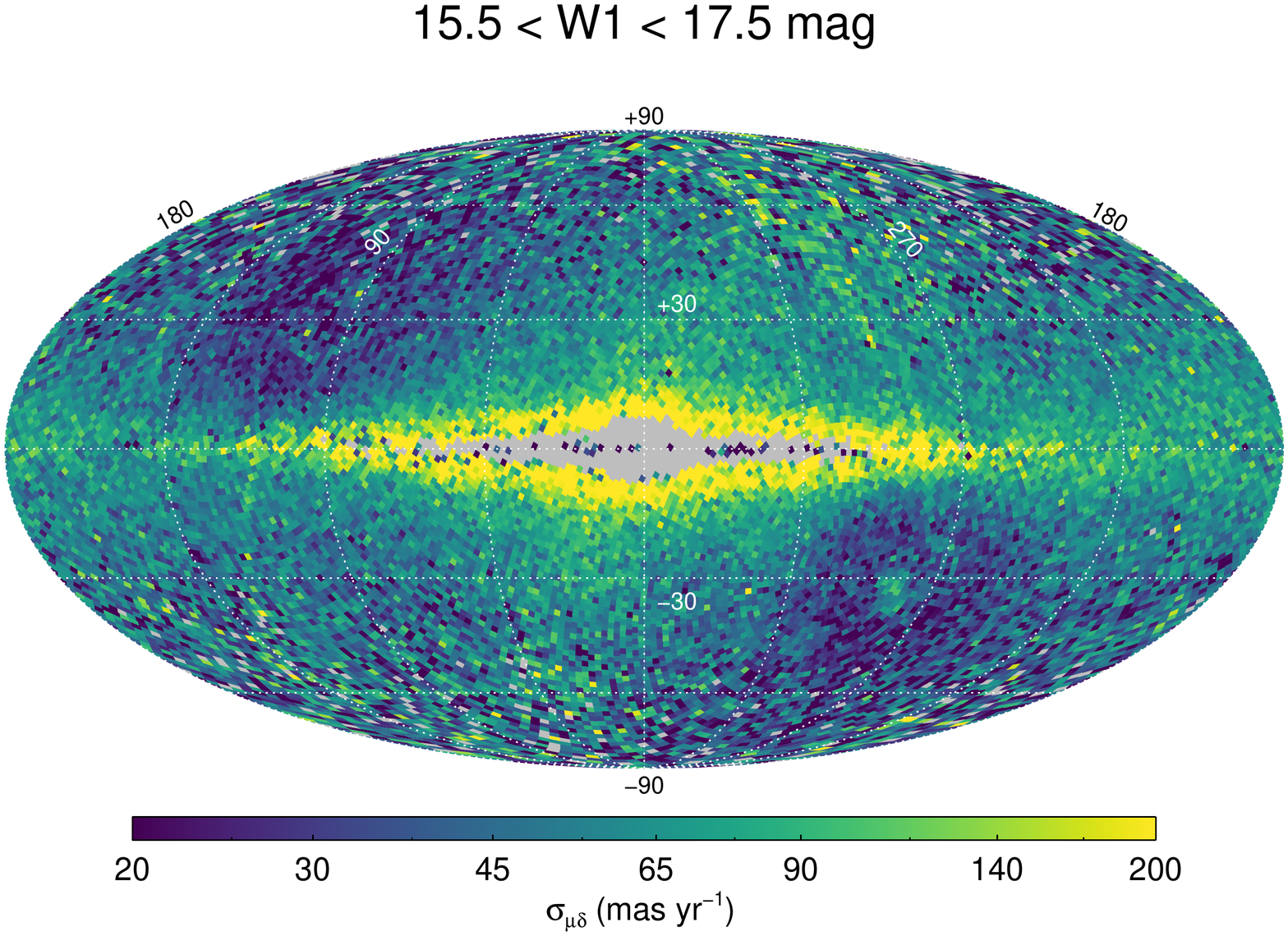}
\caption{Same as Figure~\ref{fig:map_pms}, but for three W1 magnitude ranges. Gray tiles are those where there were no sources in CatWISE in the given magnitude bin.\label{fig:map_pms_bins}}
\end{figure*}

The astrometric performance is however not uniform over the sky. Figures~\ref{fig:map_positions}--\ref{fig:map_pms_bins} show the 1-$\sigma$ dispersion in each tile with respect to \textit{Gaia} positions and motion components for the full magnitude range considered (Figures~\ref{fig:map_positions} and \ref{fig:map_pms}), and in three smaller magnitude intervals (Figures~\ref{fig:map_positions_bins} and \ref{fig:map_pms_bins}). The maps for the full magnitude range are smooth overall, indicating a fairly constant astrometric performance for CatWISE over the majority of the sky. The main features can be easily identified -- the Galactic plane (and in particular the bulge), and the Small and Large Magellanic Clouds (SMC and LMC). In those denser regions, the astrometric accuracy for the bright stars deteriorates to $\sim500$\,mas for positions and $\sim30$\,mas\,yr$^{-1}$ for motions, and to $\sim1,000$\,mas and $\sim200$\,mas\,yr$^{-1}$ (or worse) for the faint stars. 

The motion accuracy maps show additional features that appear to be related to the  \textit{WISE} survey strategy, and to the transition between the cryogenic and post-cryogenic phases of the mission. In particular, one can see better motion performance approaching the ecliptic poles (at $l=96^\circ; b=30^\circ$ and $l=276^\circ; b=-30^\circ$, respectively) compared to regions of similar Galactic latitude. This is a result of the increased coverage of the sky by \textit{WISE} at higher ecliptic latitude. Conversely, there are two strips of slightly poorer motion performance, which are coincident with the transition between cryogenic and post-cryogenic phases, where the PSF was changing rapidly as the telescope was warming up.

The maps for the faintest magnitude interval appear noisy. This is partly intrinsic, and partly an effect of statistical fluctuations due to the low number of sources with a \textit{Gaia} counterpart in each tile (as discussed above, the completeness for the \textit{Gaia} match drops to $\sim25\%$ at the faint end). Moreover, at lower Galactic latitudes, CatWISE detects only brighter sources because of confusion noise.

\subsubsection{Astrometric Assessments in Selected Tiles \label{sec:astrom_4tiles}}

We next consider the astrometric performance in more detail by focusing on four tiles:

\begin{itemize}
    \item the North Ecliptic Pole tile (NEP, tile 2709p666), a field with maximal {\it WISE} coverage and average source density;
    \item the South Ecliptic Pole tile (SEP, tile 0890m667) a field with with maximal {\it WISE} coverage and high source density (the SEP tile contains part of the LMC);
    \item the Galactic Center tile (GC, tile 2657m288) a field with average {\it WISE} coverage and maximal source density; and
    \item the COSMOS tile (tile ID 1497p015), representative of most of the sky, i.e. a field with average {\it WISE} coverage and source density.
\end{itemize}

\begin{figure*}
    \centering
    \includegraphics[width=0.49\textwidth, trim={0 3cm 2cm 3cm}, clip]{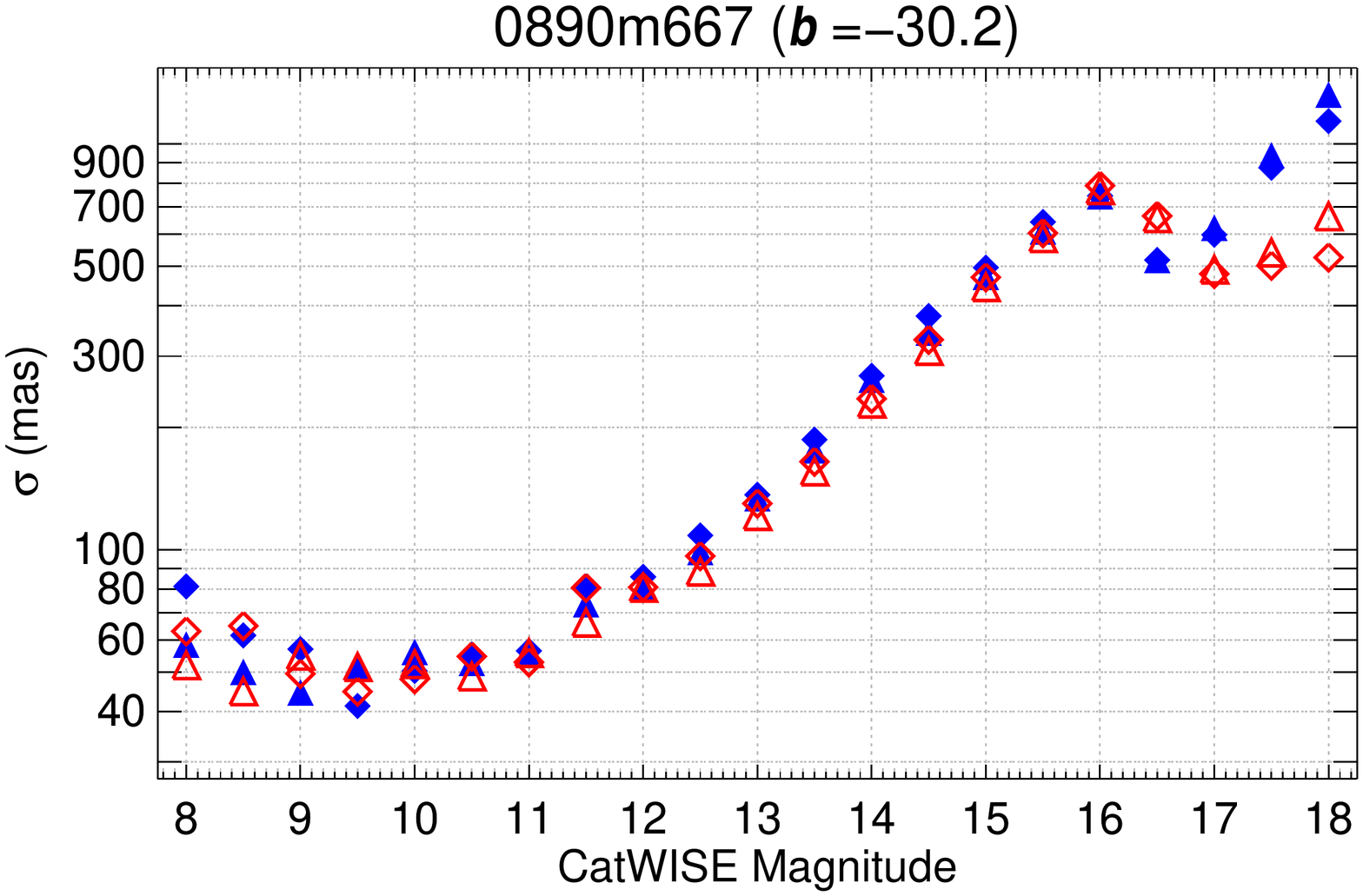}
    \includegraphics[width=0.49\textwidth, trim={0 3cm 2cm 3cm}, clip]{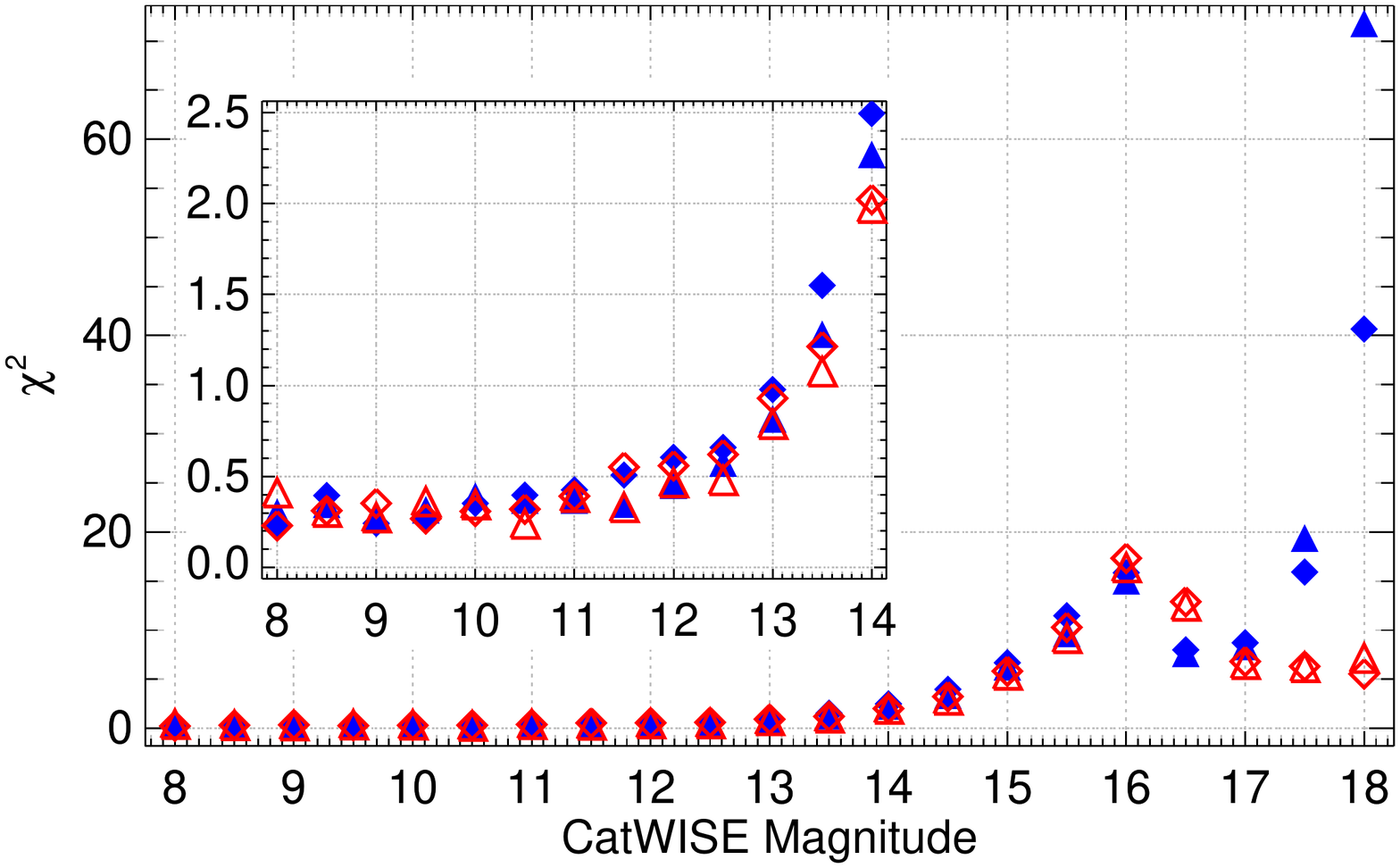}
    \includegraphics[width=0.49\textwidth, trim={0 3cm 2cm 3cm}, clip]{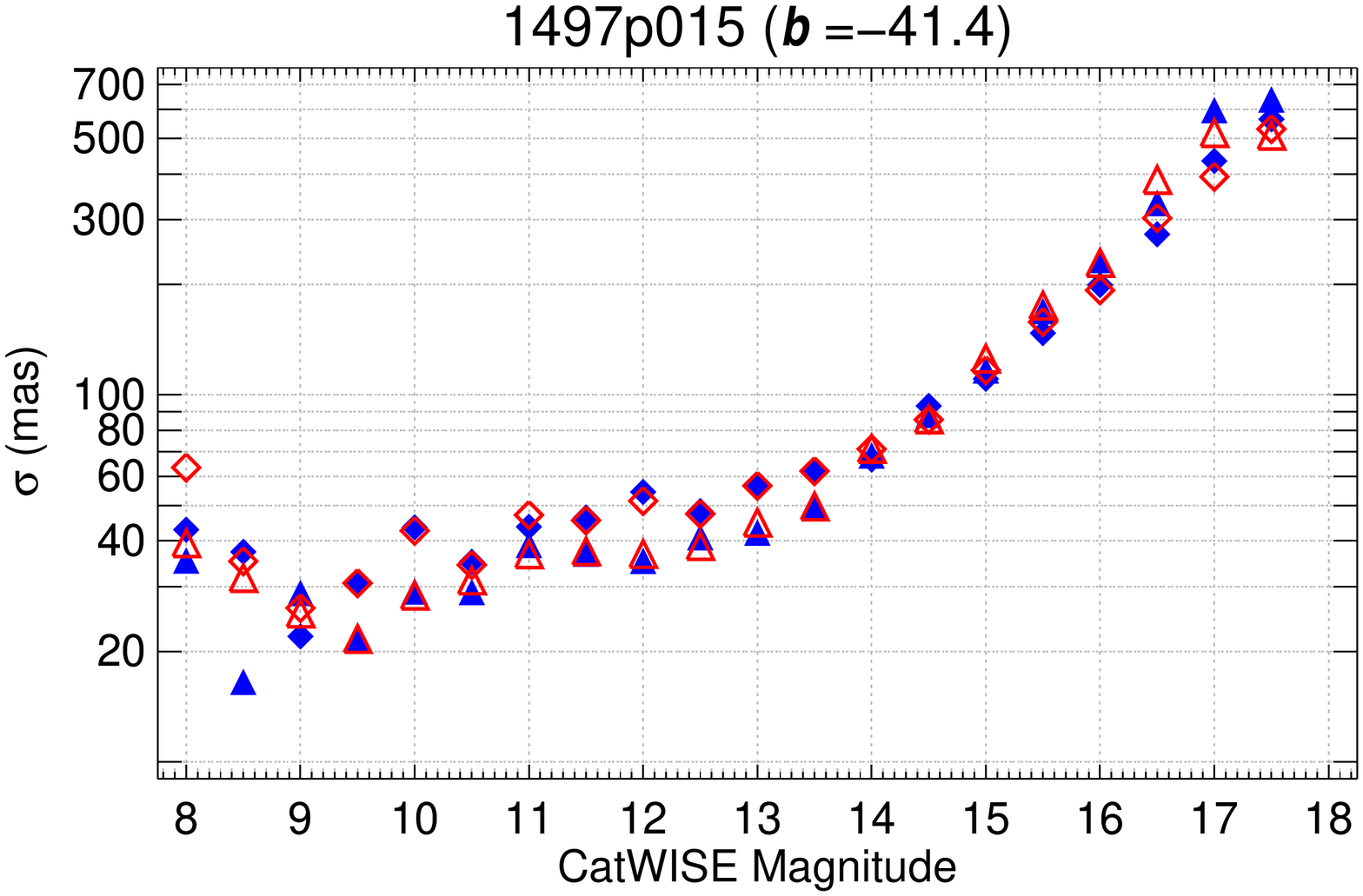}
    \includegraphics[width=0.49\textwidth, trim={0 3cm 2cm 3cm}, clip]{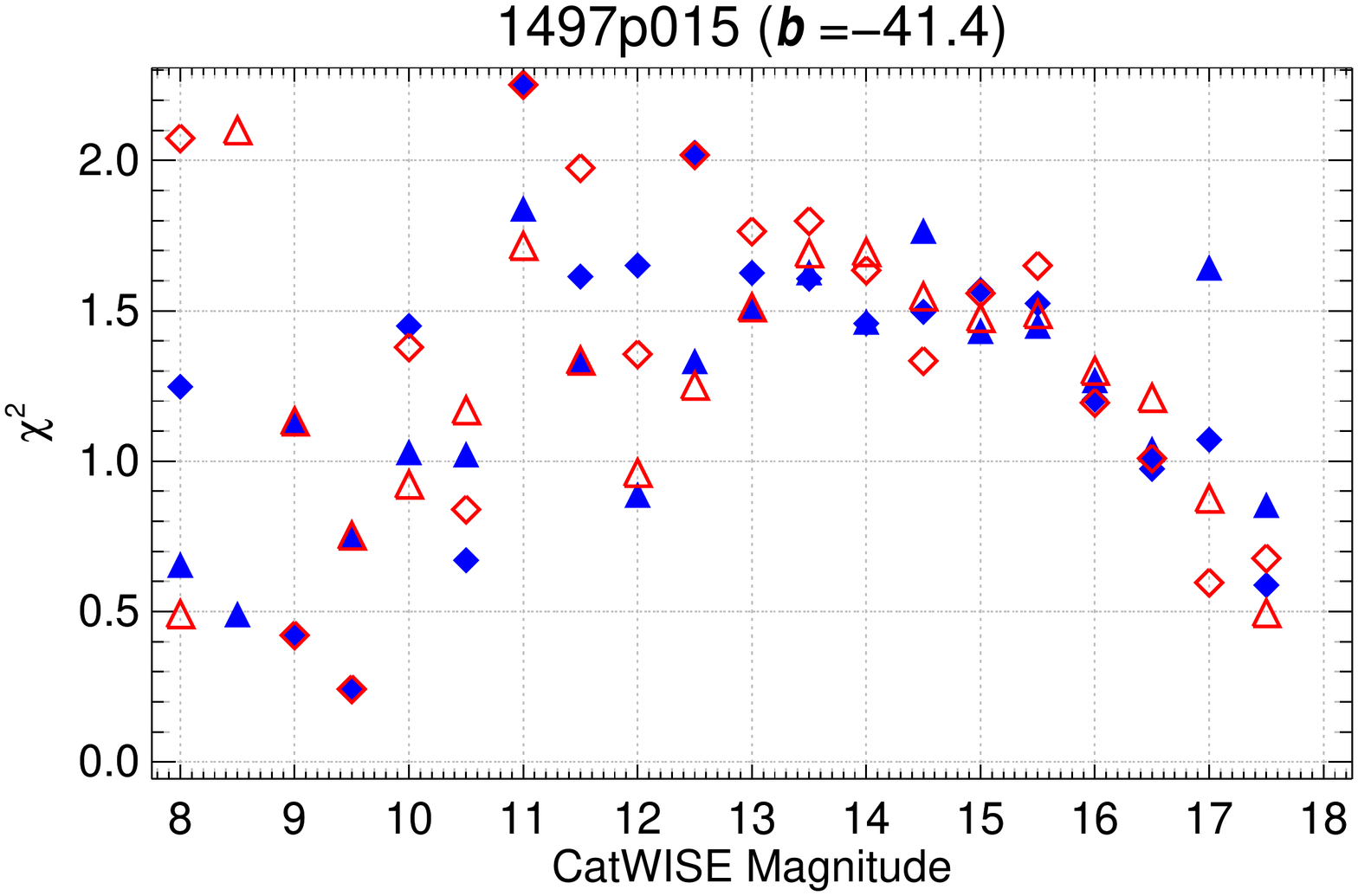}
    \includegraphics[width=0.49\textwidth, trim={0 3cm 2cm 3cm}, clip]{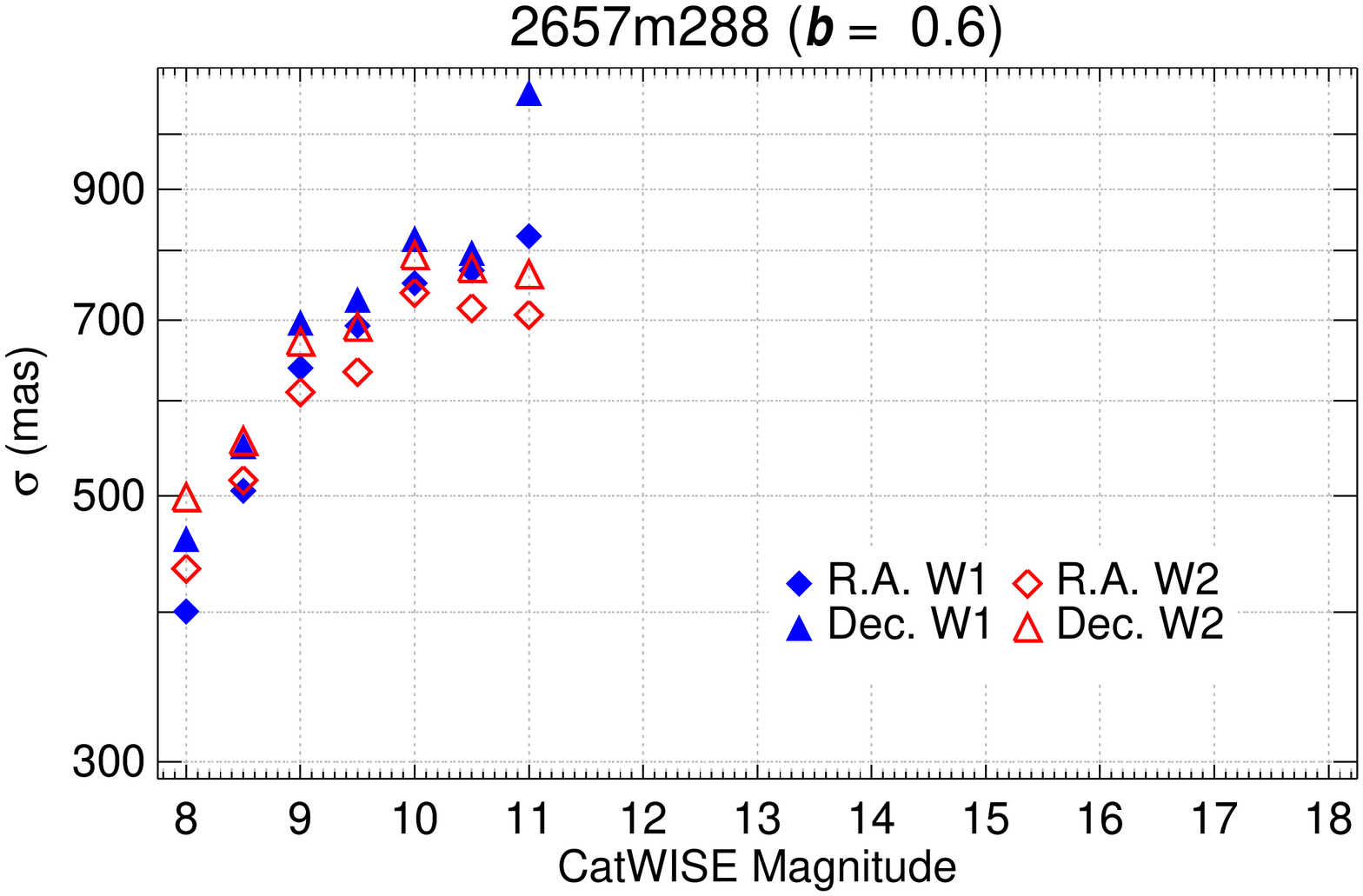}
    \includegraphics[width=0.49\textwidth, trim={0 3cm 2cm 3cm}, clip]{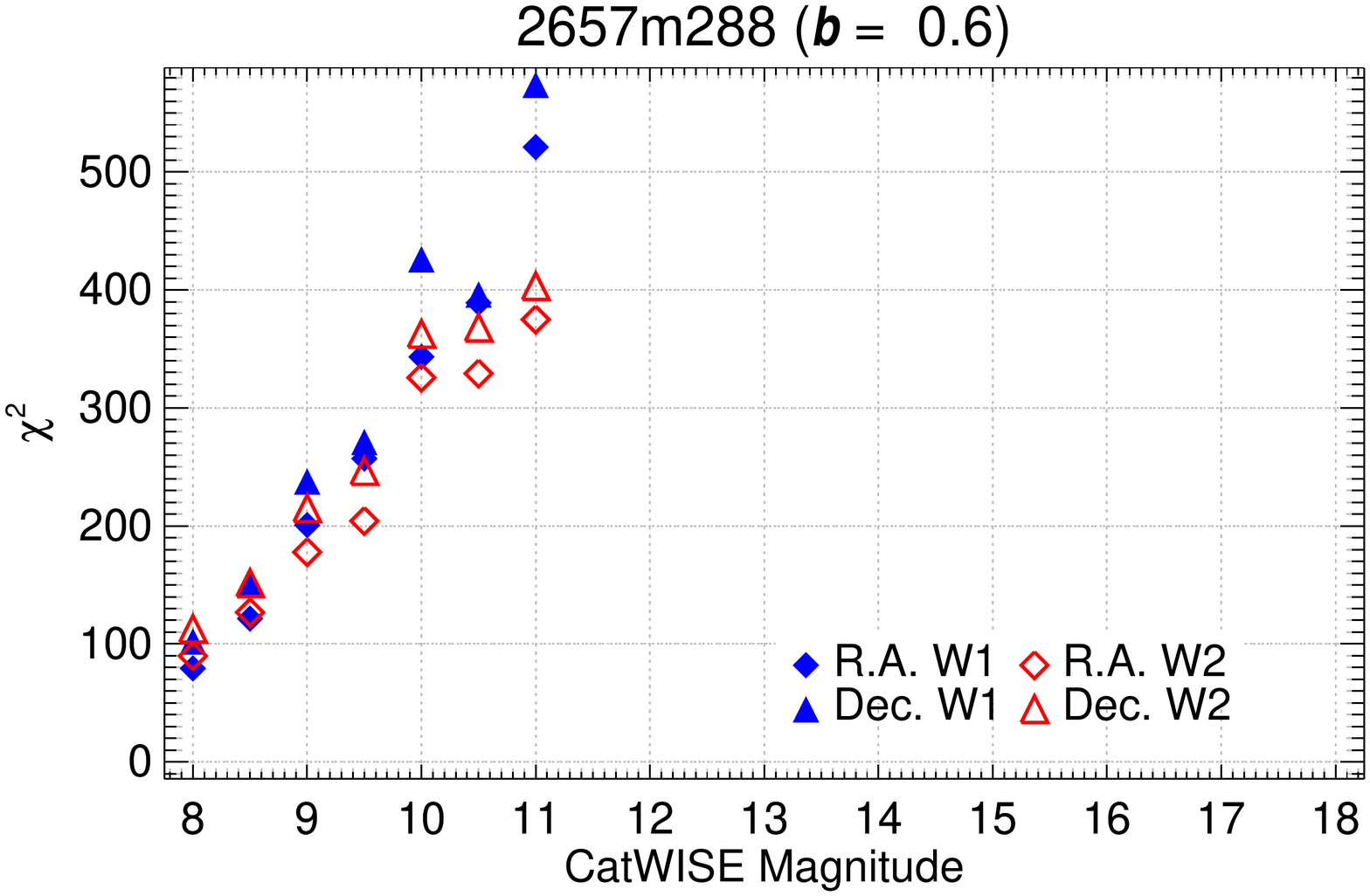}
    \includegraphics[width=0.49\textwidth, trim={0 3cm 2cm 3cm}, clip]{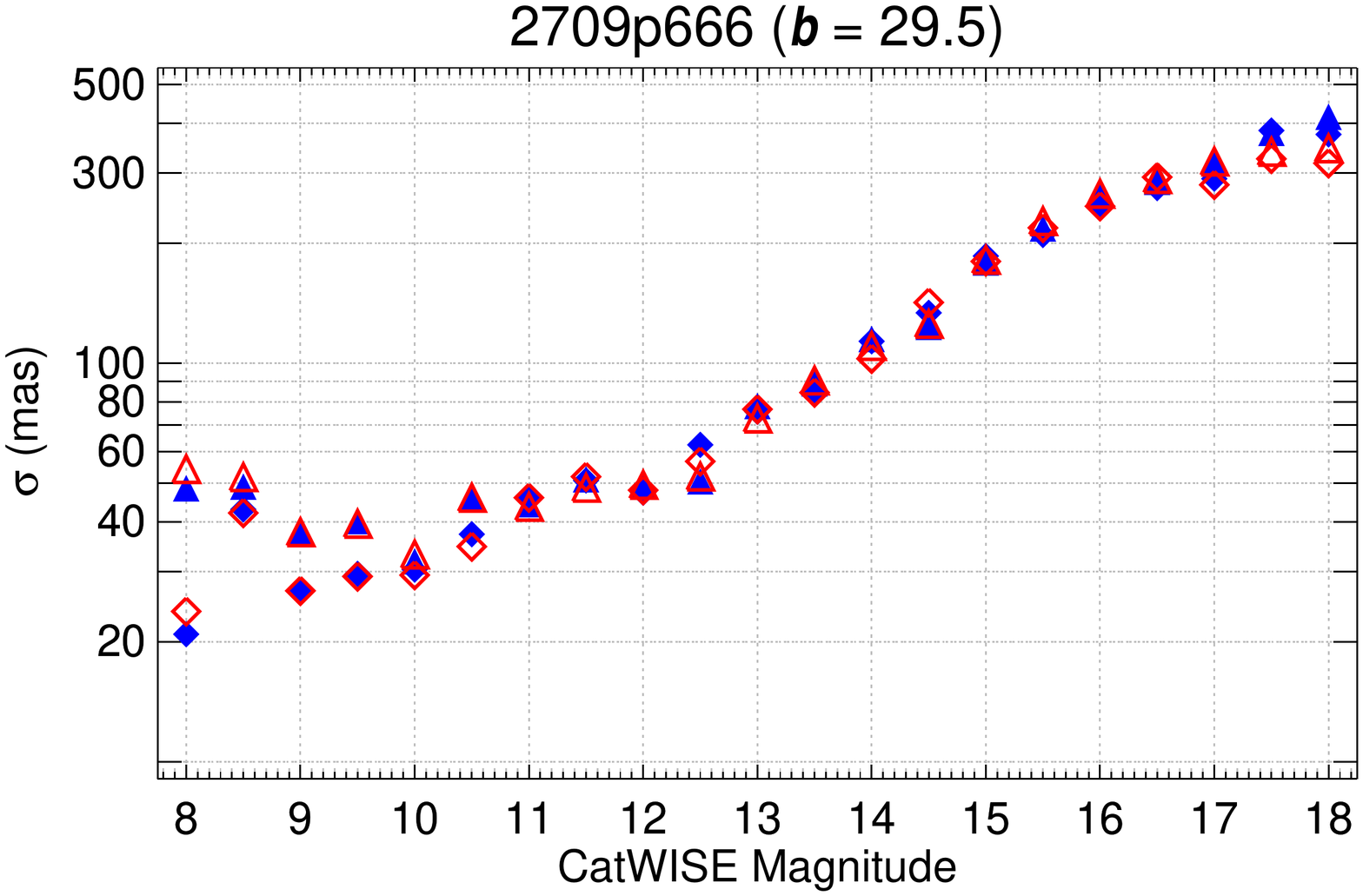}
    \includegraphics[width=0.49\textwidth, trim={0 3cm 2cm 3cm}, clip]{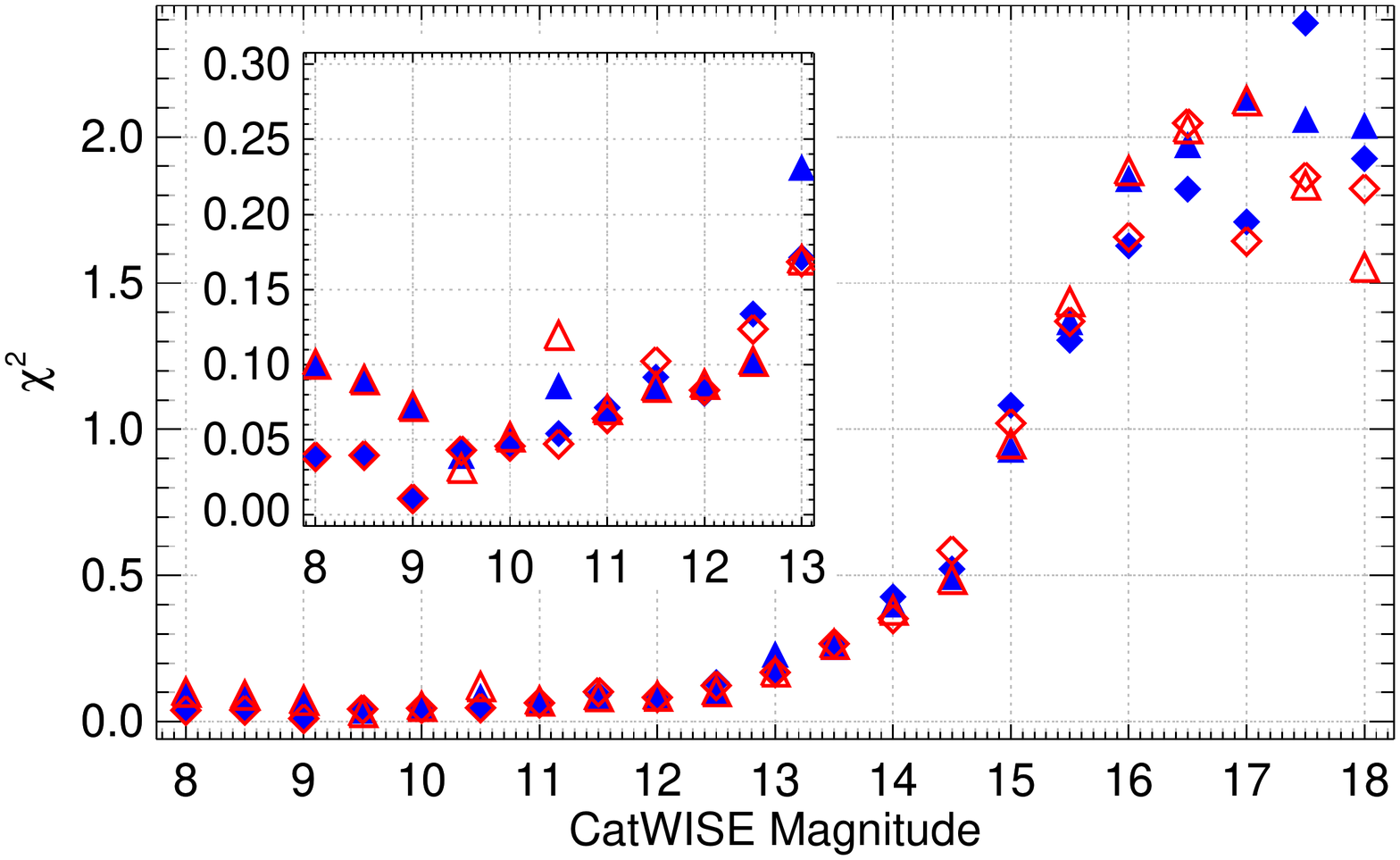}
    \caption{Same as Figure~\ref{fig:catwise_vs_gaia_fullsky}, but for four tiles: the SEP (0890m667), COSMOS (1497p015), the GC (2657m288), and the NEP (2709p666). The standard deviation between CatWISE position measurements and \textit{Gaia} DR2 position measurements, and the corresponding $\chi^2$ in 0.5 mag bins, is shown in red (R.A.) and blue (Dec.).}
    \label{fig:catwise_vs_gaia_pos}
\end{figure*}

\begin{figure*}
    \centering
    \includegraphics[width=0.49\textwidth, trim={0 3cm 2cm 3cm}, clip]{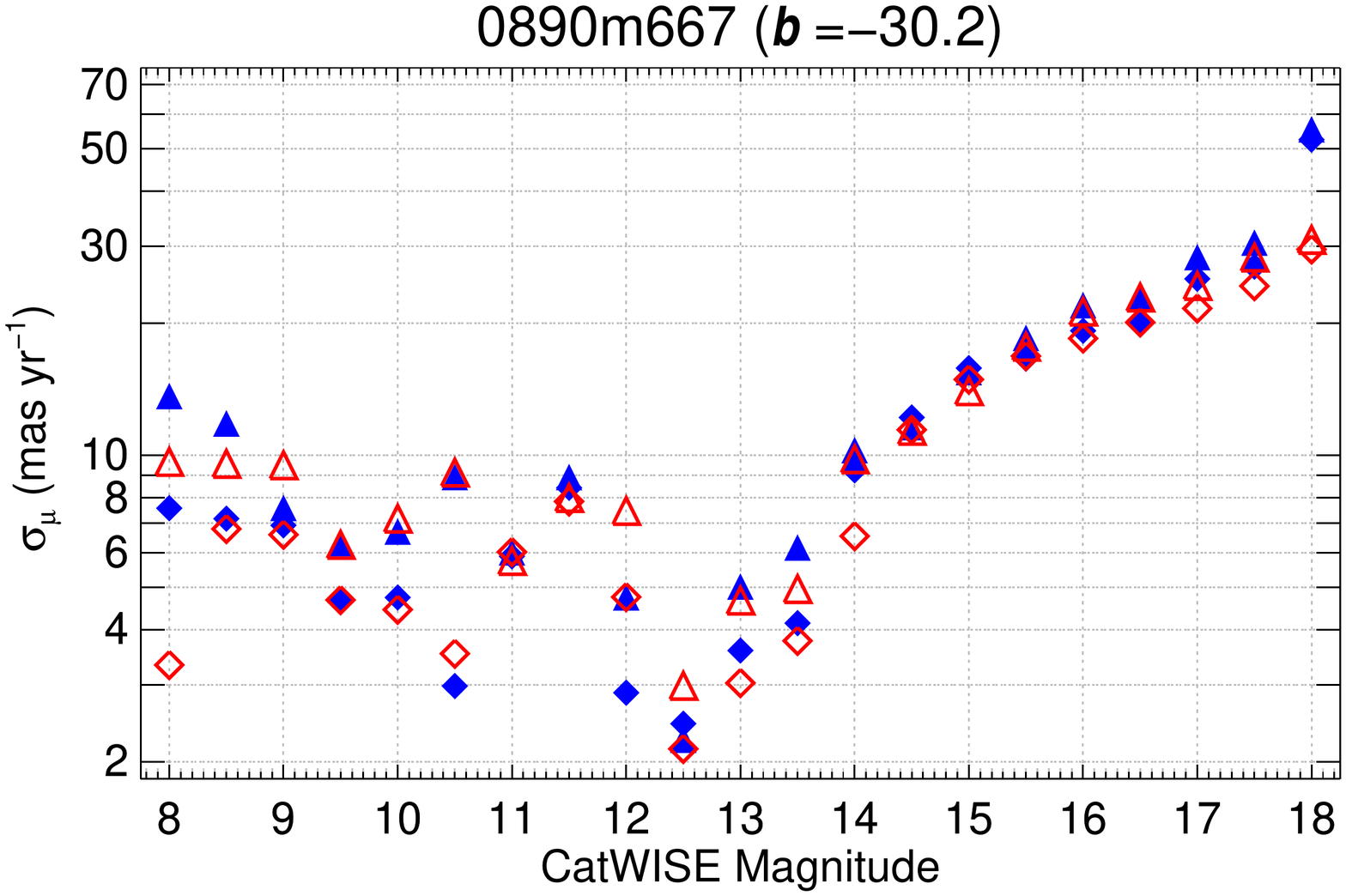}
    \includegraphics[width=0.49\textwidth, trim={0 3cm 2cm 3cm}, clip]{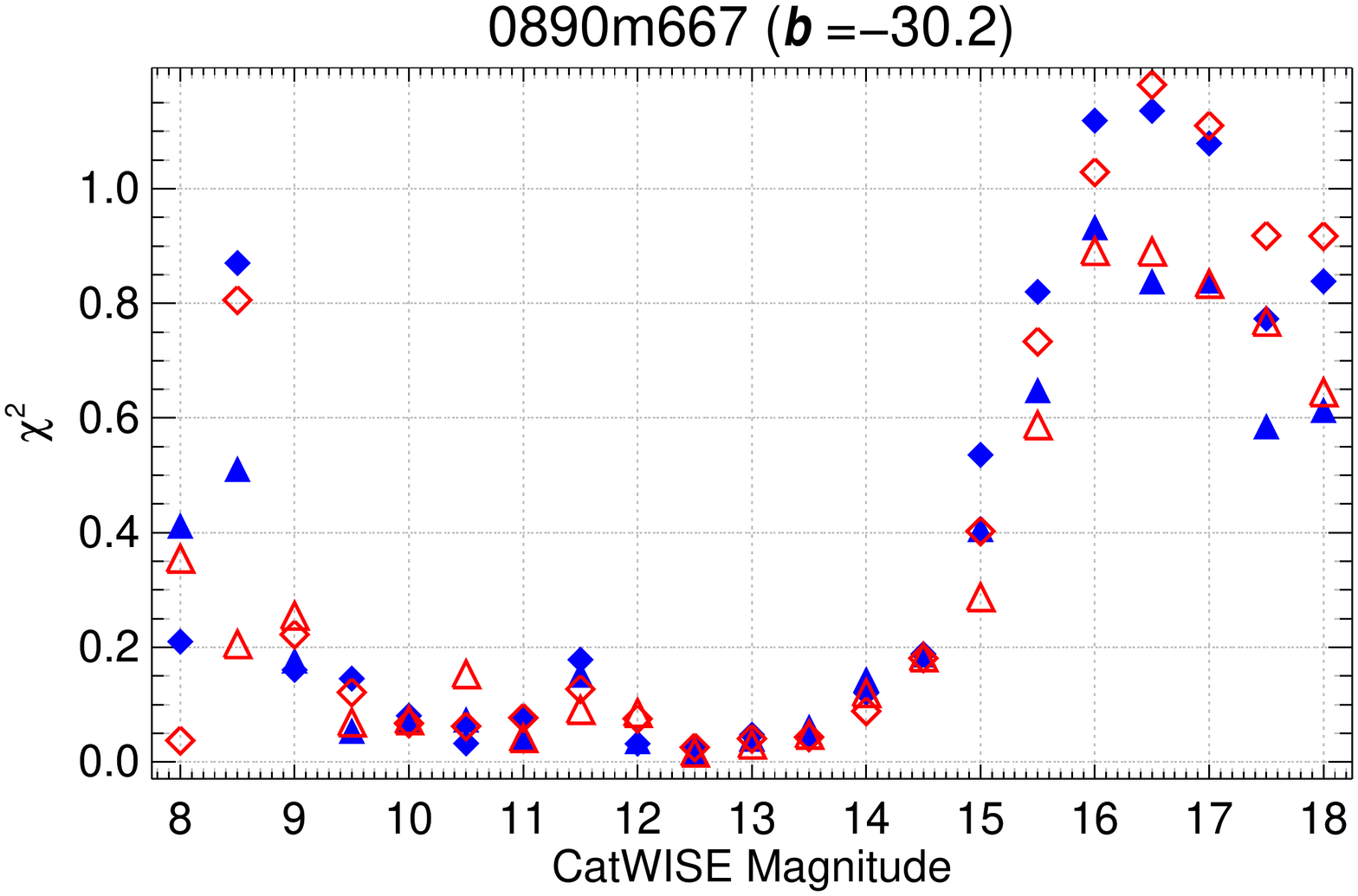}
    \includegraphics[width=0.49\textwidth, trim={0 3cm 2cm 3cm}, clip]{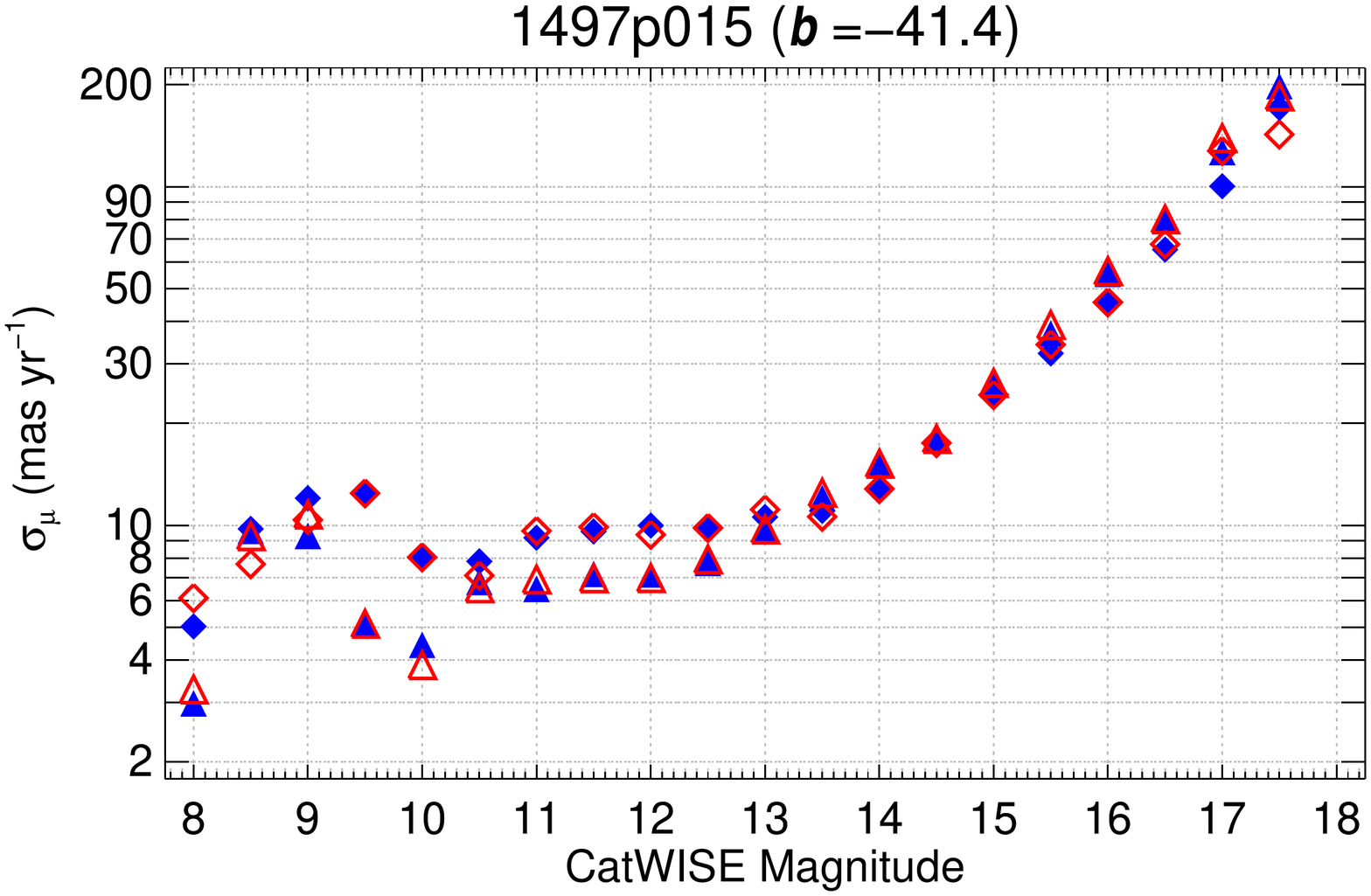}
    \includegraphics[width=0.49\textwidth, trim={0 3cm 2cm 3cm}, clip]{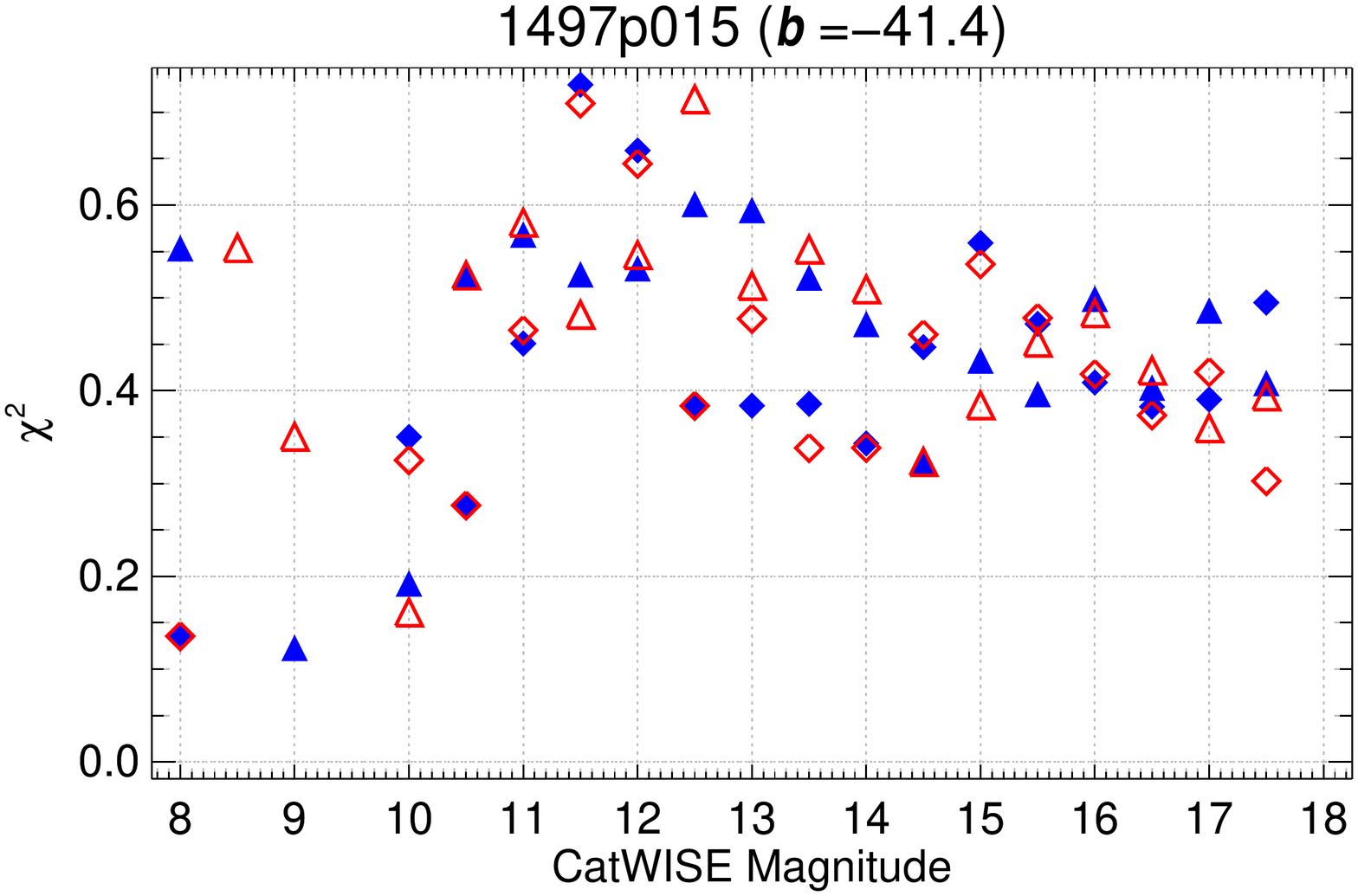}
    \includegraphics[width=0.49\textwidth, trim={0 3cm 2cm 3cm}, clip]{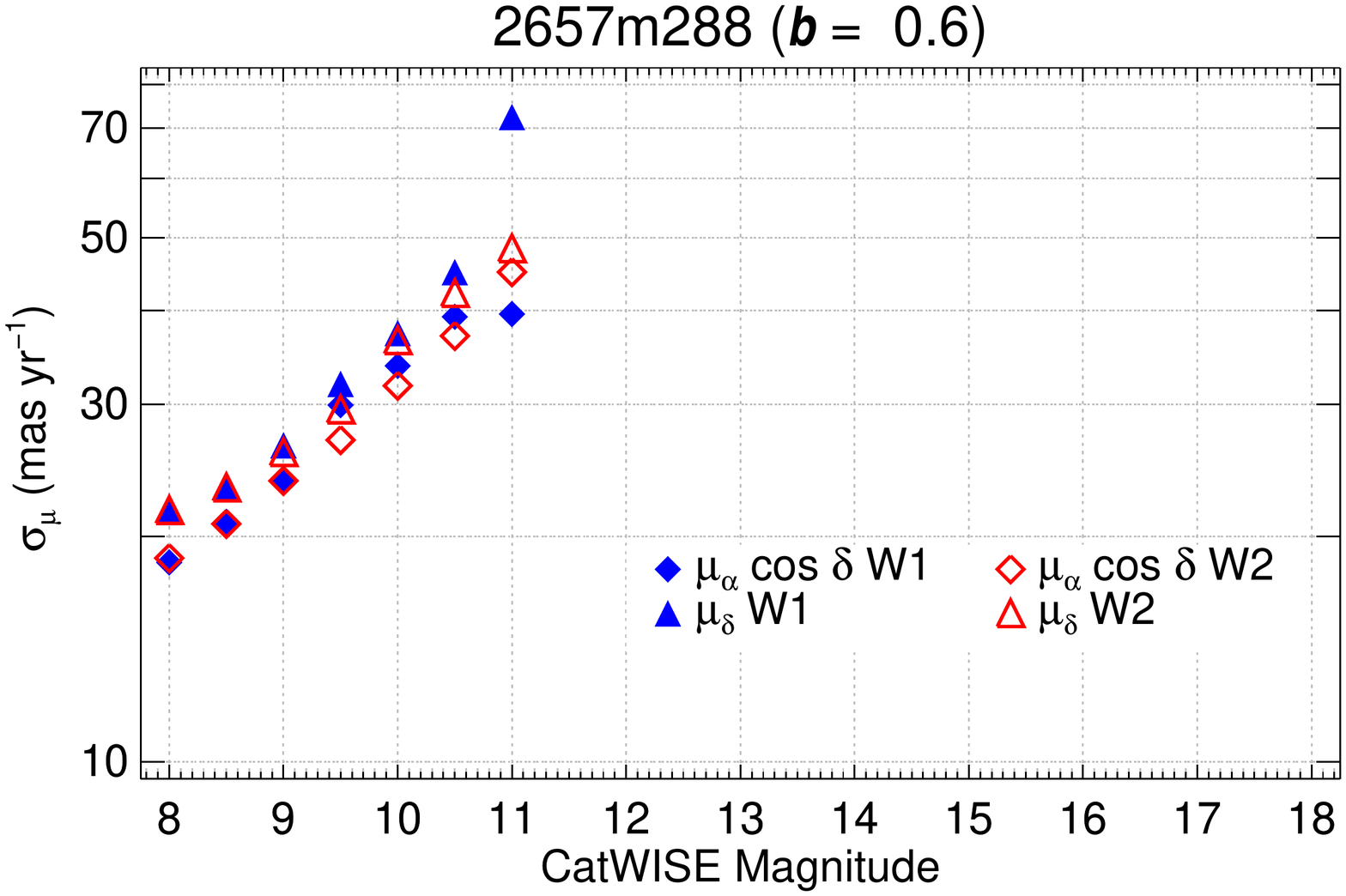}
    \includegraphics[width=0.49\textwidth, trim={0 3cm 2cm 3cm}, clip]{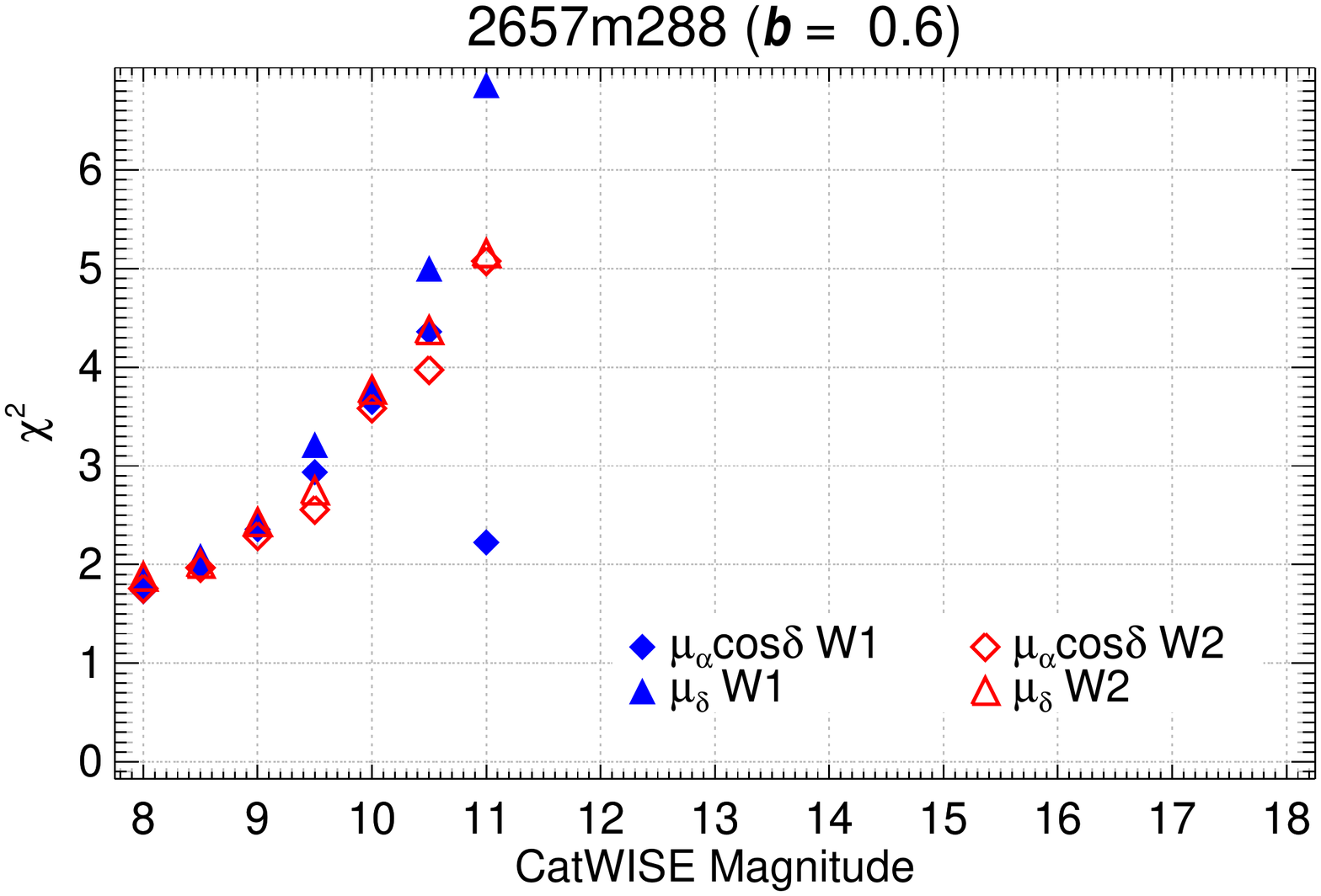}
    \includegraphics[width=0.49\textwidth, trim={0 3cm 2cm 3cm}, clip]{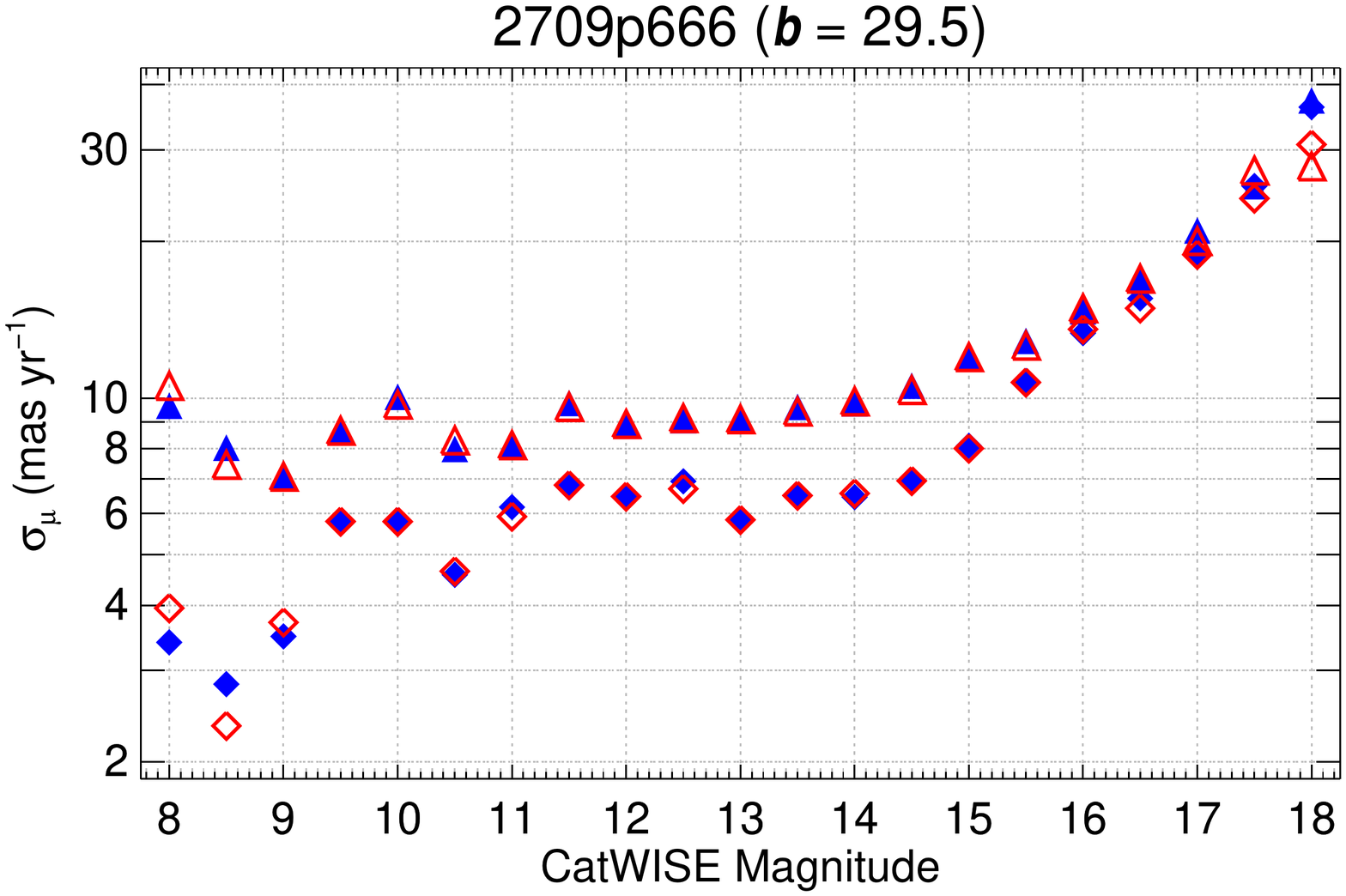}
    \includegraphics[width=0.49\textwidth, trim={0 3cm 2cm 3cm}, clip]{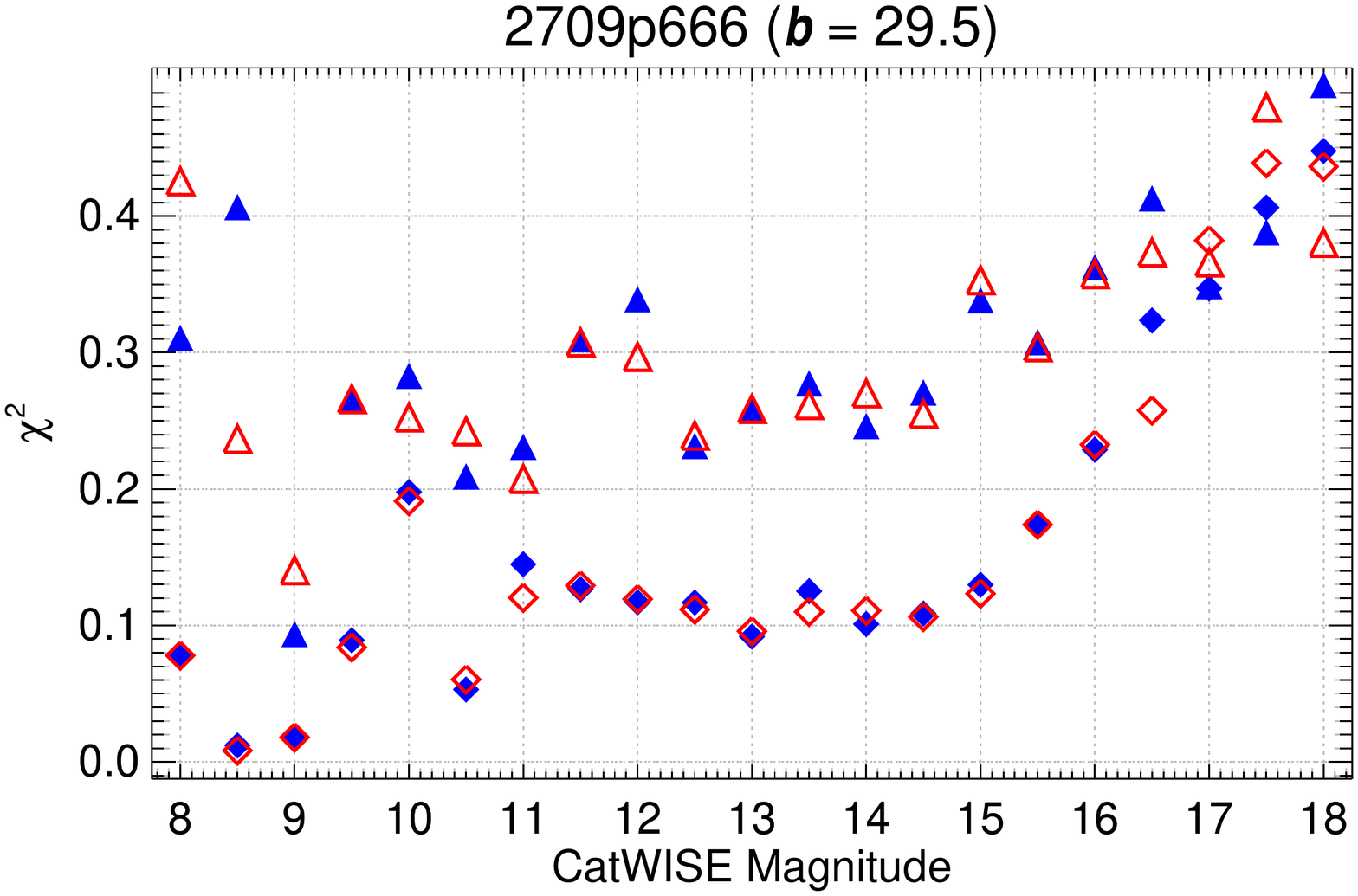}
    \caption{Same as Figure~\ref{fig:catwise_vs_gaia_pos}, but for motion measurements.}
    \label{fig:catwise_vs_gaia_pm}
\end{figure*}

The analysis of \S\ref{sec:astrom_full_sky} was repeated for these four tiles, but using all of the sources in each tile. The results are shown in Figure~\ref{fig:catwise_vs_gaia_pos} and \ref{fig:catwise_vs_gaia_pm}. We then defined ten metrics to characterize the astrometric performance of CatWISE: 
\begin{itemize}

\item $\sigma_{\rm min}$ and $\sigma_{\mu,\,\rm min}$ are the accuracy floor for positions and motions, respectively. These accuracy floors are determined as the median dispersion with respect to \textit{Gaia} in the $8<$W1,W2$<10$\,mag interval, except in the GC where we restrict to $8<$W1,W2$<9$\,mag since the astrometric accuracy starts deteriorating significantly beyond W1,W2$\sim$9\,mag (see Figure~\ref{fig:catwise_vs_gaia_pos}--\ref{fig:catwise_vs_gaia_pm}). 

\item W1$_{\rm min}$, W2$_{\rm min}$, W1$_{\mu,\,\rm min}$, and W2$_{\mu,\,\rm min}$ are the W1 and W2 mag at which $\sigma_{\rm min}$ and $\sigma_{\mu,\,\rm min}$ are exceeded by no more than 20\,mas and 5\,mas yr$^{-1}$, respectively.

\item W1$_{500}$ and W2$_{500}$ are the W1 and W2 mag at which the accuracy on positions reaches 500\,mas.

\item W1$_{\mu,\,100}$ and W2$_{\mu,\,100}$ are the W1 and W2 mag at which the accuracy on motion reaches 100\,mas\,yr$^{-1}$.

\end{itemize}

The results for the four representative tiles are summarized in Table~\ref{table_ast_fields}.

For brighter sources, $\sigma_{\rm min}$ and $\sigma_{\mu\,\rm min}$ are encouragingly small, on the order of 40\,mas and 8 mas\,yr$^{-1}$, respectively. W1$_{\rm min}$ and W2$_{\rm min}$ are $\sim12$\,mag while W1$_{\mu,\,\rm min}$ and W2$_{\mu,\,\rm min}$ are $\sim14.5$\,mag.  All the above metrics show little dependence on coverage and source density, except for the GC tile, where they are degraded to $\sigma_{\rm min}\sim530$\,mas, $\sigma_{\mu,\rm min}\sim20$\,mas, at a limiting depth of only $\sim$8\,mag and $\sim$9\,mag respectively. This is consistent with the uniformity of the maps in Figures~\ref{fig:map_positions} to \ref{fig:map_pms_bins}.

For fainter sources, W1$_{500}$, W2$_{500}$, W1$_{\mu,\,100}$ and W2$_{\mu,\,100}$ show a clearer dependence on source density and coverage. For COSMOS (i.e. a typical CatWISE tile) these metrics are $\sim$17\,mag. W1$_{500}$ and W2$_{500}$ become $\sim$9\,mag shallower in the GC, while W1$_{\mu,\,100}$ and W2$_{\mu,\,100}$ measurements are upper limits only. This is again a consequence of the fact that at lower Galactic latitudes CatWISE only detects brighter sources because of confusion noise, and for sources this bright the motion accuracy never degrades to the 100\,mas\,yr$^{-1}$ level.

W1$_{\mu,\,100}$ and W2$_{\mu,\,100}$ become $\sim$2\,mag deeper ($\sim$19\,mag) at the ecliptic poles, thanks to the higher coverage. W1$_{500}$ and W2$_{500}$ are much deeper at the NEP than the SEP, most likely an effect of the much higher source density at the SEP, and therefore the higher confusion noise. However, this does not seem to affect the motion metrics.

\begin{deluxetable*}{lcccc}
\tablecaption{CatWISE Astrometric Performance Evaluation
  Fields\label{table_ast_fields}}
\tablehead{
\colhead{} &            
\colhead{0890m667} &
\colhead{1497p015} &
\colhead{2657m288} &         
\colhead{2709p666} \\
\colhead{} & 
\colhead{SEP, LMC} & 
\colhead{COSMOS} & 
\colhead{GC} & 
\colhead{NEP}
}
\startdata
$l$ (deg) & 276.5 & 237.3 & 359.8 & 96.4 \\
$b$ (deg) & $-30.2$ & 41.4 & 0.6 & 29.5 \\
$\beta$ (deg) & $-89.6$ & $-10.2$ & $-5.4$ & 89.6 \\
Exp. & 7154 & 90 & 86 & 7839 \\
$\#$ & 71462 & 58961 & 63368 & 61702 \\
\hline
$\sigma_{\rm min}$ (mas) & 52.9 & 27.3 & 526.4 & 37.7 \\
W1$_{\rm min}$ (mas) & 11.0 & 12.5 & 8.0 & 12.0 \\
W1$_{500}$ (mag) & 15.1 & 17.0 & 8.4 & 18.5 \\
W2$_{\rm min}$ (mag) & 11.0 & 12.5 & 8.0 & 12.0 \\
W2$_{500}$ (mag) & 15.0 & 16.8 & 8.3 & 19.0 \\
\hline
$\sigma_{\mu, \rm min}$ (mas yr$^{-1}$) & 7.4 & 8.5 & 22.2 & 7.3 \\
W1$_{\mu, \rm min}$ (mag) & 14.5 & 13.5 & 9.0 & 15.5 \\
W1$_{\mu, 100}$ (mag) & 18.2 & 16.8 & $>11.0$ & $>19.0$ \\
W2$_{\mu, \rm min}$ (mag) & 14.5 & 13.5 & 9.0 & 15.5 \\
W2$_{\mu, 100}$ (mag) & $>$20.5 & 16.7 & $>$11.5 & $>$20.0 \\
\enddata
\tablecomments{ $l$, $b$, and $\beta$ are the Galactic longitude, Galactic latitude, and ecliptic latitude for the center of the tile,
in degrees. Exp. indicates the number of exposures for the tile, $\#$ the number of sources (combining catalog and reject entries). The subsequent metrics are described in detail in \S\ref{sec:astrom_perf}.}
\end{deluxetable*}

\newpage
\subsubsection{Fast Movers}
We compared the astrometric performance using merged results from  ascending and descending scans processed with separate PSF's to that using a single PSF per band for all scans processed together (see \S\ref{sec:wphot}) on a set of 19 known ultracool dwarfs, chosen from the literature. Our test set included extremely cold, very fast moving objects \citep[e.g. WISE~J085510.83--071442.5, $\mu_{\rm tot} \sim 8$ arcsec yr$^{-1}$;][]{Luhman2014}, as well as warmer, slower M dwarfs \citep[e.g. WISE J072003.20--084651.2, $\mu_{\rm tot} \sim 0.12$ arcsec yr$^{-1}$;][]{Scholz2014}. Results using separate PSF's proved superior, delivering good motion measurements for 17 out of the 19 test objects, while the single PSF option  only recovered 13 out of the 19 test objects, with a clear drop in performance for objects with $\mu_{\rm tot} \gtrsim 2.5$ arcsec yr$^{-1}$. Only one out of the five fastest moving objects in our test sample is correctly measured using the single PSF method, while using separate ascending and descending scan PSF's and merging the results measures four out of these five. 

 The two objects that our pipeline is unable to correctly measure -- WISE J163940.83--684738.6 and WISEPC J205628.90+145953.3 -- are missed at the detection stage (\S\ref{sec:mdet}) because they are partly blended with brighter nearby sources. These two sources were recovered when the detection step was run without the PSF convolution (\S\ref{sec:unwise} and \ref{sec:mdet}), yielding reasonable motions, but they are not present in the CatWISE Preliminary Catalog. 

The fastest movers in our test set have larger parallaxes \citep[up to 502 mas for WISE J104915.57-531906.1,][]{Lazorenko2018}, which will affect motion measurements when measuring ascending and descending scans together. To test this, we also processed the fastest movers using a single PSF but measuring ascending and descending scans separately and merging them. This gave better performance than processing scans together, recovering motions correctly for three of the five fastest movers. 

Comparing astrometry to \textit{Gaia} for all sources in these tiles showed that even when measuring ascending and descending scans separately and merging them, using separate PSFs further improves performance. The position sigmas were 10\% to 20\% better at all magnitudes when using separate PSFs than when using a single PSF. For brighter sources ($< 12$ mag), motion sigmas improved by $\sim20\%$ using separate PSFs, while at fainter magnitudes the single and separate PSFs yielded similar motion sigmas. Processing ascending and descending scans separately was beneficial both because it allowed the PSF asymmetry to be addressed, and because it removed the effect of parallax on measured motions.

%\newpage
\section{Example Science Applications \label{sec:science}}

An initial application of the CatWISE catalog was the first secure W2 detection for the Y dwarf WD0806-661\,B \citep{Meisner2018b}, leading to a [4.5] - W2 color consistent with the population of known Y dwarfs. CatWISE represents a major hunting ground for cold brown dwarfs. The improved depth (\S\ref{sec:photom_perf}) and motion sensitivity (\S\ref{sec:astrom_perf}) allows for a deeper, more complete search for the coldest constituents of the solar neighborhood, a crucial population if we wish to constrain the low-mass end of the mass function \citep[see e.g.][]{Kirkpatrick2019}. 

Towards this goal we are mining CatWISE using both a ``classical'' and a ``machine learning based'' approach. Here ``classical'' means a search based on color and motion cuts applied to the catalog data to select cold brown dwarf candidates, while machine learning uses the previously known population of cold brown dwarfs as a training set to develop a classifier that is then applied to the CatWISE catalog. 

Early results include the discoveries of CWISEP~J193518.59--154620.3 \citep{Marocco2019} and WISEP~J144606.62--231717.8 \citep{Marocco2020}, two of the coldest brown dwarfs identified to date, with infrared colors comparable with those of the coldest brown dwarf known, WISE J085510.83--071442.5 \citep{Luhman2014}. \textit{Spitzer} follow-up of additional cold brown dwarf candidates is   presented in \citet{Meisner2019b}.

\begin{figure}
    \centering
    \includegraphics[width=0.5\textwidth,trim={0 0cm 0 0cm}, clip]{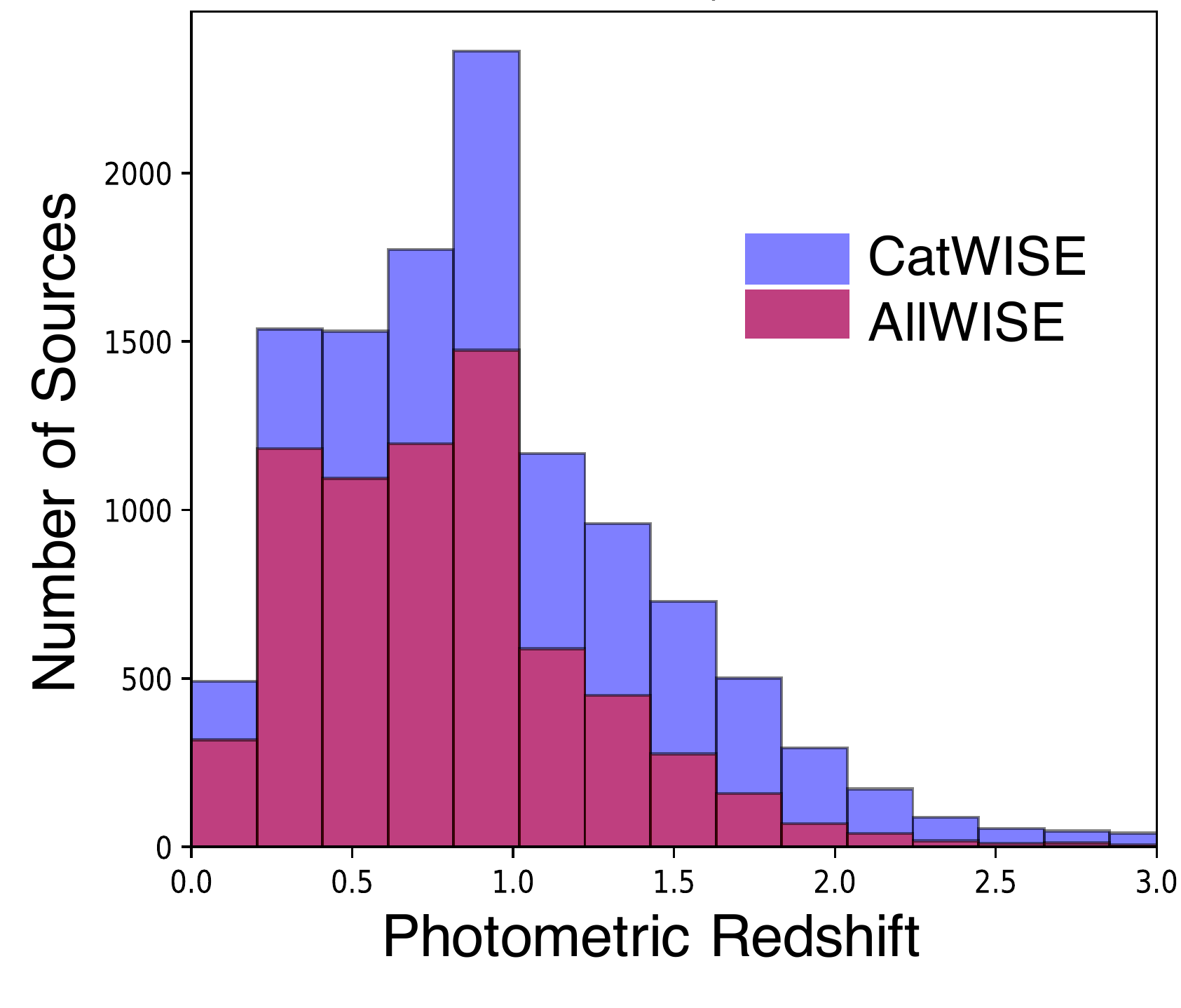}
    \caption{CatWISE detects sources $\sim0.5$ mag fainter than AllWISE, providing many times more sources at $z > 1.5$. The vertical axis is the number of COSMOS sources that were detected by CatWISE and AllWISE in the 1497p015 tile and had matches within 2\farcs5 of sources with the indicated photometric redshifts from \citet{Laigle2016}.}
    \label{fig:cosmosPhotz}
\end{figure}

Prospects are also good for discovering distant galaxy clusters using CatWISE. \citet{Gonzalez2019} used the AllWISE Catalog together with Pan-STARRS to carry out the Massive and Distant Clusters of WISE Survey (MaDCoWS), identifying over 2000 galaxy cluster candidates with photometric redshifts in the $0.7 - 1.5$ range. The additional depth of CatWISE compared to AllWISE increases the W1 detection limit for L* galaxies from $z\sim 0.8$ to $z\sim 2$. Figure \ref{fig:cosmosPhotz} illustrates the increase in the number of distant galaxies with substantial photometric redshifts in the COSMOS field \citep{Laigle2016} detected by CatWISE compared to AllWISE. Between $z=1.5$ and 1 clusters transition from being more actively star-forming than the field to being mainly quiescent in terms of star formation  \citep[e.g.,][]{Brodwin2013}, so CatWISE offers the potential of reaching the era where major formation and assembly of clusters is underway.

Finally, the AllWISE Catalog played an important role in the discovery of ULAS J1342+0928, at $z = 7.54$ the most distant quasar known \citep{Banados2018}.  The additional depth of CatWISE may enable extending this to redshifts as large as 9, when the seeds of reionization were emerging in the overwhelmingly neutral Universe. With over 3000 refereed papers making use of the existing {\it WISE} catalogs, many other uses for CatWISE can be anticipated.

\section{Data Access \label{sec:access}}

Current information about CatWISE data products is provided at 
\url{https://catwise.github.io}.  

The merged Preliminary catalog and reject files are available from IRSA (\url{https://irsa.ipac.caltech.edu}) in the WISE/NEOWISE Enhanced and Contributed Products area. IRSA's catalog search tools allow for complex search queries. IRSA also hosts the AllWISE Explanatory Supplement \citep{Cutri2013}, which provides full details on the AllWISE processing algorithms, and includes descriptions of the AllWISE Catalog columns, many of which are applicable to CatWISE. Appendix \ref{sec:columns} provides additional information about CatWISE columns.

The individual tile files have also been transferred to a data repository at the National Energy Research Scientific Computing Center (NERSC), and are available at \url{https://portal.nersc.gov/project/cosmo/data/CatWISE/prelim} in 18,240 pairs of gzipped ASCII files (one catalog and one reject file per tile) in IPAC table format, organized into 359 directories, one for each decimal degree of right ascension from 0\degree\ to 358\degree\ (there are no tiles beginning with 359). Text files providing the format and a brief description of the columns in the catalog and reject files are also provided there. The catalog and reject files for the 50 tiles near the ecliptic poles (Table \ref{table_polar_tiles}), where a single PSF per band was used for processing, include the string ``opt0" in their names. Files for tiles where different PSFs were used for ascending and descending scans (\S\ref{sec:psf}) include the string ``opt1" in their names. 

\acknowledgments

CatWISE was only possible because it uses and adapts software developed at IPAC for \textit{WISE} to the maximum extent practical. We thank Tim Conrow, Frank Masci, and Ken Marsh for many helpful discussions and assistance in developing CatWISE. We also thank the anonymous referee for a careful reading and numerous suggestions which improved this paper. CatWISE uses data products from {\it WISE}, which is a joint project of the University of California, Los Angeles, and the Jet Propulsion Laboratory (JPL)/California Institute of Technology (Caltech), funded by the National Aeronautics and Space Administration (NASA), and from {\it NEOWISE}, which is a JPL/Caltech project funded by NASA.  Characterization of CatWISE performance uses data from {\it Gaia} and from {\it Spitzer}. CatWISE is led by JPL/Caltech, with funding from NASA's Astrophysics Data Analysis Program (ADAP), and is also supported in part by ADAP grant NNH17AE75I at Lawrence Berkeley Laboratory. CatWISE is also supported by the Fellowships and Internships in Extremely Large Data Sets (FIELDS) program funded by NASA at UC Riverside. FM acknowledges support from the NASA Postdoctoral Program at the Jet Propulsion Laboratory, administered by Universities Space Research Association under a contract with NASA.

%\vspace{5mm}
\newpage
\appendix
\section{CatWISE Preliminary Catalog Column Descriptions \label{sec:columns}}

We adopt MJD 56700 (2014 Feb. 12) as the epoch for reporting positions when solving for source motion in the Preliminary Catalog. Source positions, whether incorporating source motion or not, are given in the equinox J2000 coordinate frame. Column entries involving right ascension that are in units of arcseconds have the cos(declination) term applied. The convention for designating sources from the CatWISE Preliminary Catalog and Reject Table is specified in \S\ref{sec:prelim}.

There  are 186 formatted columns of information about each source in the CatWISE Preliminary Catalog. The CatWISE Preliminary Reject Table adds a column to indicate whether the source is ``primary" (see \S\ref{sec:primary}). Descriptions of nearly all the columns can be obtained from IRSA at https://irsa.ipac.caltech.edu/data/WISE/CatWISE/gator\_docs/catwise\_colDescriptions.html.  Most of the columns have the same names as in the AllWISE Catalog, and are described in \S II.1.a of the AllWISE Explanatory Supplement \citep{Cutri2013}. Table \ref{table_newcolumns} provides information about selected columns in the CatWISE Preliminary Catalog that augments or supersedes the information provided by IRSA. Four columns ({\it w1fitr, w2fitr, glon, glat}) in the NERSC release are hidden in the IRSA release, and their names are shown in parentheses in Table \ref{table_newcolumns}. 

\begin{longrotatetable}
%\begin{deluxetable*}{lllrrrrrrll}
\begin{deluxetable*}{llllp{6in}}
\tablecaption{Additional Information on CatWISE Preliminary Catalog Columns\label{table_newcolumns}}
\tablewidth{700pt}
\tabletypesize{\footnotesize}
%\tablecaption{Observable Characteristics of Galactic/Magellanic Cloud novae with X-ray observations\label{chartable}}
%\tablewidth{700pt}
%\tabletypesize{\scriptsize}
\tablehead{
%\begin{longrotatetable}
%\startlongtable
%\begin{deluxetable}{llllL}
%\movetabledown=1in
%\begin{adjustbox}{width=10in}
%\tablewidth{12in}
%\tablecaption{\textbf{Additional Information on CatWISE Preliminary Catalog Columns}\label{table_newcolumns}}
%\tablehead{
\colhead{Name} &
\colhead{Units} & 
\colhead{Type} &
\colhead{Format} & 
\colhead{Description} \\
%\colhead{(1)} & 
%\colhead{(2)} & 
%\colhead{(3)} & 
%\colhead{(4)} &
%\colhead{(5)}
}
\startdata
 %source\_id & \nodata & char & a25 & Unique source ID, formed from a combination of the tile name, a CatWISE processing version code (always ``a1" for the Preliminary catalog and reject table), and the sequential number of the source extraction in the tile (i.e. ``src" in AllWISE. \\
 (w1fitr) & arcsec & R*4 & f7.2 & Fitting radius for W1. In the CatWISE Preliminary Catalog, w1fitr and w2fitr are always -99.99, 7.5, or 13.2. Both are hidden in the IRSA release, because they do not appear to follow the prescription given in \url{http://wise2.ipac.caltech.edu/docs/release/allsky/expsup/sec4_4c.html\#wpro} where rfit=1.25 FWHM (7\farcs5 for W1 and W2) or $2r_{sat}$ (saturation radius), whichever is larger.  \\
 (w2fitr) & arcsec & R*4 & f7.2 & Fitting radius for W2. See w1fitr. \\
 dist\_ad & arcsec & R*4 & f9.3 & Radial distance between ascending and descending apparitions. The field name in the IRSA release is dist\_ad, but in the NERSC release, this field is called dist.\\
 %dw1mag & mag & R*4 & f7.3 &  Difference between ascending and descending w1mpro values \\
 %rch2w1  & \nodata & R*4 & f7.3 & $\chi^2$ for dw1mag (1 DF) \\
 %dw2mag & mag & R*4 & f7.3 & Difference between ascending and descending w2mpro values \\
 %rch2w2 & \nodata & R*4 & f7.3 & $\chi^2$ for dw2mag (1 DF) \\
 elon\_avg & deg & R*8 & f11.6 & Averaged ecliptic longitude between ascending and descending apparitions\\
 %elonSig & arcsec & R*4 & f11.3 & $1 \sigma$ uncertainty in elon\_avg \\
 elat\_avg & deg & R*8 & f11.6 & Averaged ecliptic latitude between ascending and descending apparitions\\
 %elatSig & arcsec & R*4 & f10.3 & $1 \sigma$  uncertainty in elat\_avg \\
 %Delon & arcsec & R*4 & f11.3 & Descending - ascending ecliptic longitude. Will be null unless ka = 3. This order should have the proper sign for parallax ($\sim$Delon/2).  \\
 %DelonSig & arcsec & R*4 & f11.3 & $1 \sigma$ uncertainty in Delon \\
 %Delat & arcsec & R*4 & f11.3 & Descending - ascending ecliptic latitude \\
 %DelatSig & arcsec & R*4 & f10.3 & $1 \sigma$ uncertainty in Delat \\
 %DelonSNR & \nodata & R*4 & f11.3 & $\lvert$Delon$\rvert$/DelonSig \\
 %DelatSNR & \nodata & R*4 & f11.3 & $\lvert$Delat$\rvert$/DelatSig \\
 chi2pmra & \nodata & R*4 & 1pE10.3 & $\chi^2$ for difference between PMRA for ascending and descending apparitions (1 degree of freedom) \\
 chi2pmdec & \nodata & R*4 & 1pE10.3 & $\chi^2$ for difference between PMDec for ascending and descending apparitions (1 degree of freedom) \\
 %ka & \nodata & int & i3 & Astrometry usage code. 0: neither ascending or descending apparitions provided a solution. 1: Only the ascending apparition provided a solution. 2: Only the descending apparition provided a solution. 3: Both ascending and descending apparitions provided a solution. \\
 %k1 & \nodata & int & i3 & W1 photometry usage code. 0: neither ascending or descending apparitions provided a solution. 1: Only the ascending apparition provided a solution. 2: Only the descending apparition provided a solution. 3: Both ascending and descending apparitions provided a solution. \\
 %k2 & \nodata & int & i3 & W2 photometry usage code. 0: neither ascending or descending apparitions provided a solution. 1: Only the ascending apparition provided a solution. 2: Only the descending apparition provided a solution. 3: Both ascending and descending apparitions provided a solution. \\
 %km & \nodata & int & i3 & Proper motion usage code. 0: neither ascending or descending apparitions provided a solution. 1: Only the ascending apparition provided a solution. 2: Only the descending apparition provided a solution. 3: Both ascending and descending apparitions provided a solution. \\
 par\_pm & arcsec & R*4 & f11.3 & Parallax estimate from motion solution. Computed using PMRA and PMDec to propagate ascending and descending motion solution positions to MJD 56700, then dividing ecliptic longitude difference by 2. Because scan longitudes on opposite sides of orbit had a $5\degree$ bias from a great circle until the end of 2014, gradually increasing to $18\degree$ by the end of 2017, the actual baseline is $< 2$au, declining from 1.996 to 1.951 au over this period, although the difference is negligible given the uncertainty in parallax. Will be null unless km = 3. \\
 %par\_pmSig & arcsec & R*4 & f11.3 & $1 \sigma$ uncertainty in par\_pm  \\
 par\_stat & arcsec & R*4 & f11.3 & Parallax estimate from stationary solution. Computed using PMRA and PMDec to propagate ascending stationary-solution position from ascending observation epoch to descending observation epoch, then dividing ecliptic longitude difference by 2 (see entry for par\_pm regarding dividing by 2). Will be null unless ka = 3, km $>0$, and all w?mjdmin/max/mean values are non-null in both ascending and descending WPHOT output. \\
 %par\_sigma & arcsec & R*4 & f11.3 & $1 \sigma$ uncertainty in par\_stat \\
 dist\_cc & arcsec & R*4 & f13.3 & Maximum distance between CatWISE and AllWISE sources providing cc\_flags. See \S\ref{sec:cc-flags}. In the NERSC release, this field is called dist\_x.\\
 cc\_flags & \nodata & char & a16 & Worst case character cc\_flag from AllWISE. See \S\ref{sec:cc-flags}.  \\
 w1cc\_map & \nodata & int & i13 & %w1cc\_map from AllWISE. 
 The description in the AllWISE Explanatory Supplement is incorrect, so an updated description is provided here. 
 %\newline 
 Worst case contamination and confusion map in W1 for matching AllWISE sources. Contains integer equivalent of 12-bit binary number specifying if W1 measurement in AllWISE is believed to be contaminated by or a spurious detection of an image artifact. Elements of binary array (with bit number indicated below each code) are given in Table \ref{tab:cc_map_bits}.
 %\newline
 Bits 0, 1, 3 and 4 indicate whether band-detection is contaminated by an artifact. If brightness of detection is less than expected for artifact, source is regarded as spurious and bits 7, 8, 10, and 11 are also set accordingly. Letters denote contamination by different types of artifacts (see \S\ref{sec:artifacts}).
 %\newline 
 For example, a measurement believed to be a spurious detection of a scattered light halo and contaminated by a diffraction spike has a binary bit map value of ``010000001001'' and w1cc\_map=1033. When there is more than one AllWISE match to a CatWISE source, a logical `OR' is performed over all the matches for each bit. \\
 w1cc\_map\_str & \nodata & char & a20 &  %w1cc\_map\_str from AllWISE. See \S\ref{sec:cc-flags}. 
 The w1cc\_map\_str description in the AllWISE Explanatory Supplement is incorrect, so an updated description is provided here.  
 %\newline 
 Worst case contamination and confusion string in W1 for matching AllWISE sources. This %column is a 
 character string denotes all artifacts that may contaminate the AllWISE W1 measurement of this source, in the priority order  D,d,P,p,H,h,O,o. 
%\newline 
For example, a real detection %that is 
contaminated by a diffraction spike and a latent (persistent) image has a w1cc\_map\_str=``dp''. A spurious detection of a diffraction spike %that is 
also contaminated by a latent image has w1cc\_map\_str=``Dp''.
This string is ``null'' if there are no artifacts that affect the measurement in this band.
The value that appears in first element of the cc\_flags string is the left-most character in w1cc\_map\_str. If w1cc\_map\_str is ``null'', then the corresponding cc\_flags entry is ``0''. \\
 w2cc\_map & \nodata & int & i13 & Worst case w2cc\_map from AllWISE. %The description in the AllWISE Explanatory Supplement \textbf{is} incorrect, 
 See w1cc\_map. \\
 %\newline Contamination and confusion map in W2 for matching AllWISE source. This column contains the integer equivalent of the 12-bit binary number that specifies if the W2 measurement in AllWISE is believed to be contaminated by or a spurious detection of an image artifact. The elements of the binary array (with the bit number indicated below each code) are given in Table \ref{tab:cc_map_bits}. 
 \newline
 %Bits 0, 1, 3 and 4 indicate whether the band-detection is contaminated by an artifact. If the brightness of the detection is less than expected for that type of artifact, the source is regarded as spurious and bits 7, 8, 10, and 11 are also set accordingly. The letters denote contamination by different types of artifacts (see \S\ref{sec:artifacts}).
 %\newline For example, a measurement that is believed to be a spurious detection of a scattered light halo and contaminated by a diffraction spike has a binary bit map value of ``010000001001'' and w2cc\_map=1033. When there is more than one AllWISE match to a CatWISE source, a logical `OR' is performed over all the matches for each bit.
% \\
 w2cc\_map\_str & \nodata & char & a20 & Worst case w2cc\_map\_str from AllWISE. 
 %See \S\ref{sec:artifacts}. The w2cc\_map\_str description in the AllWISE Explanatory Supplement \textbf{is} incorrect. 
 See w1cc\_map\_str.
% \newline Contamination and confusion string in W2 for matching AllWISE source. This column is a character string that denotes all artifacts that may contaminate the AllWISE W2 measurement of this source, in the priority order  D,d,P,p,H,h,O,o. 
%\newline For example, a real detection that is contaminated by a diffraction spike and a latent (persistent) image has a w2cc\_map\_str=``dp''. A spurious detection of a diffraction spike that is also contaminated by a latent image has w2cc\_map\_str=``Dp''.
%This string is ``null'' if there are no artifacts that affect the measurement in this band.
%The value that appears in first element of the cc\_flags string is the left-most character in w2cc\_map\_str. If w2cc\_map\_str is ``null'', then the corresponding cc\_flags entry is ``0''.
\\
% n\_aw & \nodata & int & i5 & Number of AllWISE matches within 2\farcs75. See \S\ref{sec:cc-flags}. \\
 ab\_flags & \nodata & char & a9 & Two character (W1 W2) artifact flag. See \S\ref{sec:ab-flags}. \\
 w1ab\_map & \nodata & int & i9 & W1 artifact code value. See \S\ref{sec:ab-flags}. \\
 w1ab\_map\_str & \nodata & char & a13 & W1 artifact string. See \S\ref{sec:ab-flags}. \\
 w2ab\_map & \nodata & int & i9 & W2 artifact code value. See \S\ref{sec:ab-flags}. \\
 w2ab\_map\_str & \nodata & char & a13 & W2 artifact string. See \S\ref{sec:ab-flags}. \\
 (glon) & deg & R*8 & f12.6 & Galactic longitude, calculated using a pole of $\alpha = 192.85\degree\ \delta=27.13\degree$, and a zero of longitude of $\alpha = 266.4\degree\ \delta=-28.94\degree$ (both in J2000). Hidden in IRSA. See Appendix \ref{sec:caveats} for details.   \\
 (glat) & deg & R*8 & f12.6 & Galactic latitude. 
 %calculated using a pole of $\alpha = 192.85\degree\ \delta=27.13\degree$, and a zero of longitude of $\alpha = 266.4\degree\ \delta=-28.94\degree$ (both in J2000). 
 Hidden in IRSA. See glon.   \\
 p & \nodata & int & i6 & Flag to indicate if source is ``primary" in tile. See \S\ref{sec:primary}. \\
\enddata
%\end{adjustbox}
\end{deluxetable*}
\end{longrotatetable}

\newpage
\begin{deluxetable*}{c c c c c c c c c c c c}
\tablecaption{cc\_map Bit Definitions\label{tab:cc_map_bits}}
\tablehead{
\colhead{O} &
\colhead{H} &
\colhead{0} &
\colhead{P} &
\colhead{D} &
\colhead{0} &
\colhead{0} &
\colhead{o} &
\colhead{h} &
\colhead{0} &
\colhead{p} &
\colhead{d}
}
\startdata
 11 & 10 & 9 & 8 & 7 & 6 & 5 & 4 & 3 & 2 & 1 & 0 
\enddata
\end{deluxetable*}

\section{Caveats \label{sec:caveats}}

The CatWISE Preliminary Catalog contains a number of features that users should be aware of. Among these are:
\begin{itemize}
\item The number of sources per square degree has relatively small variation over the sky (Figure \ref{fig:source_density}).
\item Catalog performance is less good in high source density regions (the Galactic plane and the ecliptic poles). Figures \ref{fig:map_positions} through \ref{fig:catwise_vs_gaia_pm} and Table \ref{table_ast_fields} illustrate this.
\end{itemize}
These issues appears to arise in the detection step. Relatively few deblended sources are added in the measurement step. We are using the unWISE Catalog \citep[which contains many more sources in high density regions;][]{Schlafly2019} as a detection list for an updated version of the CatWISE catalog that is expected to be available in 2020.
\begin{itemize}
\item The completeness and reliability for bright sources is low (Figure \ref{fig:brightCompleteness}). 
\end{itemize}
 As an example, the brightest two catalog sources in W2 (CWISEP J005153.01--235140.3 and CWISEP J223327.80+060246.2) are spurious, and it is likely that other spurious bright sources are present in the catalog. Users should visually inspect interesting sources selected from the catalog using AllWISE or unWISE images before devoting significant resources to follow up observations. 
 
Additional features present in the CatWISE Preliminary Catalog include:
\begin{itemize}
\item Tabulated position uncertainties are significantly smaller than measured position scatter with respect to {\it Gaia}, as illustrated in Figure \ref{fig:catwise_vs_gaia_fullsky} (upper right) and Figure \ref{fig:catwise_vs_gaia_pos} (right).

\item A floor of 10 mas yr$^{-1}$ was imposed on the tabulated motion uncertainties, making them significantly larger than the measured scatter with respect to {\it Gaia} motion, as illustrated in Figure \ref{fig:catwise_vs_gaia_fullsky} (lower right) and Figure \ref{fig:catwise_vs_gaia_pm} (right).

\item The Galactic coordinates in the CatWISE Preliminary Catalog were calculated using a pole of $\alpha = 192.85\degree\ \delta=27.13\degree$, and a zero of longitude of $\alpha = 266.4\degree\ \delta=-28.94\degree$ (both in J2000). These directions are not perpendicular by 22\farcs663, and consequently yielded ``nan" entries for the Galactic coordinates of the 31 catalog sources and 6 reject sources within 88\farcs3 of the north Galactic pole, and the 39 catalog sources and 5 reject sources within 87\farcs8 of the south Galactic pole. The discrepancy in the glon and glat Galactic coordinates of sources compared to coordinates calculated using the IAU-defined system \citep{Blaauw1960} increases with absolute Galactic latitude from $<1"$ in the Galactic plane to $88"$ near the Galactic poles. These glon and glat values (including the 81 with glat = ``nan") appear in the NERSC release of the CatWISE Preliminary catalog and reject table files. In the IRSA release, these columns are hidden, and in addition the 81 glat ``nan" values have been replaced with $+90\degree$ or $-90\degree$ as appropriate. 
\end{itemize}

\section{Combining Ascending and Descending Scan Positions \label{sec:mergepos}}

Here we provide additional details on how ascending and descending scan positions are combined for the CatWISE Preliminary Catalog. Positions are averaged using inverse-covariance weighting. The averaging is done in a local Cartesian projection consistent with the uncertainty representation. 

A transformation matrix T is defined as follows: starting with a Cartesian $(x,y,z)$ system whose z axis points to the celestial north pole and whose x axis points to the vernal equinox, we perform two Euler rotations that place the z' axis of the rotated system on the ascending celestial $(\alpha,\delta)$ position with the y' axis aligned with the local north-south direction, and the x' axis aligned with the local east-west direction.  First, rotate about the z axis by $\phi_1 = \alpha - 90^{\circ}$, then rotate about the x' axis by $\phi_2 =  \delta - 90^{\circ}$. 
Then T is given by:
 
\begin{equation}
    T
    \equiv
    \begin{bmatrix}
    T_{11} & T_{12} & T_{13} \\
    T_{21} & T_{22} & T_{23} \\
    T_{31} & T_{32} & T_{33}
    \end{bmatrix}
\end{equation}
with elements:
\begin{equation}
\begin{aligned}
   & T_{11} = \cos \phi_1 \\
   & T_{12} = \sin \phi_1 \\
   & T_{13} = 0 \\
   & T_{21} = -\cos \phi_2 \sin \phi_1 \\
   & T_{22} = \cos \phi_2  \cos \phi_1 \\
   & T_{23} = \sin \phi_2 \\
   & T_{31} = \sin \phi_2 \sin \phi_1 \\
   & T_{32} = -\sin \phi_2 \cos \phi_1 \\
   & T_{33} = \cos \phi_2 \\
\end{aligned}
\end{equation}

\noindent This corresponds to a z' axis that looks outward from the origin, so any nearby $\alpha', \delta'$ position will have an $(x',y',z')$ vector in the rotated system whose $z'$ component will be positive and close to 1. By ``nearby" we mean within a few arcseconds of the origin, since it is very rare for an extracted source's position to vary between the ascending and descending solutions by more than that. This justifies our Cartesian approximation.

The coordinate system is computed for the ascending position, so that the ascending position of the source has coordinates $(0,0,1)$.  The descending position $(\alpha',\delta')$ is mapped into that system as follows. We construct the vector v to the descending position in the original celestial coordinate system:
\begin{equation}
\begin{aligned}
   & v_1 = \cos \alpha' \cos \delta' \\
   & v_2 = \sin \alpha'  \cos \delta' \\
   & v_3 = \sin \delta' \\
\end{aligned}
\end{equation}

\noindent and transform it into the new system, 
\begin{equation}
(x',y',z')
     \equiv
     Tv
\end{equation}

\noindent wherein its $x'$ and $y'$ coordinates  are the offsets of the descending position from the ascending position. 

The $(x',y')$ coordinates are averaged with (0,0) using inverse covariance weighting. We construct $2\times2$ error covariance matrices $\Omega_a$ and $\Omega_d$ for the ascending and descending vectors using the corresponding $\sigma_\alpha$, $\sigma_\delta$, and $\sigma_{\alpha\delta}$: 
\begin{equation}
\begin{split}
    \Omega_a = 
    \begin{bmatrix} 
    \sigma^2_\alpha & \sigma_{\alpha\delta}|\sigma_{\alpha\delta}| \\
    \sigma_{\alpha\delta}|\sigma_{\alpha\delta}| & \sigma^2_\delta 
    \end{bmatrix}_a
\\
    \Omega_d = 
    \begin{bmatrix} 
    \sigma^2_\alpha & \sigma_{\alpha\delta}|\sigma_{\alpha\delta}| \\
    \sigma_{\alpha\delta}|\sigma_{\alpha\delta}| & \sigma^2_\delta 
    \end{bmatrix}_d
\end{split}    
\end{equation}

\noindent A minimum value of $10^{-8}$ is enforced for any zeroes on the diagonal. The merged (i.e., inverse-covariance-weighted average) vector ($x_m,y_m,z_m$) and associated covariance matrix are computed as follows: 
\begin{equation}
    W_a = \Omega_a^{-1}, \;\;
    W_d = \Omega_d^{-1}
\end{equation}    
\begin{equation}
    \Omega = (W_a+W_d)^{-1} \equiv
    \begin{bmatrix} 
    \Omega_{11} & \Omega_{12} \\
    \Omega_{21} & \Omega_{22}
    \end{bmatrix}
\end{equation}    
%\begin{equation}
\begin{gather}
    \begin{bmatrix}
    x_m \\
    y_m
    \end{bmatrix} =
    \Omega \left(W_a
    \begin{bmatrix}
    0 \\
    0
    \end{bmatrix} + W_d
    \begin{bmatrix}
    x' \\
    y'
    \end{bmatrix}
    \right) = \Omega\,W_d
    \begin{bmatrix}
    x' \\
    y'
    \end{bmatrix} \\
    \notag z_m = \sqrt{1 - x^2_m - y^2_m}
\end{gather}
%\end{equation}    
\begin{equation}
    \sigma_\alpha = \sqrt{\Omega_{11}} ,\;\;
    \sigma_\delta = \sqrt{\Omega_{22}} ,\;\;
    \sigma_{\alpha\delta} = {\rm sign}(\Omega_{12})\sqrt{|\Omega_{12}|}
\end{equation}

\noindent The celestial coordinates corresponding to the $(x_m,y_m,z_m)$ vector are obtained using the inverse of the transformation matrix T described above, which is the transpose because T is orthonormal:

\begin{equation}
    \begin{bmatrix}
    v'_1 \\
    v'_2 \\
    v'_3
    \end{bmatrix}
    \equiv
    \begin{bmatrix}
    T_{11} & T_{21} & T_{31} \\
    T_{12} & T_{22} & T_{32} \\
    T_{13} & T_{23} & T_{33}
    \end{bmatrix}
    \begin{bmatrix}
    x_m \\
    y_m \\
    z_m
    \end{bmatrix}
\end{equation}

\begin{equation}
    \alpha_m = \tan^{-1}{\left(\frac{v'_2}{v'_1}\right)}, \;\; \delta_m = \sin^{-1}{v'_3}
\end{equation}

\noindent The image pixel coordinates are then computed for the merged position using the coadd WCS information.


\begin{thebibliography}{}

\bibitem[Ashby et al.(2013)]{Ashby2013} Ashby, M.~L.~N., Stanford, S.~A., Brodwin, M., et al.\ 2013, \apjs, 209, 22.

\bibitem[Bakos et al.(2002)]{Bakos2002} Bakos, G. {\'A}., Sahu, K.~C., \& N{\'e}meth, P.\ 2002, \apjs, 141, 187

\bibitem[Ba{\~n}ados et al.(2018)]{Banados2018}
Ba{\~n}ados, E.,  Venemans, B.~P., Mazzucchelli, C., et al. \nat, 553, 473B.

\bibitem[Blaauw et al.(1960)]{Blaauw1960} Blaauw, A., Gum, C.~S., Pawsey, J.~L., et al.\ 1960, \mnras, 121, 123. 

\bibitem[Brodwin et al.(2013)]{Brodwin2013} Brodwin, M., Stanford, S.~A., Gonzalez, A.~H., et al.\ 2013, \apj, 779, 138.

\bibitem[Caselden et al.(2018)]{Caselden2018} Caselden, D., Westin, P., Meisner, A., et al.\ 2018, ASCL:1806.004.

\bibitem[Connors et al.(2011)]{Connors2011} Connors, M., Wiegert, P., \& Veillet, C.\ 2011, \nat, 475, 481 

\bibitem[Cutri et al.(2012)]{Cutri2012} Cutri, R.~M., Wright, E.~L., Conrow, T., et al.\ 2012, Explanatory Supplement to the WISE All-Sky Data Release Products, { http://wise2.ipac.caltech.edu/docs/release/ allsky/expsup}

\bibitem[Cutri et al.(2013)]{Cutri2013} Cutri, R.~M., Wright, E.~L., Conrow, T., et al.\ 2013, Explanatory Supplement to the AllWISE Data Release Products,  http://wise2.ipac.caltech.edu/docs/release/ allwise/expsup 

\bibitem[Cutri et al.(2015)]{Cutri2015} Cutri, R.~M., Mainzer, A., Conrow, T., et al.\ 2015, Explanatory Supplement to the NEOWISE Data Release Products,  http://wise2.ipac.caltech.edu/docs/release/ neowise/expsup

\bibitem[Gaia Collaboration et al.(2018)]{Brown2018} Gaia Collaboration, Brown, A.~G.~A., Vallenari, A., et al.\ 2018, \aap, 616, A1

\bibitem[Gelino et al.(2011)]{Gelino2011} Gelino, C.~R., Kirkpatrick, J.~D., Cushing, M.~C., et al.\ 2011, \aj, 142, 57

\bibitem[Gonzalez et al.(2019)]{Gonzalez2019} Gonzalez, A.~H., Gettings, D.~P., Brodwin, M., et al.\ 2019, \apjs, 240, 33.

\bibitem[H{\o}g et al.(2000)]{Hog2000} H{\o}g, E., Fabricius, C., Makarov, V.~V., et al.\ 2000, \aap, 355, L27

\bibitem[Jarrett et al.(2011)]{Jarrett2011} Jarrett, T.~H., Cohen, M., Masci, F., et al.\ 2011, \apj, 735, 112 

\bibitem[Kirkpatrick et al.(2011)]{Kirkpatrick2011} Kirkpatrick, J.~D., Cushing, M.~C., Gelino, C.~R., et al.\ 2011, \apjs, 197, 19

\bibitem[Kirkpatrick et al.(2014)]{Kirkpatrick2014} Kirkpatrick, J.~D., Schneider, A., Fajardo-Acosta, S., et al.\ 2014, \apj, 783, 122 

\bibitem[Kirkpatrick et al.(2019)]{Kirkpatrick2019} Kirkpatrick, J.~D., Martin, E.~C., Smart, R.~L., et al.\ 2019, \apjs, 240, 19

\bibitem[Laigle et al.(2016)]{Laigle2016} Laigle, C., McCracken, H.~J., Ilbert, )., et al.\ 2019, \apjs, 224, 24.

\bibitem[Lang(2014)]{Lang2014} Lang, D.\ 2014, \aj, 147, 108

\bibitem[Lang, Hogg, \& Schlegel(2016)]{Lang2016} Lang, D., Hogg, D.~W., \& Schlegel, D.~J.\ 2016, \aj, 151, 36 

\bibitem[Lazorenko \& Sahlmann(2018)]{Lazorenko2018} Lazorenko, P.~F., \& Sahlmann, J.\ 2018, \aap, 618, A111

\bibitem[Lindegren et al.(2018)]{Lindegren2018} Lindegren, L., Hern{\'a}ndez, J., Bombrun, A., et al.\ 2018, \aap, 616, A2 

\bibitem[Luhman(2013)]{Luhman2013} Luhman, K.~L.\ 2013, \apjl, 767, L1 

\bibitem[Luhman(2014)]{Luhman2014} Luhman, K.~L.\ 2014, \apjl, 786, L18 

\bibitem[Marsh \& Jarrett(2012)]{MarshJarrett2012} Marsh, K.~A., \& Jarrett, T.~H.\ 2012, \pasa, 29, 269 

\bibitem[Masci(2013)]{Masci2013} Masci, F.\ 2013, arXiv:1301.2718 

\bibitem[Mainzer et al.(2011)]{Mainzer2011} Mainzer, A., Bauer, J., Grav, T., et al.\ 2011, \apj, 731, 53 

\bibitem[Mainzer et al.(2014)]{Mainzer2014} Mainzer, A., Bauer, J., Cutri, R.~M., et al.\ 2014, \apj, 792, 30

\bibitem[Marocco et al.(2019)]{Marocco2019} Marocco, F., Caselden, D., Meisner, A.~M., et al.\ 2019, \apj, 881, 17

\bibitem[Marocco et al.(2020)]{Marocco2020} Marocco, F., Kirkpatrick, J.~D., Meisner, A.~M., et al.\ 2020, \apjl, 888, L19

\bibitem[Mauduit et al.(2012)]{Mauduit2012} Mauduit, J.-C., Lacy, M., Farrah, D., et al.\ 2012, \pasp, 124, 714

%\bibitem[Meisner et al.(2018a)]{Meisner2018a} Meisner, A.~M., Lang, D., \& Schlegel, D.~J.\ 2018a, \aj, 156, 69 
%NEO2 epoch coadds - full paper

\bibitem[Meisner et al.(2018a)]{Meisner2018a} Meisner, A.~M., Lang, D., \& Schlegel, D.~J.\ 2018a, RNAAS, 2, 1
%full depth NEO3 coadds

\bibitem[Meisner et al.(2018b)]{Meisner2018b} Meisner, A.~M., Cushing, M.~C., Cutri, R., et al.\ 2018b, RNAAS, 2, 140 
%A Secure W2 Detection of WD 0806-661B from CatWISE

\bibitem[Meisner et al.(2018c)]{Meisner2018c} Meisner, A.~M., Lang, D., \& Schlegel, D.~J.\ 2018c, RNAAS, 2, 202
%NEO3 epoch coadds

\bibitem[Meisner et al.(2019a)]{Meisner2019a} Meisner, A.~M., Lang, D., Schlafly, E.~F., et al.\ 2019a, \pasp, 131, 124504
%NEO4 epoch coadds and bitmasks

\bibitem[Meisner et al.(2019b)]{Meisner2019b} Meisner, A.~M., Caselden, D., Kirkpatrick, J.~D., et al.\ 2019b, \apj, in press (arxiv:1911.12372)
%Spitzer photometry of CatWISE

\bibitem[Sanders et al.(2007)]{Sanders2007} Sanders, D.~B., Salvato, M.,  Aussel, H., et al.\ 2007, \apjs, 172, 86 

\bibitem[Schlafly et al.(2019)]{Schlafly2019} Schlafly, E.~F., Meisner, A.~M., \& Green, G.~M.\ 2019, \apjs, 240, 30 

\bibitem[Scholz(2014)]{Scholz2014} Scholz, R.-D.\ 2014, \aap, 561, A113 

\bibitem[Skrutskie et al.(2006)]{Skrutskie2006} Skrutskie, M.~F., Cutri, R.~M., Stiening, R., et al.\ 2006, \aj, 131, 1163

\bibitem[Stauffer et al.(2010)]{Stauffer2010} Stauffer, J., Tanner, A.~M., Bryden, G., et al.\ 2010, \pasp, 122, 885

\bibitem[Tsai et al.(2015)]{Tsai2015} Tsai, C.-W., Eisenhardt, P.~R.~M., Wu, J., et al.\ 2015, \apj, 805, 90 

\bibitem[van Leeuwen(2007)]{vanLeeuwen2007} van Leeuwen, F.\ 2007, \aap, 474, 653

\bibitem[Wright et al.(2010)]{Wright2010} Wright, E.~L., Eisenhardt, P.~R.~M., Mainzer, A.~K., et al.\ 2010, \aj, 140, 1868 

\end{thebibliography}
\end{document}